\documentclass[11pt]{article}
\usepackage{amssymb}
\usepackage{amsfonts}
\usepackage{graphicx}
\usepackage{epstopdf}
\usepackage{dcolumn}
\usepackage{amsmath}
\usepackage{latexsym,bm}
\usepackage{slashed}
\usepackage{cite}
\usepackage{extarrows}
\usepackage{hyperref}

\def\hybrid{
        \topmargin -20pt
        \oddsidemargin 0pt
        \headheight 0pt \headsep 0pt
        \textwidth 6.25in 
        \textheight 9.5in 
        \marginparwidth .875in
        \parskip 5pt plus 1pt \jot = 1.5ex}

\hybrid

 \csname
@addtoreset\endcsname{equation}{section}

\def\moth{\mathsurround=0pt}
\newdimen\zo \zo=0pt

\def\tick{\leaders\hrule height 0.5ex depth 0pt \hskip 0.5pt}
\def\upboxfill{$\moth \setbox\zo\hbox{\tick}%
  \hskip 3pt\hbox to 0pt{$\tick$\hss}\hrulefill \hbox to 7.5pt{$\tick$\hss}$}

\def\dtick{\leaders\hrule height .34pt depth 0.5ex \hskip 0.5pt}
\def\downboxfill{$\moth \setbox\zo\hbox{\dtick}%
  \hskip 2pt\hbox to 0pt{$\dtick$\hss}\hrulefill \hbox to 2pt{$\dtick$\hss}$}

\def \be {\begin{equation}}
\def \ee {\end{equation}}
\def \bsp {\begin{split}}
\def \esp {\end{split}}
\def \bea {\begin{eqnarray}}
\def \eea {\end{eqnarray}}

\def\mc{\mathcal}
\def\ptl{\partial}

\newcommand{\fracs}[2]{{\textstyle{#1\over #2}}}

\def\bec{\begin{center}}
\def\ec{\end{center}}

 \def\det{{\rm det\,}}
\def\be{\begin{equation}}
\def\ee{\end{equation}}
\def\bea{\begin{eqnarray}}
\def\eea{\end{eqnarray}}
\def\ba{\begin{array}}
\def\ea{\end{array}}

\begin{document}

\begin{titlepage}
\rightline{}
\rightline{\tt  MIT-CTP-4634}
\rightline{January 2015}
\begin{center}
\vskip .6cm
{\Large \bf Tensor Hierarchy and Generalized Cartan Calculus \\[1.5ex]
in {SL(3)$ \, \times\;$SL(2) 
Exceptional Field Theory}}\\
\vskip 2cm
{\large {Olaf Hohm and Yi-Nan Wang}}
\vskip 1cm
{\it {Center for Theoretical Physics}}\\
{\it {Massachusetts Institute of Technology}}\\
{\it {Cambridge, MA 02139, USA}}\\
ohohm@mit.edu, wangyn@mit.edu

\vskip 1.5cm
{\bf Abstract}
\end{center}

\vskip 0.2cm

\noindent
\begin{narrower}
We construct exceptional field theory for the duality group SL$(3)\times {\rm SL}(2)$.
The theory is defined on a space with 8 `external' coordinates and 
6 `internal' coordinates  in the $(3,2)$ fundamental representation, 
leading to a 14-dimensional generalized spacetime. 
The bosonic theory is uniquely determined by gauge invariance under 
generalized external and internal diffeomorphisms. The latter invariance can be made 
manifest by introducing higher form gauge fields and 
a so-called tensor hierarchy,  
which we  systematically develop to much higher degree than in previous
studies. To this end we introduce a novel
Cartan-like tensor calculus based on a covariant nil-potent differential, 
generalizing the exterior derivative of conventional differential geometry. 
The theory encodes the full 
$D=11$ or type IIB supergravity, respectively.

\end{narrower}

\end{titlepage}

\newpage

\tableofcontents

\newpage

\section{Introduction}

Exceptional field theory is a framework that simultaneously makes manifest the duality symmetries
of M-theory and type IIB \textit{prior} to toroidal compactification. The theory is formulated
on an extended generalized spacetime, with fields depending on coordinates transforming in a fundamental
representation of the duality group, subject to a `section constraint' or `strong constraint' that effectively
reduces the generalized spacetime to a `physical slice'. Extending the geometrical concepts of double field theory (DFT)
 \cite{Siegel:1993th,Hull:2009mi,Hull:2009zb,Hohm:2010jy,Hohm:2010pp,Hohm:2010xe}, 
exceptional field theory (EFT)
was constructed in \cite{Exceptional,E6,E7,E8,SUSYE7,Musaev:2014lna}, based on important earlier work in \cite{deWit:1986mz,Koepsell:2000xg,deWit:2000wu,West:2001as,HenryLabordere:2002dk,Damour:2002cu,West:2003fc,Hull:2007zu,Hillmann:2009ci,Berman:2010is,Berman:2011pe,Berman:2011jh,Coimbra:2011ky,Coimbra:2012af,Berman:2012vc,Cederwall:2013oaa,Hohm:2013nja,UDuality}.

While DFT encodes the fully doubled spacetime coordinates in
an $O(10,10)$ vector, the formulation of EFT is based on a split of the coordinates into `external' and
`internal' directions and a corresponding decomposition of the tensor fields, as in Kaluza-Klein compactifications. 
We stress, however, that this does not entail any truncation nor an assumption on the topology of a background. 
 After solving the section constraint, the coordinate dependence of the fields is not further 
 constrained, and therefore these theories still
 encode, in particular, the complete $D=11$ supergravity.
 Focusing on the `purely internal' components of the metric and 3-form
 of $D=11$ supergravity, it was shown 
 in \cite{Hillmann:2009ci,Berman:2010is,Berman:2011pe,Berman:2011jh,Coimbra:2011ky,Coimbra:2012af,Berman:2012vc} 
 how to combine these fields into generalized
 metrics or vielbeins that are covariant tensors under the corresponding duality group and how
 to construct actions for this subsector that are invariant under suitably generalized gauge transformations.
 Going beyond this truncation, the full EFT encodes also
 external and off-diagonal field components, as Kaluza-Klein vectors, etc., which together with their
 on-shell dual fields play an important role
 in describing the full dynamics in a duality covariant way.
 The appropriate mathematical framework
 is a generalization of the so-called `tensor hierarchy' developed in gauged supergravity 
 \cite{deWit:2005hv,deWit:2008ta}.
 It provides a generalization of Yang-Mills theory in which the gauge algebra is not governed
 by proper Lie brackets. It is based on brackets that violate the Jacobi identity in a certain `exact' way.
 In order to construct gauge covariant curvatures for the gauge fields it is then necessary to introduce
 higher $p$-form potentials in a hierarchical manner.  So far, EFTs with duality groups
 E$_{d(d)}$ have been constructed explicitly for $d=6,7,8$, in which
 case the tensor hierarchy needed for constructing an action is rather short: it ends with
 the 2-forms for $d=6,7$ and with 1-forms for $d=8$.\footnote{Moreover, for the $O(d,d)$ group of  
 DFT the tensor hierarchy ending with 2-forms is exact \cite{Hohm:2013nja}.} In this paper, we investigate the case of
 a smaller duality groups E$_{d(d)}$ (which reduce to classical Lie groups). This forces
 us to go much higher in the hierarchy and provides therefore an opportunity to investigate
 the geometrical structure of tensor hierarchies in EFT.

The smallest U-duality group is SL$(2,\mathbb{R})\times \mathbb{R}^+$ appearing for reductions 
to $D=9$, but here we
investigate the $D=8$ case, for which the duality group is
 \be\label{GIntro}
  {\rm G} \ = \ {\rm SL}(3,\mathbb{R})\times {\rm SL}(2,\mathbb{R})\;.
 \ee
(In the following we will usually mean the real continuous form of these groups, unless indicated
otherwise.)
The EFT fields in this case depend on $8$ external coordinates $x^{\mu}$ and extended
internal coordinates $Y^M$, where $M,N,\ldots=1,\ldots, 6$
label the $(3,2)$ representation of SL(3)$\times {\rm SL}(2)$.  The theory is thus defined
in $8+6=14$ dimensions, but all fields are subject to a section constraint implying that they depend
only on a subset of these coordinates.  Denoting fundamental SL$(3)$ indices by $i,j,\ldots,=1,2,3$ and 
fundamentla SL$(2)$ indices by $\alpha,\beta,\ldots=1,2$, the coordinates are $Y^M=Y^{i\alpha}$, with
conjugate derivatives $\partial_M=\partial_{i\alpha}$. The section constraint then reads
  \be\label{SectionstrIntro}
\epsilon^{ijk}\epsilon^{\alpha\beta}\ptl_{i\alpha}\otimes\ptl_{j\beta} \ = \ 0\;,
\ee
with the SL$(3)$ and SL$(2)$ invariant epsilon symbols $\epsilon^{ijk}$ and $\epsilon^{\alpha\beta}$,
respectively. This constraint projects out the $(3,1)$ sub-representation in the tensor product
$(3,2)\otimes (3,2)$ given by $\ptl_{i\alpha}\otimes\ptl_{j\beta}$. This means that quadratic derivatives as in (\ref{SectionstrIntro}), acting on arbitrary objects, are consistently set to zero,
in particular  $\epsilon^{ijk}\epsilon^{\alpha\beta}\ptl_{i\alpha}A\,\ptl_{j\beta}B=0$
for any fields and gauge parameters $A,B$. While somewhat unconventional,
the use of fields depending on extended coordinates subject to a section constraint in this way is well motivated
by string theory: in string field theory on toroidal backgrounds, the string field depends both on
momentum and winding coordinates, transforming covariantly under the T-duality group, subject
to the level-matching constraint. The section constraint above is a natural extension of this constraint.
In fact, the duality group G in (\ref{GIntro}) contains the subgroup
 \be
  {\rm G} \, \supset \, {\rm SL}(2,\mathbb{R})\times {\rm SL}(2,\mathbb{R}) \ \cong \ {\rm SO}(2,2)\;,
 \ee
which is the T-duality group of string theory on a 2-torus.  It is easy to see (and will be displayed in
the main text) that reducing the duality group accordingly by eliminating the dependence on two
of the six coordinates, the constraint (\ref{SectionstrIntro}) reduces to $\tilde{\partial}\cdot\partial=0$,
with momentum derivatives $\partial$ and winding derivatives $\tilde{\partial}$, which is the
strong form of the level-matching constraint for the massless string fields.

The SL(3)$\times {\rm SL}(2)$ covariant section constraint (\ref{SectionstrIntro}) can be naturally solved
in order to obtain $D=11$ supergravity. If we pick a particular SL$(2)$ direction, say $1$, the constraint
is solved by fields depending only on the three coordinates $y^i\equiv Y^{i1}$.
This gives rise to a theory with eleven coordinates $(x^{\mu},y^i)$ that is on-shell fully
equivalent to $D=11$ supergravity. Intriguingly, as pointed out in \cite{Exceptional} and
in analogy to type II DFT \cite{Hohm:2011zr,Hohm:2011dv}, this constraint allows for a second, inequivalent solution.
Picking now a particular SL$(3)$ direction, say $1$, the constraint is also solved
by fields depending only on the two coordinates $y^{\alpha}\equiv Y^{1\alpha}$.
This leads to a theory in $10=8+2$ dimensions
that is on-shell equivalent to type IIB supergravity. It has an unbroken SL$(2)\times {\rm SL}(2)$ symmetry,
whose first factor is the S-duality group of type IIB and the second factor is the surviving subgroup of
the internal diffeomorphisms.
In this sense the EFT unifies
M-theory and type IIB, thereby geometrizing the S-duality group of type IIB.
It is thus tempting to interpret EFT as an implementation of F-theory. 
We will comment on such an interpretation in the main text.

\medskip

\textit{Summary of results:}
As the results of this paper are somewhat technical, for the 
reader's convenience we summarize here the main results and our notation. The generalized diffeomorphisms of 
the internal space (coordinatized by the six $Y^M$) are infinitesimally 
given by generalized Lie derivatives $\mathbb{L}_{\Lambda}$ w.r.t.~a parameter   
$\Lambda^M(x,Y)$, acting on a generic SL(3)$\times {\rm SL}(2)$ tensor $V$, 
 \be\label{genLieTRANS}
  \delta_{\Lambda}V \ = \ \mathbb{L}_{\Lambda}V\;. 
 \ee 
The explicit action of $\mathbb{L}_{\Lambda}$ will be given in sec.~2. Let us stress 
that generally $V$ carries a non-trivial density weight, denoted by $\lambda(V)$, 
entering the Lie derivative via the term $\lambda\partial_N\Lambda^N V$. 
The crucial property of the generalized Lie derivative is that it preserves the 
structure of the U-duality group. For instance, the internal (generalized) metric 
encodes a field ${\cal M}_{ij}\in {\rm SL}(3)$, satisfying $\det {\cal M}=1$, 
which constraint is invariant under generalized Lie derivatives
 for ${\cal M}$ carrying 
density weight zero. This is in contrast to conventional geometry, 
where the determinant (the volume) necessarily enters as an independent 
degree of freedom. 

The Lie derivative is defined on arbitrary SL(3)$\times {\rm SL}(2)$ representations, 
in such a way that it is compatible with the natural algebraic tensor operations 
that relate different representations. For instance, denoting the space of (3,2) tensors $A^{m\alpha}$
 with weight $\lambda$ as $\mathfrak{A}(\lambda)$ and the corresponding 
space of (3,1) tensors $B_m$ as $\mathfrak{B}(\lambda)$, we can define 
 \be
  \bullet:\; \mathfrak{A}(\lambda_1)\times\mathfrak{A}(\lambda_2)\, \rightarrow\,
   \mathfrak{B}(\lambda_1+\lambda_2)\;:\qquad
    (A_1\bullet A_2)_m \ \equiv \ \epsilon_{ijm}\epsilon_{\alpha\beta}A_1^{i\alpha}A_2^{j\beta}\,, 
 \ee  
and similarly for other representations. 
The Lie derivative then satisfies the Leibniz property 
\be\label{distribution}
 \mathbb{L}_\Lambda(V\bullet W) \ = \ (\mathbb{L}_\Lambda V)\bullet W +
 V\bullet(\mathbb{L}_\Lambda W)\;, 
\ee
for arbitrary tensors $V,W$. 

Most importantly, we will also develop a differential calculus that makes the construction 
of the tensor hierarchy feasible. Given a generalized tensor that transforms as 
(\ref{genLieTRANS}), in general its partial derivative will not transform covariantly 
(i.e.~with the Lie derivative). For tensors in specific representations and with specific density weights, 
however, there are certain projections of the derivative that do transform covariantly. 
For instance, we define a differential operator acting between the following spaces 
 \be
  \widehat\partial\,:\;\; \mathfrak{B}(\fracs{1}{3})\;\, \longrightarrow\;\, \mathfrak{A}(\fracs{1}{6})\;,\qquad
    (\widehat\partial B)^{i\alpha} \ \equiv \ \epsilon^{ijk}\epsilon^{\alpha\beta}\ptl_{j\beta}B_k\;.
 \ee
We will prove that $\widehat\partial B$ indeed transforms covariantly 
as a tensor of weight $\frac{1}{6}$. This proof uses in an essential 
way the section constraint (\ref{SectionstrIntro}) and the precise weight of $B_m$. 
More generally, we will define the action of $\widehat\partial$ on an entire chain 
of representation spaces with definite weights, acting as 
\be\label{partialHIERINTRO}
\mathfrak{A}(\fracs{1}{6})\xlongleftarrow{\widehat\partial}\mathfrak{B}(\fracs{1}{3})\xlongleftarrow{\widehat\partial}\mathfrak{C}(\fracs{1}{2})\xlongleftarrow{\widehat\partial}\mathfrak{D}(\fracs{2}{3})
\xlongleftarrow{\widehat\partial}\mathfrak{E}(\fracs{5}{6})
\;, 
\ee
where the definition of the additional tensor spaces will be given in sec.~2.3. 
Note that the arrows indicate descending density weights: the action of $\widehat\partial$ 
decreases the weight by $\frac{1}{6}$.  
A crucial property of $\widehat\partial$ is that it squares to zero, 
 \be
  \widehat\partial\,^2 \ \equiv \ \widehat\partial \, \circ \,\widehat\partial \ = \ 0\;, 
 \ee 
again as a consequence of the section constraint (\ref{SectionstrIntro}). 
An intriguing feature of this calculus is that the generalized Lie derivative,  
acting on tensors in the above spaces, can be expressed in terms of 
$\bullet$ and $\widehat\partial$ as follows
 \be\label{MAGICintro}
  \mathbb{L}_{\Lambda}V \ = \ \Lambda\bullet \widehat\partial V+\widehat\partial(\Lambda\bullet V)\;, 
 \ee
where the gauge parameter $\Lambda$ takes values in $\mathfrak{A}(\fracs{1}{6})$. 
We will see that a gauge parameter that is $\widehat\partial$ exact, i.e., 
$\Lambda=\widehat\partial\chi$, is trivial in the sense that the corresponding Lie derivative 
acts trivially on fields, $\mathbb{L}_{\widehat\partial\chi}=0$.  This is important because 
although the generalized Lie derivatives close on fields satisfying the section constraint
according to an antisymmetric bracket (`the E-bracket'), 
$\big[\mathbb{L}_{\Lambda_1},\mathbb{L}_{\Lambda_2}\big]=\mathbb{L}_{[\Lambda_1,\Lambda_2]_{\rm E}}$, 
this bracket does not define a Lie algebra in that the Jacobi identity is violated.  
The non-vanishing `Jacobiator', however, is $\widehat\partial$ exact: for $U,V,W\in\mathfrak{A}(\fracs{1}{6})$ 
one finds 
 \be
  \big[\big[U,V\big]_{\rm E},W\big]_{\rm E} + {\rm cycl.} 
  \ = \ \tfrac{1}{6}\;
  \widehat\partial\Big(\big[U,V\big]_{\rm E}\bullet W+{\rm cycl.} \Big) \;. 
 \ee
Hence the Jacobiator does not act on fields and so the symmetry variations $\delta_{\Lambda}$
do satisfy the Jacobi identity, as it should be. 

The striking similarity between the above calculus and the conventional Cartan calculus of differential forms
should be evident:\footnote{`Cartan calculus' denotes the formalism in differential geometry 
with exterior derivative d, Lie derivative ${\cal L}_X$ and contraction operator $i_{X}$, see 
 {\tt http://planetmath.org/cartancalculus}, which here all have direct analogs.} 
The operator $\widehat\partial$ is the analogue of the differential d acting (covariantly) 
on forms, which maps 
spaces $\Lambda^pT^*$ into $\Lambda^{p+1}T^*$ (the form degree is the analogue of the density weight) 
 and also squares to zero, d$^2=0$. 
It should be noted that this property of d (as well as its covariance) 
are consequences of $\partial_{[m}\,\partial_{n]}=0$ satisfied by conventional partial derivatives 
acting on sufficiently smooth functions. Thus, $\partial_{[m}\,\partial_{n]}=0$ 
is the analogue of the section constraint (\ref{SectionstrIntro}). 
The latter is much stronger, of course, in that it is symmetric in the derivatives and hence needs 
to be imposed on products by hand. 
Finally, the formula (\ref{MAGICintro}) is the analogue of Cartan's `magic formula' 
${\cal L}_{\Lambda}=i_{\Lambda}{\rm d}+{\rm d}\,i_{\Lambda}$, with ${\cal L}_{\Lambda}$ 
the conventional Lie derivative and $i_{\Lambda}$ the contraction of a form with the vector $\Lambda$, 
which is the analogue of the action by $\Lambda\, \bullet$\,.

With the above calculus the tensor hierarchy enters very naturally as follows. 
We introduce gauge fields  $A^{(1)}\in \mathfrak{A}(\fracs{1}{6})$, which are one-forms w.r.t.~the \textit{external}
8-dimensional space, in order to define external covariant derivatives  
${\cal D}\equiv {\rm d}-\mathbb{L}_{A^{(1)}}$. These are covariant 
under generalized Lie derivatives with parameters $\Lambda=\Lambda(x,Y)$. 
Next we define a covariant 2-form curvature ${\cal F}^{(2)}\in \mathfrak{A}(\fracs{1}{6})$ 
for the gauge vectors,\footnote{In the introduction we employ differential 
form notation in order not to clutter the equations. More explicit formulas will be given in the main text. 
Here, for instance,  
$A^{(1)}\wedge_{\rm E}A^{(1)}\equiv \frac{1}{2}\big[A_{\mu},A_{\nu}\big]_{\rm E}{\rm d}x^{\mu}\wedge {\rm d}x^{\nu}$.} 
 \be
  {\cal F}^{(2)} \ \equiv \ {\rm d}A^{(1)}-A^{(1)}\wedge_{\rm E}A^{(1)}
  +\widehat\partial B^{(2)}\;. 
 \ee 
Here we introduced a 2-form $B^{(2)}\in  \mathfrak{B}(\fracs{1}{3})$, which is needed in order 
to make this curvature gauge covariant. Indeed, the first part looks formally like the 
curvature of a Yang-Mills connection, but since the underlying E-bracket leads to 
a non-vanishing Jacobiator (that is $\widehat\partial$ exact) we have to add the 
2-form potential and assign to it suitable gauge transformations. 
Moreover, there is an additional redundancy in the above definition, corresponding to 
the new (one-form) gauge symmetry associated with $B^{(2)}$. 
This scheme can be continued, defining next a curvature ${\cal H}^{(3)}\in  \mathfrak{B}(\fracs{1}{3})$
for $B^{(2)}$, 
 \be\label{H3Intro}
  {\cal H}^{(3)} \ \equiv \ {\cal D}B^{(2)}-\omega_{\rm CS}^{(3)}(A)+\widehat\partial C^{(3)}\;, 
 \ee
with a newly introduced 3-form $C^{(3)}$ that according to (\ref{partialHIERINTRO}) 
takes values in $\mathfrak{C}(\fracs{1}{2})$, and $\omega_{\rm CS}^{(3)}$ denotes 
the non-abelian Chern-Simons 3-form of $A^{(1)}$ (but based on the E-bracket 
rather than the Lie bracket). These curvatures satisfy the non-trivial Bianchi identity 
 \be
  {\cal D}{\cal F}^{(2)} \ = \ \widehat\partial{\cal H}^{(3)}\;. 
 \ee 
This identity determines the form of ${\cal H}^{(3)}$ in (\ref{H3Intro}), but only up 
to $\widehat\partial$ closed terms. Next we may introduce curvatures 
${\cal J}^{(4)}\in \mathfrak{C}(\fracs{1}{2})$ for the 3-form, which in turn requires the 
introduction of a 4-form, whose curvature we denote by ${\cal K}^{(5)}\in  \mathfrak{D}(\fracs{2}{3})$. 
They satisfy Bianchi identities 
 \be
  \begin{split}
   {\cal D}{\cal H}^{(3)}+\frac{1}{2}{\cal F}^{(2)}\wedge \bullet\, {\cal F}^{(2)} \ &= \ 
   \widehat\partial{\cal J}^{(4)}\;, \\
   {\cal D}{\cal J}^{(4)}+{\cal F}^{(2)}\wedge\bullet\,  {\cal H}^{(3)} \ &= \ \widehat\partial{\cal K}^{(5)}\;. 
  \end{split}
 \ee  
It should be evident from these relations that the consistency of the full theory 
hinges on the precise interplay between the exterior derivative d of the external space, 
which raises the form degree, and the derivative $\widehat\partial$ of  the internal generalized space, 
which lowers the density weight.  
We could continue the construction of the hierarchy further but for the purposes 
of the SL$(3)\times {\rm SL}(2)$ EFT it is sufficient to stop here. 

We can now give the full EFT in a form that is manifestly invariant under the internal generalized diffeomorphisms 
and the higher $p$-form gauge symmetries emerging in the hierarchy. The bosonic fields 
are the 8-dimensional metric $g_{\mu\nu}$, the internal 6-dimensional generalized metric 
${\cal M}_{MN}\equiv {\cal M}_{ij}{\cal M}_{\alpha\beta}$, 
and $p$-forms with $p=1,\ldots, 5$  
in the representation spaces 
(\ref{partialHIERINTRO}) (although, as we shall discuss below, the 5-form is not strictly needed). 
The dynamics is then encoded in the (pseudo-)action
 \be\label{actionIntro}
  \begin{split}
   S \ = \ \int {\rm d}^6Y\,\Big[\,&{\rm d}^8x\,\sqrt{g}\,\Big(\widehat{R}
   +{\cal R}({\cal M},g)+\tfrac{1}{4}{\cal M}^{MN}\nabla_Mg^{\mu\nu}\nabla_{N}g_{\mu\nu}\Big) \\[0.5ex]
  &- \tfrac{1}{4}\, \mc{D}\mc{M}^{ij}\wedge\star \mc{D}\mc{M}_{ij}
  -\tfrac{1}{4}\, \mc{D}\mc{M}^{\alpha\beta}\wedge\star \mc{D}\mc{M}_{\alpha\beta}\\[1ex]
  &+\tfrac{1}{2}\,{\cal M}_{MN}\,{\cal F}^{(2)M}\wedge\star {\cal F}^{(2)N}
  +\tfrac{1}{2}\,{\cal M}^{mn}\,{\cal H}^{(3)}_m\wedge\star {\cal H}^{(3)}_n\\[1ex]
  &+\tfrac{1}{4}\,{\cal M}_{\alpha\beta}\,{\cal J}^{(4)\alpha}\wedge \star {\cal J}^{(4)\beta}+{\cal L}_{\rm top}\, \Big]\;. 
  \end{split}
 \ee  
In the first line, $\widehat{R}$ denotes the covariantized Ricci scalar for $g_{\mu\nu}$, ${\cal R}$ 
is a generalized Ricci scalar for the generalized metric ${\cal M}_{MN}$, which necessarily 
also depends on $\det g$, and $\nabla_M$ denotes a covariant internal derivative. 
In the last line, ${\cal L}_{\rm top}$ denotes a topological Chern-Simons-like action, 
whose precise form is defined in (\ref{topological}) below. This term is needed for consistency
with the self-duality constraint on the 3-form 
 \be
  {\cal M}_{\alpha\beta}{\cal J}^{(4)\beta} \ = \ -\epsilon_{\alpha\beta}\star {\cal J}^{(4)\beta}\;, 
 \ee
which has to be imposed by hand after varying the action.   
In the above action every term is manifestly gauge invariant under the internal 
generalized diffeomorphisms and  their higher-form descendants in the 
tensor hierarchy, being written in terms of the covariant derivatives and curvatures discussed above. 
Importantly, the theory is also invariant under external 8-dimensional diffeomorphisms 
generated by a parameter $\xi^{\mu}(x,Y)$. Unless $\xi^{\mu}$ is independent of $Y$
this symmetry is not manifest but rather relates all terms in the action. 
In fact, the bosonic theory is completely fixed by the invariance under combined (in total 14-dimensional) 
external and internal generalized diffeomorphisms. 
The above theory takes the structural form of 8-dimensional gauged supergravity,  
but we stress again that the non-abelian gauge structure encodes the additional coordinate dependence. 
Upon picking one of the solutions of the section constraint discussed above, this reduces 
to a theory that is on-shell fully equivalent to $D=11$ or type IIB supergravity \textit{without}
any compactification and/or truncation.

\medskip

This rest of this paper is organized as follows. In sec.~2 we develop the generalized Lie derivatives
acting on fields living in some tensor product power
of the fundamental representation $(3,2)$. Due to the product structure of the duality group,
we also need Lie derivatives acting on pure SL$(3)$ or pure SL$(2)$ tensors.
In particular, we develop the tensor or Cartan calculus that relates tensor
fields in different representations in a covariant manner. In sec.~3 we apply these results
by developing the tensor hierarchy including forms up to degree 5.
With these results we are ready in sec.~4 to define the EFT dynamics and to prove
gauge invariance under generalized internal diffeomorphisms.
In sec.~5 we will discuss the gauge structure and invariance under the 
external diffeomorphisms in somewhat more detail than in previous papers. 
In particular, we will discuss the gauge algebra which becomes field-dependent. 
In sec.~6 we outline the explicit embedding of $D=11$ supergravity and type IIB.
We conclude with an outlook in sec.~7.

\section{Generalized Lie derivatives and gauge algebra}
In this section we define the generalized Lie derivatives governing generalized (internal)
diffeomorphisms and their `E-bracket' gauge algebra. In the first subsection this will be done for fields in the fundamental
$(3,2)$ representation, while in the second subsection we specialize to tensors in smaller
representations, which is a novelty possible due the product structure of the duality group.
In the third subsection, we develop a new tensor calculus.

\subsection{Tensors in fundamental representation}
We begin by summarizing some aspects of the section constraint  (\ref{SectionstrIntro}). In general,
the second-order derivatives $\ptl_{i\alpha}\otimes\ptl_{j\beta}$, where the tensor product sign
indicates that the partial derivatives may act on different arbitrary objects, lives in the
tensor product
 \be
  (3,2)\otimes (3,2) \ = \
  \big[\, (3,3)\oplus (6,1)\,\big]_{\rm{anti}}\quad \oplus \quad \big[\,(6,3)+\underline{(3,1)}\,\big]_{\rm{sym}}\;,
 \ee
where we indicated the symmetric and anti-symmetric parts. Here $6$ denotes the symmetric
SL$(3)$ representation labeled by $(ij)$, and $3$ denotes the symmetric SL$(2)$ representation
labeled by $(\alpha\beta)$. The underlined representation $(3,1)$ is set to zero 
by the section constraint (\ref{SectionstrIntro}). Explicitly, the section constraint simply amounts to
 \be\label{EXPLSEc}
  \partial_{i\alpha}\otimes \partial_{j\beta}-\partial_{j\alpha}\otimes \partial_{i\beta}
  -\partial_{i\beta}\otimes \partial_{j\alpha}
  +\partial_{j\beta}\otimes \partial_{i\alpha} \ = \ 0\;,
 \ee
setting to zero the projection antisymmetric both in $i,j$ and $\alpha,\beta$.
Note that when acting on a single object the constraint simplifies,
 \be\label{simplifiedSEction}
  \epsilon^{\alpha\beta}\partial_{i\alpha}\partial_{j\beta}A \ = \ 0\;, \qquad
  \epsilon^{kij}\partial_{i\alpha}\partial_{j\beta}A \ = \ 0\;,
 \ee
because by the commutativity of partial derivatives antisymmetry in one index type
implies antisymmetry in the other.

For completeness let us show that the section constraint (\ref{SectionstrIntro}) reduces to
the strong constraint (i.e.~the stronger version of the level-matching constraint in DFT) upon 
reducing the U-duality group to the corresponding T-duality group. 
To this end we split the SL(3) index as $i=(i',3)$ and then drop the 
dependence on the two coordinates $Y^{3\alpha}$. The remaining 
four coordinates $Y^{M'}\equiv Y^{i'\alpha}$ then live in the vector representation 
of the surviving group ${\rm SO}(2,2)={\rm SL}(2)\times {\rm SL}(2)$, 
which is the T-duality group for compactification on a 2-torus. 
The ${\rm SO}(2,2)$ invariant metric is given by 
 \be 
  \eta^{M'N'} \ \equiv  \ \eta^{i'\alpha, j'\beta} \ \equiv \ \epsilon^{i'j'}\epsilon^{\alpha\beta}\;, 
 \ee
so that the section constraint (\ref{SectionstrIntro})  reduces to 
 \be
  \eta^{M'N'}\partial_{M'}\otimes \partial_{N'} \ = \ 0\;, 
 \ee
which indeed is the strong constraint in DFT, as we wanted to show.

We now turn to the definition of generalized Lie derivatives  $\mathbb{L}_{\Lambda}$
that govern internal generalized diffeomorphisms
generated by a gauge parameter $\Lambda^M=\Lambda^{i\alpha}$ in the $(3,2)$ representation.
Generalized Lie derivatives are defined in analogy to standard Lie derivatives, with the crucial difference
that they preserve the group structure, say of the generalized metric ${\cal M}\in {\rm G}$ to be used below.
This is achieved by defining the Lie derivative so that it contains a projector onto the adjoint representation
\cite{Coimbra:2011ky,Berman:2012vc}
 \be\label{Liedecompose}
  \mathfrak{g} \ \cong \ {\bf (8,1)}\oplus {\bf (1,3)}\;.
 \ee
  A novel feature here is that due to the product structure of the duality group
the adjoint decomposes into two sub-representations, whose contributions a priory could appear with
independent coefficients.
Acting on a vector, i.e., a tensor in the $(3,2)$ representation, the generalized Lie derivative is given by
\be\label{projlie}
  \mathbb{L}_{\Lambda} V^M \ = \ \Lambda^N\partial_NV^M
   -2  (\mathbb{P}_{(8,1)})^{M}{}_{N}{}^{P}{}_{Q}\partial_P\Lambda^Q V^N
   -3 (\mathbb{P}_{(1,3)})^{M}{}_{N}{}^{P}{}_{Q}\partial_P\Lambda^Q V^N
   +\lambda \,\partial_P\Lambda^P V^M\;,
 \ee
where $ \mathbb{P}_{(8,1)}$ and $\mathbb{P}_{(1,3)}$ are the projectors corresponding to (\ref{Liedecompose}).
Moreover, we included an arbitrary density weight term proportional to $\lambda$.
The projectors are given by\footnote{In order to verify this explicitly, write
the SL$(3)$ and SL$(2)$ generators in the $(3,2)$ representation as \be\label{generators}
 \begin{split}
  (t_{i}{}^{j})_M{}^{N} \ = \ (t_{i}{}^{j})_{k\alpha}{}^{l\beta} \ = \ \delta_{\alpha}^{\beta}
  \Big(\delta_i^l\delta_k^j-\frac{1}{3}\delta_{i}^j\delta^l_k\Big)\;,\quad
   (t_{\alpha}{}^{\beta})_M{}^{N} \ = \ (t_{\alpha}{}^{\beta})_{i\gamma}{}^{j\delta}
   \ = \  \delta_{i}^{j} \Big(\delta_{\alpha}^{\delta}\delta_{\gamma}^{\beta}-
   \frac{1}{2}\delta_{\alpha}^{\beta}\delta^{\delta}_{\gamma}\Big)\;.
 \end{split}
 \ee
The Cartan-Killing form $\kappa_{AB}  \equiv  {\rm tr}(t_A t_B)$, where $A,B$ label the total
Lie algebra $\mathfrak{sl}(3)\oplus \mathfrak{sl}(2)$, can be computed explicitly and
shown to take a block-diagonal form. The projectors are then given by
$\mathbb{P}^{M}{}_{N}{}^{K}{}_{L} =  \kappa^{ab}(t_{a})_N{}^M (t_b)_{L}{}^{K}$,
where
$a,b$ label the indices either of the $(8,1)$ or the $(1,3)$ block.}
  \be\label{PROJ8}
   \begin{split}
   (\mathbb{P}_{(8,1)})^{M}{}_{N}{}^{K}{}_{L} \ = \  (\mathbb{P}_{(8,1)})^{i\alpha}{}_{l\delta}{}^{j\beta}{}_{k\gamma}
   \ = \ \fracs{1}{2}\delta^i_k\delta^j_l\delta^{\alpha}_{\delta}\delta^{\beta}_{\gamma}
   -\fracs{1}{6}\delta^j_k\delta^i_l\delta^{\alpha}_{\delta}\delta^{\beta}_{\gamma}\;, \\[1ex]
  (\mathbb{P}_{(1,3)})^{M}{}_{N}{}^{K}{}_{L} \ = \  (\mathbb{P}_{(1,3)})^{i\alpha}{}_{l\delta}{}^{j\beta}{}_{k\gamma}
   \ = \ \fracs{1}{3}\delta^i_l\delta^j_k\delta^{\alpha}_{\gamma}\delta^{\beta}_{\delta}
   -\fracs{1}{6}\delta^j_k\delta^i_l\delta^{\alpha}_{\delta}\delta^{\beta}_{\gamma}\;.
 \end{split}
 \ee
The coefficients in (\ref{projlie}) are determined by closure, as we will discuss momentarily.
Using the projectors inside the generalized Lie derivative (\ref{projlie})
one obtains
 \be\label{EXplGenLie}
  \mathbb{L}_{\Lambda}V^{i\alpha} \ = \ \Lambda^{j\beta}\partial_{j\beta}V^{i\alpha}
  -V^{j\alpha}\partial_{j\beta}\Lambda^{i\beta}-V^{i\beta}\partial_{j\beta}\Lambda^{j\alpha}
  +\big(\lambda+\fracs{5}{6}\big)\partial_{j\beta}\Lambda^{j\beta}\,V^{i\alpha}\;.
 \ee
A generalized Lie derivative can also be defined for a tensor $W_{i\alpha}$ in the representation $(\bar{3},\bar{2})$,
 \be\label{LIEVECTORDOWN}
  \mathbb{L}_{\Lambda}W_{i\alpha} \ = \ \Lambda^{j\beta}\partial_{j\beta}W_{i\alpha}
  +\partial_{i\beta}\Lambda^{j\beta}\,W_{j\alpha}+\partial_{j\alpha}\Lambda^{j\beta}\,W_{i\beta}
  +\big(\lambda-\fracs{5}{6}\big)\partial_{j\beta}\Lambda^{j\beta}\,W_{i\alpha}\;.
 \ee
This definition is such that the singlet $V^{i\alpha}W_{i\alpha}$ transforms as a scalar density whose
weight is the sum of the weights of $V$ and $W$.
The generalized Lie derivative, say in the form (\ref{EXplGenLie}), can
also be written in terms of epsilon tensors as follows
\be\label{GenLieIndexform}
 \mathbb{L}_\Lambda V^{i\alpha} \ = \ \Lambda^{j\beta}\ptl_{j\beta} V^{i\alpha}-V^{j\beta}\ptl_{j\beta}\Lambda^{i\alpha}+\epsilon^{ijn}\epsilon_{kln}\epsilon^{\alpha\beta}\epsilon_{\gamma\delta}\ptl_{j\beta}\Lambda^{k\gamma} V^{l\delta}
+\big(\lambda-\fracs{1}{6}\big)\ptl_{j\beta}\Lambda^{j\beta}V^{i\alpha}\;,
\ee
as can be verified straightforwardly using standard epsilon tensor identities.\footnote{Our conventions are
as follows: The SL(2) tensors $\epsilon_{\alpha\beta}$ and $\epsilon^{\alpha\beta}$ are related by
$\epsilon^{\alpha\beta}\epsilon_{\gamma\delta}=2\delta^{\alpha}_{[\gamma}\delta^{\beta}_{\delta]}$ and therefore
$\epsilon^{\alpha\gamma}\epsilon_{\beta\gamma}=\delta^{\alpha}_{\beta}$. Similarly, the SL$(3)$ tensors $\epsilon^{ijk}$ and
$\epsilon_{ijk}$ are related
by $\epsilon^{ijm}\epsilon_{klm}=2\delta^i_{[k}\delta^j_{l]}$.}
 A useful alternative form of the generalized Lie derivative can then be obtained by introducing
the tensor
 \be
\bsp
Z^{MN}{}_{KL} \ = \
Z^{i\alpha,j\beta}{}_{k\gamma,l\delta} \ \equiv \ &\epsilon^{ijm}\epsilon_{klm}\epsilon^{\alpha\beta}\epsilon_{\gamma\delta}\\
\ =\ &\delta^i_k\delta^j_l\delta^\alpha_\gamma\delta^\beta_\delta-\delta^i_k\delta^j_l\delta^\alpha_\delta\delta^\beta_\gamma-\delta^i_l\delta^j_k\delta^\alpha_\gamma\delta^\beta_\delta+
\delta^i_l\delta^j_k\delta^\alpha_\delta\delta^\beta_\gamma\,, \label{Ztensor}
\end{split}
\ee
so that (\ref{GenLieIndexform}) becomes in somewhat more covariant notation
\be\label{COVVar}
 \mathbb{L}_\Lambda V^M \ = \ \Lambda^N\ptl_N V^M-V^N\ptl_N\Lambda^M+Z^{MN}{}_{PQ}\ptl_N \Lambda^P V^Q+\big( \lambda-\fracs{1}{6}\big)\ptl_N \Lambda^N V^M\;.
\ee
This form is instructive, because it shows that $Z$ measures the deviation from the standard Lie derivative of a
vector(-density), and it also shows that vectors of weight $\tfrac{1}{6}$ are special, which will be important below.
Similarly, the generalized Lie derivative on a vector with lower index reads
\be\label{COVVarDown}
 \mathbb{L}_\Lambda V_M \ = \ \Lambda^N\ptl_N V_M+\ptl_M\Lambda^N V_N
 -Z^{PQ}{}_{MN}\ptl_P \Lambda^N V_Q+\big( \lambda+\fracs{1}{6}\big)\ptl_N \Lambda^N V_M\;.
\ee
The tensor $Z$ defined above has the useful property that due to the section constraint (\ref{SectionstrIntro})
\be\label{Zsection}
 Z^{KL}{}_{MN}\,   \ptl_K\otimes\ptl_L \ = \ 0\;,
\ee
as is manifest from the definition in the first line of (\ref{Ztensor}).
Let us note the following consequence of the constraint in the form (\ref{Zsection}). First note that we can 
also define a `generalized' scalar (of weight zero) transforming as $\delta_{\Lambda}S \ = \ \Lambda^N\partial_NS$.
Its partial derivative then transforms covariantly,
 \be\label{covPArDE}
  \delta_{\Lambda}(\partial_MS) \ = \ \Lambda^N\partial_N(\partial_MS)+\partial_M\Lambda^N\partial_NS
  \ = \ \mathbb{L}_{\Lambda}(\partial_M S)\;,
 \ee
which, thanks to (\ref{Zsection}), equals the generalized Lie derivative (\ref{COVVarDown}) for $\lambda=-\fracs{1}{6}$.
Thus, $\partial_MS$
is a generalized (co-)vector of weight $-\frac{1}{6}$.

We now turn to the closure of the gauge transformations governed by the
generalized Lie derivatives (\ref{projlie}). An explicit computation shows that, up to
the section constraint (\ref{SectionstrIntro}), particularly used in the form (\ref{Zsection}),
the generalized Lie derivatives close,
 \be\label{AcLosure}
  \big[\mathbb{L}_{\Lambda_1},\mathbb{L}_{\Lambda_2}\big] \ = \ \mathbb{L}_{[\Lambda_1,\Lambda_2]_{\rm E}}\;,
 \ee
according to the `E-bracket'
 \be\label{EBracket}
  \big[ \Lambda_1,\Lambda_2\big]_{\rm E}^M \ = \
  \Lambda_1^N\ptl_N \Lambda_2^M+\fracs{1}{2}Z^{MN}{}_{PQ}\ptl_N \Lambda_1^P \Lambda_2^Q-(1\leftrightarrow 2)\;.
\ee
As is common in EFT, the E-bracket does not define a Lie bracket in that the Jacobi identity is violated.
The resulting Jacobiator is, however, of a certain `trivial' form. This means that the generalized Lie derivative
w.r.t.~the corresponding gauge parameter does not act on fields as a consequence of the section constraint (\ref{SectionstrIntro}). The non-trivial Jacobiator is therefore consistent with the Jacobi identity
satisfied by the gauge variations on fields, $[\delta_{\Lambda_1},[\delta_{\Lambda_2},\delta_{\Lambda_3}]]+\cdots=0$.

Let us pause and discuss the form of gauge parameters that are trivial
in this sense. Specifically, we claim that a gauge parameter of the form
 \be\label{trivPARA}
  \Lambda^M \ \equiv \ Z^{MN}{}_{PQ}\,\partial_N\chi^{PQ}\;,
 \ee
for arbitrary symmetric $\chi^{PQ}$, does not generate a generalized Lie derivative.
In order to verify this we compute with (\ref{COVVar}) for this parameter
 \be
\bsp
\delta_{\Lambda}V^M \ = \  \mathbb{L}_{\Lambda}V^M \ = \
-V^L Z^{MN}{}_{PQ}\ptl_L\ptl_N\chi^{PQ}
+Z^{ML}{}_{RS}Z^{RN}{}_{PQ}\ptl_L\ptl_N\chi^{PQ}V^S\;. \label{trivialgauge}
\end{split}
\ee
Here we used that the transport and density terms (i.e.~the first and last terms in (\ref{COVVar}))
vanish due to the section constraint in the form (\ref{Zsection}).
The remaining two terms in here cancel due to the identity
\be
Z^{M(L}{}_{PQ}\,Z^{N)P}{}_{RS}\,\ptl_L\otimes\ptl_N \ = \ Z^{M(L}{}_{RS}\,\delta^{N)}_Q\, \ptl_L\otimes\ptl_N\;, \label{ZZ}
\ee
which can be confirmed with the explicit form (\ref{Ztensor}) and use of the section constraint
(\ref{SectionstrIntro}). 
Let us determine the number of independent trivial parameters according to (\ref{trivPARA}). The tensor
$\chi^{PQ}$ has $\tfrac{1}{2}6\cdot 7$ independent components, but many of them are projected out,
as can be seen by inserting the $Z$ tensor,
\be
\Lambda^{i\alpha} \ = \
\epsilon^{ijm}\epsilon_{klm}\epsilon^{\alpha\beta}\epsilon_{\gamma\delta}\ptl_{j\beta}\chi^{k\gamma,l\delta}\;.
\ee
In fact, in $\chi^{k\gamma,l\delta}$ both the SL$(2)$ and SL$(3)$ indices
are contracted with their respective epsilon tensors, reducing the number of independent
components to $3$. Parametrizing $\chi$ in terms of such a vector,
\be
\chi_m \ = \ \epsilon_{ijm}\epsilon_{\alpha\beta}\chi^{i\alpha,j\beta}\;, \qquad
\chi^{i\alpha,j\beta} \ = \ \fracs{1}{4}\epsilon^{ijm}\epsilon^{\alpha\beta}\chi_m\;,
\ee
we obtain for the trivial gauge parameter
\be
 \Lambda^{i\alpha} \ = \ \epsilon^{ijm}\epsilon^{\alpha\beta}\ptl_{j\beta}\chi_m\;.
\ee

We now return to the Jacobiator and show that it is of the trivial form (\ref{trivPARA}).
As we will discuss below, the gauge parameters $\Lambda^M$ have to be thought of as
generalized vectors of weight $\fracs{1}{6}$ and so we
generally define the Jacobiator for generalized vectors of weight $\fracs{1}{6}$. One finds
\be\label{JACID}
\begin{split}
\big[\big[U,V\big]_{\rm E},W\big]_{\rm E}+\mathrm{cycl.}\ =\ &\fracs{1}{3}\big(\big[U,V\big]_{\rm E},W\big)+\mathrm{cycl.}\;,
\end{split}
\ee
with the symmetric pairing
\be
\begin{split}
(U,V)^M \ &\equiv \ \fracs{1}{2}(\mathbb{L}_U V^M+\mathbb{L}_V U^M) \\
\ &= \ \fracs{1}{2}Z^{MN}{}_{PQ}\ptl_N U^P V^Q+\fracs{1}{2}Z^{MN}{}_{PQ}\ptl_N V^P U^Q \ = \ \fracs{1}{2}Z^{MN}{}_{PQ}\ptl_N \big(U^P V^Q \big)\;.
\end{split}
\label{sym}
\ee
This follows in complete analogy to the discussions in \cite{E6} and so we conclude that the
Jacobiator is indeed of the trivial form (\ref{trivPARA}). This symmetric pairing also encodes
the difference between the E-bracket (\ref{EBracket}) and the generalized Lie derivative of a vector of
weight $\fracs{1}{6}$, c.f.~(\ref{COVVar}):
 \be\label{bracketDerDiff}
  \mathbb{L}_{U}V \ = \ \big[U,V\big]_{\rm E}+\big(U,V\big)\;.
 \ee
Put differently, the generalized Lie derivative differs from the E-bracket by a term that does
not generate Lie derivatives.   As the pairing is symmetric we can also conclude that the E-bracket
equals the antisymmetrized generalized Lie derivative,
 \be\label{ASYmmStuff}
  \big[U,V\big]_{\rm E} \ = \ \fracs{1}{2}\big(\mathbb{L}_{U}V-\mathbb{L}_{V}U\big)\;.
 \ee
Both these relations will be instrumental below.
Using (\ref{ASYmmStuff}) and the algebra (\ref{AcLosure}) it is a straightforward computation to show that
\be
\mathbb{L}_\Lambda\big[U,V\big]_{\rm E}-\big[\mathbb{L}_\Lambda U,V\big]_{\rm E}
-\big[U,\mathbb{L}_\Lambda V\big]_{\rm E}
\ = \ \fracs{1}{2}\left(\mathbb{L}_{\mathbb{L}_U\Lambda}V+\mathbb{L}_{\mathbb{L}_\Lambda U}V\right)-(U\leftrightarrow V)
\ = \ 0\;,
\ee
where we last step follows from the triviality of the symmetric pairing (\ref{sym}).
We thus proved
\be
\mathbb{L}_\Lambda\big[U,V\big]_{\rm E} \ = \ \big[\mathbb{L}_\Lambda U,V\big]_{\rm E}
+\big[U,\mathbb{L}_\Lambda V\big]_{\rm E}\;,
\ee
which means that the E-bracket is covariant under the generalized diffeomorphisms generated
by generalized Lie derivatives.

\subsection{Tensors in general representations of the duality group}
We now define the action of generalized Lie derivatives on tensors living
in more general representations than the fundamental $(3,2)$ and its
higher tensor powers.
We start with a fundamental SL$(3)$ vector, more specifically a field in the $(\bar{3},1)$ representation
of the duality group, where the bar indicates that the index is a lower, covariant index.
There is a natural way to relate such a vector $B_m$ to two vectors $A_{1,2}$ in
the $(3,2)$:
 \be\label{BmDEF}
 B_m \ = \ \epsilon_{ijm}\epsilon_{\alpha\beta}A_1^{i\alpha}A_2^{j\beta}\;.
 \ee
We now require that an SL$(3)$ vector such as $B_m$ transforms under
generalized Lie derivatives so as to be compatible with this equation assuming a Leibniz property, i.e.,
 \be
  \mathbb{L}_\Lambda B_m \ = \
   \epsilon_{ijm}\epsilon_{\alpha\beta}\,\big(\mathbb{L}_\Lambda A_1^{i\alpha}\,A_2^{j\beta}
   +A_1^{i\alpha}\,\mathbb{L}_\Lambda A_2^{j\beta}\big) \;.
 \ee
Evaluating this with (\ref{GenLieIndexform}), it amounts to
 \be
  \begin{split}
\mathbb{L}_\Lambda B_m
\ = \ &\epsilon_{ijm}\epsilon_{\alpha\beta}\big[A_1^{i\alpha}(\Lambda^{k\gamma}\ptl_{k\gamma}A_2^{j\beta}-A_2^{k\gamma}\ptl_{k\gamma}\Lambda^{j\beta}+
\epsilon^{jks}\epsilon_{pqs}\epsilon^{\beta\gamma}\epsilon_{\sigma\phi}\ptl_{k\gamma}\Lambda^{p\sigma}A_2^{q\phi}
\\&\qquad \qquad
+(\lambda_1-\fracs{1}{6})\ptl_{k\gamma}\Lambda^{k\gamma}A_2^{j\beta})+(1\leftrightarrow 2)\big]\,.
\end{split}
\ee
Using standard identities for the epsilon symbols it is straightforward to rewrite the right-hand side
in terms of $B_m$ as defined in (\ref{BmDEF}). This yields
 \be\label{GENLieB}
  \mathbb{L}_\Lambda B_m \ = \ \Lambda^{k\gamma}\ptl_{k\gamma} B_m+\ptl_{m\gamma}\Lambda^{k\gamma}B_k
  +\big(\lambda-  \fracs{1}{3}\big)\ptl_{k\gamma}\Lambda^{k\gamma}B_m\;,
\ee
where $\lambda=\lambda_1+\lambda_2$ for the weights $\lambda_{1,2}$ of the vectors $A_{1,2}$.
An alternative form is given by
 \be
  \mathbb{L}_\Lambda B_m \ = \ \Lambda^{k\gamma}\ptl_{k\gamma} B_m
  +\epsilon_{mnk}\,\epsilon^{pqk}\partial_{p\gamma}\Lambda^{n\gamma}\,B_q
  +\big(\lambda+  \fracs{2}{3}\big)\ptl_{k\gamma}\Lambda^{k\gamma}B_m\;.
\ee
Generally, we take this, or equivalently (\ref{GENLieB}),
as the definition of the Lie derivative on a $(\bar{3},1)$ vector of weight $\lambda$.
We define the generalized Lie derivative on tensors with an arbitrary number of (lower)
fundamental SL$(3)$ indices analogously.  Note that writing the generalized Lie derivative as in (\ref{GENLieB})
each index contributes an extra $-\tfrac{1}{3}$. For instance, on a 2-tensor $B_{mn}$ of weight $\lambda$ it reads
  \be\label{LBmn}
 \mathbb{L}_\Lambda B_{mn} \ = \ \Lambda^{k\gamma}\ptl_{k\gamma}B_{mn}+\ptl_{m\gamma}\Lambda^{k\gamma}B_{kn}+\ptl_{n\gamma}\Lambda^{k\gamma}B_{mk}+\big(\lambda-\fracs{2}{3}\big)\ptl_{k\gamma}\Lambda^{k\gamma} B_{mn}\;.
 \ee
Specializing to ${\cal M}_{mn}\in {\rm SL}(3)$ the gauge transformations then preserve
${\rm det}\, {\cal M}=1$ for $\lambda=0$, as will be instrumental below.

Given the action of the generalized Lie derivative on a vector $B_m$, we can determine the
action on a vector $D^m$ in the dual $(3,1)$ representation from the requirement that
the resulting singlet $D^mB_m$ transforms as a scalar if both $D$ and $B$ have weight zero
(and otherwise as a scalar density whose weight is the sum of the weights of  $D$ and $B$).
This yields the form of the generalized Lie derivative on a vector $D^m$ of weight $\lambda$,
 \be\label{GenLie31}
  \mathbb{L}_{\Lambda} D^m \ = \ \Lambda^{k\gamma}\partial_{k\gamma}D^m
  -D^k\partial_{k\gamma}\Lambda^{m\gamma}+\big(\lambda+\fracs{1}{3}\big)\partial_{k\gamma}\Lambda^{k\gamma}
  \,D^m\;.
 \ee
As before, the generalized Lie derivative acts analogously on tensors with an arbitrary number
of upper SL$(3)$ indices, with each index adding $+\fracs{1}{3}$ to the density term.

Next we discuss tensors in the fundamental of SL$(2)$, i.e., transforming as $(1,2)$ under the
full duality group. Such a tensor $C^{\alpha}$ can be constructed from a $(\bar{3},1)$ vector $B_m$
and a fundamental $(3,2)$ vector $A^{m\alpha}$ as follows
 \be
  C^{\alpha} \ \equiv \ B_{m}A^{m\alpha}\;.
 \ee
The action of the generalized Lie derivative on $C^{\alpha}$ is then determined by postulating
the Leibniz property
 \be
  \mathbb{L}_{\Lambda}C^{\alpha} \ = \ \mathbb{L}_{\Lambda}B_m\,A^{m\alpha}
  +B_m\,\mathbb{L}_{\Lambda}A^{m\alpha}\;.
 \ee
Using the form of the generalized Lie derivatives in (\ref{GenLieIndexform}),
(\ref{GENLieB}) and employing epsilon tensor relations, it is straightforward to
show that the right-hand side can be written in terms of $C^{\alpha}$,
 \be\label{GenLie21}
  \mathbb{L}_{\Lambda}C^{\alpha} \ = \ \Lambda^{n\beta}\partial_{n\beta}C^{\alpha}
  -\partial_{n\beta}\Lambda^{n\alpha}\,C^{\beta}+\big(\lambda+\fracs{1}{2}\big)
  \partial_{n\beta}\Lambda^{n\beta}\,C^{\alpha}\;,
 \ee
where $\lambda=\lambda(A)+\lambda(B)$ is the sum of the density weights of $A^{m\alpha}$ and $B_m$.
This equation can equivalently be written as
 \be\label{GenLie22}
   \mathbb{L}_{\Lambda}C^{\alpha} \ = \ \Lambda^{n\beta}\partial_{n\beta}C^{\alpha}
   -\epsilon^{\alpha\beta}\epsilon_{\gamma\delta}\,\partial_{n\beta}\Lambda^{n\gamma}\,C^{\delta}
   +\big(\lambda-\fracs{1}{2}\big)\partial_{n\beta}\Lambda^{n\beta}\,C^{\alpha}\;.
 \ee
We take (\ref{GenLie21}), or equivalently (\ref{GenLie22}),
to be the definition of the generalized Lie derivative on a vector $C^{\alpha}$ of weight $\lambda$.
Its action on a higher tensor power is defined analogously. When written in the form analogous to (\ref{GenLie21})
this adds a $\fracs{1}{2}$ to the weight term for each index. This definition is then such that
for a 2-tensor ${\cal M}^{\alpha\beta}\in {\rm SL}(2)$ the condition $\det {\cal M}=1$ is gauge invariant
for $\lambda=0$. Moreover, one may verify that $\epsilon^{\alpha\beta}$ is a gauge invariant tensor
of weight $\lambda=0$, $\mathbb{L}_{\Lambda}\epsilon^{\alpha\beta}=0$.\footnote{Note that this is different
from conventional differential geometry, where the epsilon tensor is invariant under Lie derivatives as a
tensor of weight one.}
Given this SL(2) and gauge invariant tensor $\epsilon^{\alpha\beta}$ the $2$ representation
is equivalent to the contragredient or dual $\bar{2}$ representation. Thus, we can define the
generalized Lie derivative on a vector $C_{\alpha}$ by using
 \be
   C_\alpha\ = \ C^\beta\epsilon_{\beta\alpha}\;, \qquad
   C^\alpha \ = \ \epsilon^{\alpha\beta} C_\beta\;.
 \ee
We obtain
\be\label{GENLIEalphadown}
 \mathbb{L}_\Lambda C_\alpha \ = \
 \Lambda^{k\beta}\ptl_{k\beta}C_\alpha
 +\ptl_{k\alpha}\Lambda^{k\beta} C_\beta+(\lambda-\fracs{1}{2})\ptl_{k\beta}\Lambda^{k\beta}C_\alpha\;.
\ee
Again, this definition extends straightforwardly to tensors with an arbitrary number of lower SL$(2)$
indices, each index adding $-\fracs{1}{2}$ to the density term.

We close this subsection by noting that the transformation behavior of the various tensors introduced
here is mutually compatible and defined in such a way that the weights add up naturally. For instance,
given tensors $A^{m\alpha}$ and $C_{\alpha}$ in the $(3,2)$ and $(1,\bar{2})$ representations, respectively,
the vector
 \be
  D^m \ \equiv \ A^{m\alpha}C_{\alpha}\;,
 \ee
transforms as (\ref{GenLie31}) with the weight $\lambda$ that is the sum of the weights of
$A$ and $C$. Similarly, any SL$(3)\times {\rm SL}(2)$ invariant contraction of fields
will transform according to the respective generalized Lie derivatives with a total weight that is
given by the sum of the `component' weights. This will be instrumental in the next subsection.

 \subsection{Generalized Cartan calculus}
So far we discussed the covariant transformation of tensors in various representations of the
duality group SL$(3)\times {\rm SL}(2)$ and how to construct new covariant tensors algebraically,
i.e., by means of various contractions of indices.
In this subsection we now introduce a differential or Cartan-like calculus that allows us to
take certain projected derivatives of tensor fields (of specific density weights) that lead to new covariant tensors.
This is closely analogous to the calculus of differential forms, in which the exterior
derivative ${\rm d}$ maps a covariant $p$-form to a covariant $(p+1)$-form
and satisfies ${\rm d}^2=0$. In fact, we will introduce a differential operator $\widehat\partial$
that is also nilpotent, so that $\widehat\partial\,^2=0$, and which satisfies relations very analogous
to those of the standard Cartan calculus.

To begin, we introduce a useful notation for various algebraic operations mapping
tensor representations into each other. We start with tensors $A^{m\alpha}$ in the
$(3,2)$ representation, carrying an arbitrary weight $\lambda$, and denote the space of such tensors as
 \be
  \mathfrak{A}(\lambda)\; :\quad \text{space of vectors $A^{m\alpha}$ of weight $\lambda$}\;.
 \ee
 Similarly, we denote the space of vectors $B_m$ in the $(\bar{3},1)$ representation as
 \be
  \mathfrak{B}(\lambda)\; :\quad \text{space of vectors $B_m$ of weight $\lambda$}\;.
 \ee
Then there is a natural operation (or contraction), in the following denoted by $\bullet$, that maps
 \be\label{ContrMap}
   \bullet\;:\quad \mathfrak{A}(\lambda_1)\times\mathfrak{A}(\lambda_2)\quad \longrightarrow\quad
   \mathfrak{B}(\lambda_1+\lambda_2)\;,
 \ee
defined by
 \be\label{vecvecpair}
  (A_1\bullet A_2)_m \ \equiv \ \epsilon_{ijm}\epsilon_{\alpha\beta}A_1^{i\alpha}A_2^{j\beta}\;.
 \ee
As this is the operation used in (\ref{BmDEF}) to define the generalized Lie derivative on
$\mathfrak{B}$, this operation indeed maps tensors of the indicated weights into each other.

More generally,  we define
spaces $\mathfrak{C}(\lambda)$, $\mathfrak{D}(\lambda)$ and $\mathfrak{E}(\lambda)$
of tensors $C^{\alpha}$, $D^m$ and $E_{m\alpha}$, respectively. Fields taking values in these spaces (of
specific density weights) are appearing in the tensor hierarchy to be developed in the
next section, and for the reader's convenience we collected them in the above table.
For completeness, we also introduce the notation $\mathfrak{S}(\lambda)$ for scalar densities, 
although there will be no $p$-form potentials in this representation. 
\begin{table}
\centering
\begin{tabular}{|c|c|c|c|c|c|}
\hline
space&$\mathfrak{A}(\fracs{1}{6})$&$\mathfrak{B}(\fracs{1}{3})$&$\mathfrak{C}(\fracs{1}{2})$&$\mathfrak{D}(\fracs{2}{3})$&$\mathfrak{E}(\fracs{5}{6})$\\
\hline
representation&$A^{i\alpha}$&$B_m$&$C^\alpha$&$D^m$&$E_{m\alpha}$\\
\hline
\end{tabular}
\caption{SL(3)$\times$ SL(2) representations with density weights as appearing in the tensor hierarchy}\label{t:reps}
\end{table}

For the reader's convenience, 
we summarize the action of 
generalized Lie derivatives on the objects listed in Table \ref{t:reps}, with the specific weights indicated:
\bea
\mathbb{L}_\Lambda A^{i\alpha} &=& \Lambda^{j\beta}\ptl_{j\beta} A^{i\alpha}-A^{j\beta}\ptl_{j\beta}\Lambda^{i\alpha}+\epsilon^{ijn}\epsilon_{kln}\epsilon^{\alpha\beta}\epsilon_{\gamma\delta}\ptl_{j\beta}\Lambda^{k\gamma} A^{l\delta}\;,\label{genLieA}\\
 \mathbb{L}_\Lambda B_m &=& \Lambda^{k\alpha}\ptl_{k\alpha} B_m+\ptl_{m\alpha}\Lambda^{k\alpha}\,B_k\;,\label{genLieB}\\
  \mathbb{L}_{\Lambda}C^{\alpha} &=& \Lambda^{k\beta}\partial_{k\beta}C^{\alpha}
   -\epsilon^{\alpha\beta}\epsilon_{\gamma\delta}\,\partial_{k\beta}\Lambda^{k\gamma}\,C^{\delta}\;,\label{genLieC}\\
   \mathbb{L}_{\Lambda} D^m &=& \Lambda^{k\gamma}\partial_{k\gamma}D^m
  -D^k\partial_{k\gamma}\Lambda^{m\gamma}+\partial_{k\gamma}\Lambda^{k\gamma}
  \,D^m\;,\label{genLieD}\\
  \mathbb{L}_\Lambda E_{m\alpha} &=& \Lambda^{k\beta}\ptl_{k\beta}E_{m\alpha}+E_{k\alpha}\ptl_{m\beta}\Lambda^{k\beta}
  +E_{m\beta}\ptl_{k\alpha}\Lambda^{k\beta}\;\label{genLieE}.
\eea

The contraction operation (\ref{ContrMap}) can be extended naturally to maps between
various of the spaces introduced.
For instance,
 \be\label{Collcontracs}
 \begin{split}
  &\bullet\,:\quad
  \mathfrak{A}(\lambda_1)\times\mathfrak{C}(\lambda_2)\,\longrightarrow\,\mathfrak{D}(\lambda_1+\lambda_2):
  \qquad\quad  (A\bullet C)^m \ \equiv \ \epsilon_{\alpha\beta} C^\alpha A^{m\beta}\;, \\[0.5ex]
   &\bullet\,:\quad \mathfrak{B}(\lambda_1)\times\mathfrak{B}(\lambda_2)\;\longrightarrow\;
   \mathfrak{D}(\lambda_1+\lambda_2):\quad\;\; (B_1\bullet B_2)^m \ \equiv \ \epsilon^{ijm}B_{1i}B_{2j}\;,\\[0.5ex]
   &\bullet\,:\quad \mathfrak{A}(\lambda_1)\times\mathfrak{B}(\lambda_2)\;\longrightarrow\;
   \mathfrak{C}(\lambda_1+\lambda_2):\qquad\quad  (A\bullet B)^\alpha \ \equiv \ A^{m\alpha}B_{m}\;,\\[0.5ex]
   &\bullet\,:\quad \mathfrak{A}(\lambda_1)\times\mathfrak{D}(\lambda_2)\;\longrightarrow\;
   \mathfrak{E}(\lambda_1+\lambda_2):\qquad\;  (A\bullet D)_{m\alpha} \ \equiv \
   \epsilon_{mnk}\epsilon_{\alpha\beta}A^{n\beta}D^k\;,\\[0.5ex]
   &\bullet\,:\quad \mathfrak{B}(\lambda_1)\times\mathfrak{C}(\lambda_2)\;\longrightarrow\;
   \mathfrak{E}(\lambda_1+\lambda_2):\qquad\;  (B\bullet C)_{m\alpha} \ \equiv \
   \epsilon_{\alpha\beta}B_m C^\beta\\[0.5ex]
   &\bullet\,:\quad \mathfrak{C}(\lambda_1)\times\mathfrak{C}(\lambda_2)\;\longrightarrow\;
   \mathfrak{S}(\lambda_1+\lambda_2):\qquad \;\;\; \, (C_1\bullet C_2) \ \equiv \
   \epsilon_{\alpha\beta}C_1^{\alpha} C_2^\beta\\[0.5ex]
   &\bullet\,:\quad \mathfrak{D}(\lambda_1)\times\mathfrak{B}(\lambda_2)\;\longrightarrow\;
   \mathfrak{S}(\lambda_1+\lambda_2):\qquad \;\;\; \, (D\bullet B) \ \equiv \
  D^m B_m
   \;.
 \end{split}
 \ee
We will use the notation $\bullet$ universally, as it is always clear from the context
which projection is applied.
As most $\bullet$ operations involve tensors in two different spaces, there is 
in general no symmetry or antisymmetry property. For the following special cases, however, 
we have the symmetry property 
\be
A_1,A_2\, \in \, \mathfrak{A} \,: \qquad A_1\bullet A_2 \ = \ A_2\bullet A_1\;, 
\ee
and the antisymmetry properties for  
$B_1,B_2\in\mathfrak{B}$ and  $C_1,C_2\in\mathfrak{C}$
\bea
B_1\bullet B_2&=&-B_2\bullet B_1\;, \\
C_1\bullet C_2&=&-C_2\bullet C_1\; .
\eea
It is also convenient to define the $\bullet$ operator to be always commutative when acting on two different spaces, for instance, 
\be
A\, \in \, \mathfrak{A},\  \, B\, \in \, \mathfrak{B}\,: \qquad A\bullet B\equiv B\bullet A\;.
\ee
We have summarized the results of the tensor operations denoted by $\bullet$ in table 2. 
\begin{table}\label{BINARYop}
\centering
\begin{tabular}{|c|c|c|c|c|c|c|}
\hline
$\bullet$&$\mathfrak{S}(0)$&$\mathfrak{A}(\fracs{1}{6})$&$\mathfrak{B}(\fracs{1}{3})$&$\mathfrak{C}(\fracs{1}{2})$&$\mathfrak{D}(\fracs{2}{3})$&$\mathfrak{E}(\fracs{5}{6})$\\
\hline
$\mathfrak{S}(0)$&$\mathfrak{S}(0)$&$\mathfrak{A}(\fracs{1}{6})$&$\mathfrak{B}(\fracs{1}{3})$&$\mathfrak{C}(\fracs{1}{2})$&$\mathfrak{D}(\fracs{2}{3})$&$\mathfrak{E}(\fracs{5}{6})$\\
\hline
$\mathfrak{A}(\fracs{1}{6})$&$\mathfrak{A}(\fracs{1}{6})$&$\mathfrak{B}(\fracs{1}{3})$&$\mathfrak{C}(\fracs{1}{2})$&$\mathfrak{D}(\fracs{2}{3})$&$\mathfrak{E}(\fracs{5}{6})$&$\mathfrak{S}(1)$\\
\hline
$\mathfrak{B}(\fracs{1}{3})$&$\mathfrak{B}(\fracs{1}{3})$&$\mathfrak{C}(\fracs{1}{2})$&$\mathfrak{D}(\fracs{2}{3})$&$\mathfrak{E}(\fracs{5}{6})$&$\mathfrak{S}(1)$&\\
\hline
$\mathfrak{C}(\fracs{1}{2})$&$\mathfrak{C}(\fracs{1}{2})$&$\mathfrak{D}(\fracs{2}{3})$&$\mathfrak{E}(\fracs{5}{6})$&$\mathfrak{S}(1)$& &\\
\hline
$\mathfrak{D}(\fracs{2}{3})$&$\mathfrak{D}(\fracs{2}{3})$&$\mathfrak{E}(\fracs{5}{6})$&$\mathfrak{S}(1)$& & &\\
\hline
$\mathfrak{E}(\fracs{5}{6})$&$\mathfrak{E}(\fracs{5}{6})$&$\mathfrak{S}(1)$& & & &\\
\hline
\end{tabular}
\caption{The result of the binary operation $\bullet$}
\end{table}

It follows from the discussion in the previous subsection that the operation $\bullet$ is covariant
in the sense that
\be\label{distribution}
 \mathbb{L}_\Lambda(X\bullet Y) \ = \ (\mathbb{L}_\Lambda X)\bullet Y +
 X\bullet(\mathbb{L}_\Lambda Y)\;,
\ee
for any tensors $X$ and $Y$ belonging to the spaces listed above.

We are now ready to introduce the covariant differential operator $\widehat\partial$ mapping between
the spaces of the specific weights indicated.
More specifically, the operator $\widehat\partial$ acts on the spaces in the above table in descending order,
\be\label{partialHIER}
\mathfrak{A}(\fracs{1}{6})\xlongleftarrow{\widehat\partial}\mathfrak{B}(\fracs{1}{3})\xlongleftarrow{\widehat\partial}\mathfrak{C}(\fracs{1}{2})\xlongleftarrow{\widehat\partial}\mathfrak{D}(\fracs{2}{3})
\xlongleftarrow{\widehat\partial}\mathfrak{E}(\fracs{5}{6})
\;.
\ee
We see that $\widehat\partial$ in each step lowers the density weight by $-\fracs{1}{6}$
(as did the partial derivative on a scalar, c.f.~(\ref{covPArDE}) above).
Let us now define the action of $\widehat\partial$ on the various tensors.
We start with the highest space in the above sequence and work our way down, starting with
\be
\widehat\partial\;:\quad \mathfrak{E}(\fracs{5}{6})\quad \longrightarrow\quad \mathfrak{D}(\fracs{2}{3})\;,
\ee
which is defined by
\be\label{barpartialE}
 (\widehat\partial E)^m \ \equiv \ \epsilon^{mnk}\epsilon^{\alpha\beta}\ptl_{n\alpha}E_{k\beta}\;.
\ee
Our task is to prove that $\widehat\partial E$ so defined transforms covariantly, i.e., with the
generalized Lie derivative (\ref{GenLie31}) of weight $\lambda=\fracs{2}{3}$, or (\ref{genLieD}).
To this end we first compute the general gauge transformation of the un-projected
partial derivative of $E_{m\alpha}$, using (\ref{genLieE}),
 \be\label{COVSTEPPP}
 \begin{split}
  \delta_{\Lambda}\big(\partial_{n\alpha}E_{k\beta}\big) \ = \ \,&
  \Lambda^{l\gamma}\partial_{l\gamma}\big(\partial_{n\alpha}E_{k\beta}\big) + \partial_{n\gamma}\Lambda^{l\gamma}
  \partial_{l\alpha}E_{k\beta}+\partial_{l\alpha}\Lambda^{l\gamma}\partial_{n\gamma}E_{k\beta}
  +\partial_{k\gamma}\Lambda^{l\gamma}\partial_{n\alpha} E_{l\beta}\\
  &+\partial_{l\beta}\Lambda^{l\gamma}\partial_{n\alpha}E_{k\gamma}-\partial_{l\gamma}\Lambda^{l\gamma}
  \partial_{n\alpha}E_{k\beta}+\partial_{n\alpha}\partial_{k\gamma}\Lambda^{l\gamma} E_{l\beta}
  +\partial_{n\alpha}\partial_{l\beta}\Lambda^{l\gamma} E_{k\gamma}\;.
 \end{split}
 \ee
Here we used the section constraint in the form (\ref{EXPLSEc}) in order to rewrite
the term $\partial_{n\alpha}\Lambda^{l\gamma}\partial_{l\gamma} E_{k\beta}$ that arises in
this computation in terms of three other terms.
Next, we have to compare this result with the expected generalized Lie derivative of a tensor
with the index structure of $\partial_{n\alpha}E_{k\beta}$. Comparing, say, with (\ref{GENLieB}) and
(\ref{GENLIEalphadown}), we infer that all expected $\partial\Lambda\partial E$ terms rotating the indices are present.
According to the rules spelled out in the previous subsection, for the density term
we have to add $-\fracs{1}{3}$ for every SL$(3)$ index and $-\fracs{1}{2}$ for every SL$(2)$ index,
for which here there are two each, 
implying that the density term contains the factor $(\lambda-\fracs{5}{3})$.
The density term in (\ref{COVSTEPPP}) has coefficient $-1$ and so we learn that $\lambda=\fracs{2}{3}$.
The last two terms in the second line of (\ref{COVSTEPPP}) are non-covariant terms, and so
we finally conclude that $\partial_{n\alpha}E_{k\beta}$ transforms with the generalized Lie derivative
of weight $\lambda=\fracs{2}{3}$, up to anomalous terms given by
 \be
  \Delta^{\rm nc}_{\Lambda}\big(\partial_{n\alpha}E_{k\beta}\big) \ \equiv \
  \big(\delta_{\Lambda}-\mathbb{L}_{\Lambda}\big)\big(\partial_{n\alpha}E_{k\beta}\big) \ = \
  \partial_{n\alpha}\partial_{k\gamma}\Lambda^{l\gamma} E_{l\beta}
  +\partial_{n\alpha}\partial_{l\beta}\Lambda^{l\gamma} E_{k\gamma}\;.
 \ee
(Here and in the following we use the notation $\Delta^{\rm nc}_{\Lambda}$ for the non-covariant variation of any term.)
Thus, as expected, the partial derivative does not transform covariantly. However, once we project it as in
(\ref{barpartialE}) and use that the epsilon tensors are gauge invariant, we obtain
 \be
  \Delta^{\rm nc}_{\Lambda}\big(\widehat\partial E\big)^m \ = \ \epsilon^{mnk}\epsilon^{\alpha\beta}
  \big(\partial_{n\alpha}\partial_{k\gamma}\Lambda^{l\gamma} E_{l\beta}
  +\partial_{n\alpha}\partial_{l\beta}\Lambda^{l\gamma} E_{k\gamma} \big) \ = \ 0\;,
 \ee
where in the last step we used the section constraint in the form (\ref{simplifiedSEction}).
Thus, $\widehat\partial E$ transforms covariantly, as we wanted to prove.

In the next step of the sequence the operator $\widehat\partial$,
  \be
\widehat\partial:\quad  \mathfrak{D}(\fracs{2}{3})\quad \longrightarrow\quad \mathfrak{C}(\fracs{1}{2})\;,
\ee
is defined by
\be\label{barpartialD}
(\widehat\partial D)^\alpha \ \equiv \ \epsilon^{\alpha\beta}\ptl_{m\beta}D^m\;.
\ee
Let us confirm that with this definition $\widehat\partial$ is nilpotent
in that
 \be
  \widehat\partial \ \circ \ \widehat\partial:\quad
  \mathfrak{E}(\fracs{5}{6})\quad \longrightarrow\quad \mathfrak{C}(\fracs{1}{2})
 \ee
acts trivially. Indeed, with (\ref{barpartialD}) and (\ref{barpartialE}) we compute its
action on a tensor $E\in \mathfrak{E}(\fracs{5}{6})$,
 \be
  (\widehat\partial\,\widehat\partial\,E)^{\alpha} \ = \ \epsilon^{\alpha\beta}\partial_{m\beta}
  \big(\epsilon^{mnk}\epsilon^{\gamma\delta}\partial_{n\gamma}E_{k\delta}\big) \ = \ 0\;,
 \ee
 which vanishes as a consequence of the section constraint in the form (\ref{simplifiedSEction}).
 It remains to show that the derivative operation defined in (\ref{barpartialD}) is covariant.
 Analogously to our proof around (\ref{COVSTEPPP}) this can be verified by an explicit
 computation. One uses the generalized Lie derivative (\ref{genLieD}) compute the gauge variation of  (\ref{barpartialD}) and then verifies
 that, upon using the section constraint, it agrees with the generalized Lie derivative
 (\ref{genLieC}) acting on the tensor $(\widehat\partial D)^\alpha$ of weight $\fracs{1}{2}$.
Let us stress again that this covariance property crucially hinges on the precise weights indicated here.

Next we define the action
\be
\widehat\partial\,:\quad \mathfrak{C}(\fracs{1}{2})\quad \longrightarrow\quad\mathfrak{B}(\fracs{1}{3})\;,
\ee
given by
 \be\label{bardelC}
  (\widehat\partial C)_m \ = \ \ptl_{m\alpha}C^\alpha\;.
 \ee
In combination with (\ref{barpartialD}) it is again easy to see that $\widehat\partial^2=0$,
 \be
  (\widehat\partial\,\widehat\partial\,D)_m \ = \ \partial_{m\alpha}\big(\epsilon^{\alpha\beta}\partial_{n\beta}D^n\big)
   \ = \ 0\;,
 \ee
using the section constraint (\ref{simplifiedSEction}).
The proof of covariance is again straightforward.

The final map
 \be
  \widehat\partial\,:\quad \mathfrak{B}(\fracs{1}{3})\quad \longrightarrow\quad \mathfrak{A}(\fracs{1}{6})\;,
 \ee
acts as
 \be\label{parderB}
  (\widehat\partial B)^{i\alpha} \ = \ \epsilon^{ijk}\epsilon^{\alpha\beta}\ptl_{j\beta}B_k\;.
\ee
It is again straightforward to verify that it leads to a nilpotent operator, satisfying $\widehat\partial^2=0$,
and that this differential operator is gauge covariant.

This concludes our definition of the action of the covariant differentials. An obvious question is whether we
can extend (\ref{partialHIER}) even further, for instance: can we define a covariant action of $\widehat\partial$
on $\mathfrak{A}(\fracs{1}{6})$? One may convince oneself that this is not possible. In fact,
we are supposed to find a projection or contraction of $\partial_{m\alpha}A^{n\beta}$ that
transforms covariantly. The only possibilities are to take the trace either over the SL$(2)$ or the SL$(3)$ indices,
but if any of these is covariant then certainly the full trace $\partial_{m\alpha}A^{m\alpha}$ is covariant.
Writing the latter as $\partial_MA^M$ it is easy to see with (\ref{COVVar}),
however, that is does not transform covariantly for $\lambda=\frac{1}{6}$.
Thus, there is no covariant extension of $\widehat\partial$
to $\mathfrak{A}(\fracs{1}{6})$. Because of this, let us note as a cautionary remark that in general
\be
\widehat\partial(A\bullet B) \ \neq \ \widehat\partial A\bullet B+A\bullet \widehat\partial B\;,
\ee
because the $\widehat\partial$ in some terms may not even be defined.
However, for special cases there are relations of this type: for $B_1,B_2\in\mathfrak{B}(\fracs{1}{3})$
one may verify
\be
\widehat\partial B_1\bullet B_2-\widehat\partial B_2\bullet B_1 \ = \ \widehat\partial(B_1\bullet B_2)\label{BBprop}\;.
\ee
Also, for $B\in\mathfrak{B}(\fracs{1}{3}),C\in\mathfrak{C}(\fracs{1}{2})$,
\be
\widehat\partial B\bullet C+B\bullet\widehat\partial C\ =\ \widehat\partial(B\bullet C)\label{BCprop}\;.
\ee

It is also important to point out that if we view the operation $\bullet$ as a product
this product is not associative in general. We have, however, the following relations
for any $A,B,C\in\mathfrak{A}$,
\be
A\bullet (B\bullet C)+B\bullet (A\bullet C)+C\bullet (A\bullet B) \ = \ 0 \;, \label{AAAprop}
\ee
and for any $A_1,A_2\in\mathfrak{A},B\in\mathfrak{B}$,
\be
A_1\bullet(A_2\bullet B)+A_2\bullet (A_1\bullet B)+B\bullet(A_1\bullet A_2)\ =\ 0 \;. \label{AABprop}
\ee
Moreover, if $A_1,A_2\in\mathfrak{A},B_1,B_2,B_3\in\mathfrak{B},C\in\mathfrak{C},D\in\mathfrak{D}$, the following
associativity properties hold:
\be
\bsp
A_1\bullet(A_2\bullet D) \ &= \ (A_1\bullet A_2)\bullet D\\
A_1\bullet(B_1\bullet C) \ &= \ (A_1\bullet B_1)\bullet C \ = \ -B_1\bullet(A_1\bullet C) \ = \ -C\bullet(A_1\bullet B_1)\\
B_1\bullet(B_2\bullet B_3) \ &= \ B_2\bullet(B_3\bullet B_1) \ = \ B_3\bullet(B_1\bullet B_2)\\
\ &= \ (B_1\bullet B_2)\bullet B_3
\ = \ (B_2\bullet B_3)\bullet B_1 \ = \ (B_3\bullet B_1)\bullet B_2\;. \label{associative}
\end{split}
\ee
They can be easily verified by explicitly writing out the index based defintions.

Let us next discuss a curious interplay between the derivative operator $\widehat\partial$
and the generalized Lie derivative that is very reminiscent to the Cartan calculus of differential forms.
Of course, the operator $\widehat\partial$ commutes with the Lie derivative $\mathbb{L}_{\Lambda}$
in the sense that it is gauge covariant. Put differently, the following diagram is commutative:
\be
\begin{array}{ccccccccccc}\mathfrak{A}(\fracs{1}{6})&\xlongleftarrow{\widehat\partial}&\mathfrak{B}(\fracs{1}{3})&\xlongleftarrow{\widehat\partial}&\mathfrak{C}(\fracs{1}{2})&
\xlongleftarrow{\widehat\partial}&\mathfrak{D}(\fracs{2}{3})&\xlongleftarrow{\widehat\partial}&\mathfrak{E}(\fracs{5}{6})\\
\Big{\downarrow}{\mathbb{L}_\Lambda}&&\Big{\downarrow}{\mathbb{L}_\Lambda}&&\Big{\downarrow}{\mathbb{L}_\Lambda}&&\Big{\downarrow}{\mathbb{L}_\Lambda}&&\Big{\downarrow}{\mathbb{L}_\Lambda}\\
\mathfrak{A}(\fracs{1}{6})&\xlongleftarrow{\widehat\partial}&\mathfrak{B}(\fracs{1}{3})&\xlongleftarrow{\widehat\partial}&\mathfrak{C}(\fracs{1}{2})&
\xlongleftarrow{\widehat\partial}&\mathfrak{D}(\fracs{2}{3})&\xlongleftarrow{\widehat\partial}&\mathfrak{E}(\fracs{5}{6})
\end{array}
\ee
In addition, one can express the generalized Lie derivative acting on tensors in $\mathfrak{B}$,
$\mathfrak{C}$ and $\mathfrak{D}$ in terms of $\widehat\partial$ and the contraction operation $\bullet$.
Specifically, for any tensor $X$ taking values in these spaces we have
\be
 \mathbb{L}_\Lambda X \ = \ \Lambda\bullet\widehat\partial X+\widehat\partial(\Lambda\bullet X)\;.  \label{magic}
\ee
Equivalently, if we denote the operation of acting with $\Lambda\,\bullet$ on a tensor by $i_{\Lambda}$
this relation becomes
\be
\mathbb{L}_\Lambda \ = \ i_\Lambda\circ\widehat\partial+\widehat\partial\circ i_\Lambda\; ,
\ee
which is completely analogous to the familiar ${\cal L}_{X}=i_X \,{\rm d}+{\rm d}\, i_X$
that holds for Lie derivative and exterior derivative ${\rm d}$ acting on differential forms
(sometimes denoted Cartan's `magic formula').
The relation (\ref{magic}) can be verified by an explicit computation, which we briefly illustrate
for a tensor $X=B_m\in \mathfrak{B}(\fracs{1}{3})$. We compute for the two terms on the right-hand side
 \be
 \begin{split}
   (\Lambda\bullet\widehat\partial B)_m  \ &= \ \epsilon_{mnk}\epsilon_{\alpha\beta}\Lambda^{n\alpha}(\widehat\partial B)^{k\beta}
   \ = \ \epsilon_{mnk}\epsilon_{\alpha\beta}\Lambda^{n\alpha}\epsilon^{kpq}\epsilon^{\beta\gamma}
   \partial_{p\gamma}B_{q} \\
   \ &= \ \Lambda^{n\alpha}\partial_{n\alpha}B_m-\Lambda^{n\alpha}\partial_{m\alpha}B_n\;,
 \end{split}
 \ee
where we used  (\ref{vecvecpair}) and (\ref{parderB}), and
 \be
  \widehat\partial(\Lambda\bullet B)_m \ = \ \partial_{m\alpha}(\Lambda\bullet B)^{\alpha}
  \ = \ \partial_{m\alpha}(\Lambda^{n\alpha}B_n) \ = \ \partial_{m\alpha}\Lambda^{n\alpha} B_n
  +\Lambda^{n\alpha}\partial_{m\alpha} B_n\;,
 \ee
where we used (\ref{bardelC}) and the third definition in (\ref{Collcontracs}).
Combing these two results we obtain
 \be
  (\Lambda\bullet\widehat\partial B)_m+  \widehat\partial(\Lambda\bullet B)_m
  \ = \ \Lambda^{n\alpha}\partial_{n\alpha}B_m+\partial_{m\alpha}\Lambda^{n\alpha} B_n\;,
 \ee
which agrees with the generalized Lie derivative (\ref{genLieB}) acting on a vector
$B_m$ of weight $\lambda=\fracs{1}{3}$.
The validity of (\ref{magic}) for tensors in $\mathfrak{C}$ and $\mathfrak{D}$ is verified analogously.
Let us note that for $V,W\in \mathfrak{A}(\fracs{1}{6})$
 \be\label{roundbracketexact}
  \big(V,W\big) \ \equiv \ \frac{1}{2}( \mathbb{L}_{V}W+\mathbb{L}_{W}V) \ = \ \frac{1}{2}\widehat\partial(V\bullet W)\;,
 \ee
which follows by using  (\ref{sym}), (\ref{vecvecpair}) and (\ref{parderB}).
This implies an alternative writing for the relation (\ref{bracketDerDiff}) between the
Lie derivative and the E-bracket,
 \be\label{FINALLieE}
  \mathbb{L}_{V}W \ = \ \big[V,W\big]_{\rm E}+\frac{1}{2}\widehat\partial (V\bullet W)\;.
 \ee

Given the analogies between the differential $\widehat\partial$ and the exterior derivative of 
differential forms, one may wonder whether there is an 
analogue of de Rham cohomology. In particular, one may wonder whether 
there is a version of the Poincar\'e lemma according to which locally a $\widehat\partial$ closed form 
is $\widehat\partial$ exact, 
 \be\label{POINcare}
  \psi\,\in\,\mathfrak{B}(\fracs{1}{3})\,,\; \mathfrak{C}(\fracs{1}{2}) \;\;{\rm or}\;\;
  \mathfrak{D}(\fracs{2}{3})\,: \qquad 
  \widehat\partial\psi \ = \ 0 \quad \Rightarrow \quad  \psi \ = \ \widehat\partial\chi\;\;?
 \ee 
In fact, one may give a straightforward argument for this statement, reducing it to the conventional Poincar\'e lemma. 
For instance, let $B\in \mathfrak{B}(\fracs{1}{3})$ with $\widehat\partial B=0$, i.e., 
 \be
  (\widehat\partial B)^{i\alpha} \ = \ \epsilon^{ijk}\epsilon^{\alpha\beta}\partial_{j\beta}B_k \ = \ 0\;. 
 \ee 
We split the derivatives as $\partial_i\equiv \partial_{i1}$ and $\partial_{i}'\equiv \partial_{i2}$, 
after which this equation gives two relations, 
 \be
  \begin{split}
      (\widehat\partial B)^{i1} \ = \ \epsilon^{ijk}\partial_j'B_k \ &= \ 0\qquad \Rightarrow \qquad 
      B_k \ = \ \partial_k'\chi'\;, \\
      (\widehat\partial B)^{i2} \ = \ -\epsilon^{ijk}\partial_jB_k \ &= \ 0\qquad \Rightarrow \qquad 
      B_k \ = \ \partial_k\chi\;.   
   \end{split}
  \ee     
Together these two equations imply $B_k(y,y')=\partial_k\chi(y)+\partial_k'\chi'(y')$, 
so that setting $C^{\alpha}\equiv (\chi,\chi')$ this becomes, upon restoring SL$(3)\times {\rm SL}(2)$ 
covariant notation,  
 \be
  B_k \ = \ \partial_{k\alpha}C^{\alpha}\qquad \Leftrightarrow \qquad
  B \ = \ \widehat\partial C\;, 
 \ee
showing that $B$ is exact. This argument proceeds analogously for the other 
two spaces in (\ref{POINcare}).
However, there is a subtlety with the above alleged proof. 
It is only valid if we keep all six coordinates, i.e., 
before restricting to a particular solution of the section constraint. For instance, in the example discussed 
the proof goes through for the M-theory solution but not for the IIB solution. Similarly, 
for each of the three spaces in (\ref{POINcare}) for precisely one of the M-theory or type IIB solutions 
does the proof go through. Thus, the Poincar\'e lemma is not generally true in the strongly constrained theory, 
but for a given representation space it is only true for a particular solution of the section constraint. 
We return to this 
issue in sec.~5.

We close this section by briefly discussing invariant integration over the $Y$-space 
and the notion of integration by parts with the differentials $\widehat\partial$. 
There is an invariant integral of the $\bullet$ product of two 
tensors if and only if it results in a scalar whose weight is $1$. 
For instance, for $C_1,C_2\in \mathfrak{C}(\frac{1}{2})$ we have 
$C_1\bullet C_2\in \mathfrak{S}(1)$, see (\ref{Collcontracs}),  and hence 
 \be\label{STEPInt}
  \delta_{\Lambda}\big(C_1\bullet C_2\big) \ = \ 
  \Lambda^N\partial_N\big(C_1\bullet C_2\big)+\partial_N\Lambda^N\big(C_1\bullet C_2\big)
  \ = \ \partial_N\big(\Lambda^N(C_1\bullet C_2)\big)\;. 
 \ee 
Since $C_1\bullet C_2$ thus varies into a total derivative it follows that\footnote{It should be stressed 
that, given the section constraint, 
this integration over the 6-dimensional $Y$-space is somewhat formal. 
For the M-theory or type IIB solution, fields depend either only on three or two of these coordinates, 
and we assume that the redundant integrals $\int {\rm d}^3y$ or $\int {\rm d}^4y$ 
simply give an overall constant (which may be 
absorbed into a rescaling of the Newton constant multiplying the action).} 
 \be
  \int {\rm d}^6Y\,C_1\bullet C_2 \ \equiv \ \int {\rm d}^6Y\,\epsilon_{\alpha\beta}
  \,C_1^{\alpha}\,C_2^{\beta}
 \ee
is gauge invariant. Note that this invariance does not require an explicit volume density 
because the involved tensors already carry non-trivial weights. 
Let us now consider the special case that $C_1$ is $\widehat\partial$ exact, 
 \be
  C_1 \ = \ \widehat\partial D \quad \Leftrightarrow \quad C_1^{\alpha} \ = \ \epsilon^{\alpha\beta}\partial_{m\beta}D^m\;. 
 \ee
We then compute 
 \be
 \begin{split}
  \int {\rm d}^6Y\,\widehat\partial D\bullet C_2 \ &= \ \int {\rm d}^6Y\,\epsilon_{\alpha\beta}\,
  \epsilon^{\alpha\gamma}\,\partial_{m\gamma}D^m\,C_2^{\beta}
  \ = \  \int {\rm d}^6Y\,\partial_{m\beta}D^m\,C_2^{\beta} \\[0.5ex] 
  \ &= \  -\int {\rm d}^6Y\,D^m \partial_{m\beta}C_2^{\beta} \ \equiv \ 
  -\int {\rm d}^6Y\, D\bullet \widehat\partial C_2\;, 
 \end{split}
 \ee
where we integrated by parts, employed  
$\bullet :\;\mathfrak{D}(\frac{2}{3})\times \mathfrak{B}(\frac{1}{3})\rightarrow \mathfrak{S}(1)$
defined in (\ref{Collcontracs}) and used (\ref{bardelC}). 
We thus have 
 \be
    \int {\rm d}^6Y\,\widehat\partial D\bullet C \ = \  -\int {\rm d}^6Y\, D\bullet \widehat\partial C\;,
 \ee
showing that we can integrate by parts with $\widehat\partial$.      
It should be emphasized, however, that in contrast to the standard Cartan calculus 
of differential forms the operation $\bullet$ has two different interpretations 
on both sides of this equation. As a particular corollary we have 
 \be\label{totalbarpart}
  \int {\rm d}^6Y\,\widehat\partial D_1\bullet \widehat\partial D_2 \ = \ 
  - \int {\rm d}^6Y\, D_1 \bullet \widehat\partial\,^2D_2 \ = \ 0\;, 
 \ee
by $\widehat\partial\,^2=0$. This will be instrumental below when checking properties of invariant actions.

\section{The tensor hierarchy}
We have now developed enough technology in order to construct the tensor hierarchy efficiently.
We start by introducing covariant derivatives ${\cal D}_{\mu}$ that covariantize the gauge symmetries
given by generalized internal diffeomorphisms spanned by $\Lambda^M$.
This is necessary because the gauge parameter will be a function of the internal $Y^M$ and
the external $x^{\mu}$, $\Lambda^M=\Lambda^M(x,Y)$.
We introduce gauge connection one-forms $A_{\mu}{}^{M}$, which then, by consistency,
requires the introduction of an entire hierarchy of forms.

\subsection{Covariant derivatives, gauge connections and 2-forms}
We introduce gauge connection one-forms $A_{\mu}{}^{M}\in {\mathfrak {A}}(\fracs{1}{6})$ and define
the covariant derivative by
 \be\label{COvDER}
  \mathcal{D}_\mu \ = \ \ptl_\mu -\mathbb{L}_{A_\mu}\;,
\ee
where the generalized Lie derivative acts in the appropriate representation of
the object on which ${\cal D}_{\mu}$ acts. Here $A_{\mu}{}^{M}$ carries density weight $\lambda=\frac{1}{6}$,
the same as the gauge parameter.
The covariant derivative transforms covariantly if the gauge field transforms as
\be\label{deltaAgauge}
 \delta_\Lambda A_\mu{}^M \ = \ \mathcal{D}_\mu \Lambda^M\;.
\ee
This follows by a straightforward calculation of the gauge transformations of
the covariant derivative of a generic tensor $V$,
\be
\bsp
\delta_\Lambda(\mathcal{D}_\mu V) \ = \ &\,
\delta_{\Lambda}\big(\partial_{\mu}V-\mathbb{L}_{A_{\mu}}V\big) \ = \
\ptl_\mu(\mathbb{L}_\Lambda V)-\mathbb{L}_{A_\mu}\mathbb{L}_\Lambda V-\mathbb{L}_{\ptl_\mu\Lambda-\mathbb{L}_{A_\mu}\Lambda}V\\
\ = \ &\,\mathbb{L}_{\partial_{\mu}\Lambda}V+\mathbb{L}_{\Lambda}(\partial_{\mu} V)
-\mathbb{L}_{A_\mu}\mathbb{L}_\Lambda V-\mathbb{L}_{\partial_{\mu}\Lambda}V
+\mathbb{L}_{\mathbb{L}_{A_\mu}\Lambda}V\\
\ = \ &\,
\mathbb{L}_{\Lambda}\big(\partial_{\mu}V-\mathbb{L}_{A_{\mu}}V\big)
+\big[\mathbb{L}_{\Lambda},\mathbb{L}_{A_{\mu}}\big]V
+\mathbb{L}_{[A_\mu,\Lambda]_{\rm E}+(A_{\mu},\Lambda)}V\\
\ = \ &\, \mathbb{L}_{\Lambda}({\cal D}_{\mu}V)\;.
\end{split}
\ee
Here we used the relation (\ref{bracketDerDiff})
implying that the difference between Lie derivative and E-bracket is
of a trivial form that is immaterial in the argument of a Lie derivative, and we used
the E-bracket algebra (\ref{AcLosure}).

Our next task is to construct a gauge covariant field strength for the connections $A_{\mu}{}^{M}$.
The naive field strength as in Yang-Mills theory, based on the E-bracket, reads
\be
 F_{\mu\nu}{}^M \ = \ 2\,\ptl_{[\mu}A_{\nu]}{}^M-\big[A_\mu,A_\nu\big]_{\rm E}^M\;.
\ee
However, since the E-bracket, having a non-trivial Jacobiator,
does not define a Lie algebra, this does not define a gauge covariant object.
More generally, the variation of $F_{\mu\nu}$ under an arbitrary variation $\delta A_{\mu}$
does not take the expected covariant form $2\mathcal{D}_{[\mu}\,\delta A_{\nu]}$. Let us compute the anomalous part.  Thanks to the calculus introduced in the
previous section, this can be done
in a completely index-free fashion:
\be
\bsp
\delta F_{\mu\nu} \ &= \ 2\Big(\partial_{[\mu}\,\delta A_{\nu]}-\big[A_{[\mu},\delta A_{\nu]}\big]_{\rm E}\Big) \\[0.5ex]
\ &= \ 2\Big(\partial_{[\mu}\,\delta A_{\nu]} - \mathbb{L}_{A_{[\mu}}\delta A_{\nu]}
+\big(A_{[\mu},\delta A_{\nu]}\big) \Big) \\[0.5ex]
\ &= \ 2\,{\cal D}_{[\mu}\,\delta A_{\nu]}+\widehat\partial\big(A_{[\mu}\bullet \delta A_{\nu]}\big)\;. \label{fieldcov2}
\end{split}
\ee
Here we used (\ref{bracketDerDiff}) in the second line and (\ref{roundbracketexact})
in the last line. We infer that the variation differs from the expected covariant result by a $\widehat\partial$ exact
term. In the spirit of the tensor hierarchy this can now be repaired by
introducing 2-form potentials $B_{\mu\nu}\in \mathfrak{B}(\fracs{1}{3})$ and defining the improved field strength
\be\label{improvedF}
 \mc{F}_{\mu\nu} \ \equiv \  F_{\mu\nu} \,+ \, \widehat\partial B_{\mu\nu} \;,
\ee
or, restoring explicit index notation,
\be\label{explicitF}
\mc{F}_{\mu\nu}{}^{i\alpha} \  = \ 2\,\ptl_{[\mu}A_{\nu]}{}^{i\alpha}
-\big[A_\mu,A_\nu\big]_{\rm E}^{i\alpha} \, + \, \epsilon^{ijk}\epsilon^{\alpha\beta}\ptl_{j\beta}B_{\mu\nu\, k}\; .
\ee
Defining the covariant variation $\Delta B_{\mu\nu}$ of the 2-forms by
 \be\label{DELTAB}
  \Delta B_{\mu\nu} \ \equiv \ \delta B_{\mu\nu}+A_{[\mu}\bullet\delta A_{\nu]}\;,
\ee
we see with (\ref{fieldcov2}) that the improved field strength then varies as
\be\label{deltacalF}
   \delta\mc{F}_{\mu\nu} \ = \ 2\,\mc{D}_{[\mu}\, \delta A_{\nu]} \, + \, \widehat\partial(\Delta B_{\mu\nu}) \;.
\ee
Next we turn to the $\Lambda$ gauge variation of ${\cal F}_{\mu\nu}$. We first note
that, as usual, the commutator of covariant derivatives yields the field strength,
\be
 \big[ \mc{D}_\mu,\mc{D}_\nu\big ] \ = \ -\mathbb{L}_{F_{\mu\nu}} \ = \ -\mathbb{L}_{\mc{F}_{\mu\nu}}\,,\label{commDA}
\ee
which follows by a straightforward explicit computation. Note that in this relation the difference
between the naive and the improved field strength is immaterial, as they differ by a trivial exact
term that does not generate a Lie derivative. We then compute the $\Lambda$ gauge variation with
(\ref{deltacalF}),
 \be
 \begin{split}
  \delta_{\Lambda}{\cal F}_{\mu\nu} \ &= \ \big[{\cal D}_{\mu},{\cal D}_{\nu}\big]\Lambda
  +\widehat\partial(\Delta_{\Lambda}B_{\mu\nu}) \ = \ -\mathbb{L}_{{\cal F}_{\mu\nu}}\Lambda
  + \widehat\partial(\Delta_{\Lambda}B_{\mu\nu}) \\[0.5ex]
  \ &= \ \mathbb{L}_{\Lambda}{\cal F}_{\mu\nu}-\widehat\partial(\Lambda\bullet {\cal F}_{\mu\nu} )
  + \widehat\partial(\Delta_{\Lambda}B_{\mu\nu})\;,
 \end{split}
 \ee
using (\ref{roundbracketexact}) in the second line. Thus, the field strength transforms covariantly,
\be
 \delta_\Lambda\mc{F}_{\mu\nu} \ = \ \mathbb{L}_\Lambda\mc{F}_{\mu\nu}\;,
\ee
provided we assign the following gauge transformation to the 2-form:
 \be\label{DeltaBLam}
  \Delta_{\Lambda}B_{\mu\nu} \ = \
  \Lambda\bullet {\cal F}_{\mu\nu}\;.
 \ee
Note that there is no contribution of the naive covariant form $\mathbb{L}_{\Lambda}B_{\mu\nu}$.
The 2-form also comes with its own gauge symmetry with 1-form parameter
$\Xi_{\mu}\in \mathfrak{B}(\fracs{1}{3})$,
 \be\label{DELXIB}
   \Delta_\Xi B_{\mu\nu} \ = \ 2\,\mc{D}_{[\mu}\Xi_{\nu]}\;.
\ee
In order for this transformation to leave the field strength ${\cal F}_{\mu\nu}$ invariant,
we need to assign an extra gauge transformation to the 1-forms $A_{\mu}$.  Using the triviality
of generalized Lie derivatives w.r.t~$\widehat\partial$ exact arguments it is easy to see that
${\cal D}_{\mu}$ commutes with $\widehat\partial$. It then follows with
(\ref{deltacalF}) that ${\cal F}_{\mu\nu}$ is invariant under (\ref{DELXIB}), provided
the gauge vectors transform as
 \be\label{DELXIA}
  \delta A_{\mu} \ = \ -\widehat\partial\, \Xi_{\mu}\;.
 \ee
We have to verify that this assignment is consistent with the earlier determination of
the gauge transformation of $A_{\mu}$ so that the covariant derivative
transforms covariantly.  This follows because in the definition (\ref{COvDER})
the shift of $A_{\mu}$ by a $\widehat\partial$ exact term drops out of the covariant derivative.

So far we have determined the gauge transformation of $B_{\mu\nu}$ so that
the improved field strength ${\cal F}_{\mu\nu}$ is gauge covariant, but this requirement
actually does not uniquely determine the gauge transformation of $B_{\mu\nu}$.
In fact, from the definition (\ref{improvedF}) of ${\cal F}_{\mu\nu}$ we see that we may
shift $B_{\mu\nu}$ by an arbitrary $\widehat\partial$ exact term, which will drop out
by $\widehat\partial\,^2=0$. Thus, there is an additional redundancy, or gauge invariance,
that in fact turns out to be gauged by the next higher form in the hierarchy, the 3-form,
to which we turn now.

\subsection{3-form potentials}
The most direct way to introduce the 3-form is via the field strength of the 2-form.
In turn, this field strength can be conveniently introduced by
requiring a Bianchi identity for the field strength ${\cal F}_{\mu\nu}$. The conventional
Bianchi identity ${\cal D}{\cal F}=0$ does not hold because the gauge algebra
is not a Lie algebra. Rather, we compute
 \be
\bsp
\mc{D}_{[\mu} F_{\nu\rho]} \ = \ &\,
\ptl_{[\mu}F_{\nu\rho]}-\mathbb{L}_{A_{[\mu}}F_{\nu\rho]} \\[0.5ex]
\ = \ &-\ptl_{[\mu}\big[A_\nu,A_{\rho]}\big]_{\rm E}-2\, \mathbb{L}_{A_{[\mu}}\ptl_\nu A_{\rho]}+\mathbb{L}_{A_{[\mu}}
\big[A_\nu,A_{\rho]}\big]_{\rm E} \\[0.5ex]
\ = \ &\, -\partial_{[\mu}\big(\mathbb{L}_{A_{\nu}}A_{\rho]}\big)-2\, \mathbb{L}_{A_{[\mu}}\ptl_\nu A_{\rho]}
+\big[A_{[\mu},
\big[A_\nu,A_{\rho]}\big]_{\rm E}+\frac{1}{2}\widehat\partial\big(A_{[\mu}\bullet \big[A_{\nu},A_{\rho]}\big]_{\rm E}\big)
\\
\ = \ &\, -\mathbb{L}_{\partial_{[\mu}A^{}_{\nu}} A_{\rho]}-\mathbb{L}_{A_{[\rho}}\partial_{\mu}A_{\nu]}+\big[A_{[\mu},
\big[A_\nu,A_{\rho]}\big]_{\rm E}+\frac{1}{2}\widehat\partial\big(A_{[\mu}\bullet \big[A_{\nu},A_{\rho]}\big]_{\rm E}\big)
 \\[0.5ex]
\ = \ &\,\widehat\partial\big(-\ptl_{[\mu}A_\nu\bullet A_{\rho]}+\fracs{1}{3}A_{[\mu}\bullet\big[A_\nu,A_{\rho]}\big]_{\rm E}\big)\;.
\end{split}
\ee
Here we used (\ref{ASYmmStuff}) and (\ref{FINALLieE}) in the step from the second to the third line
and the form of the Jacobiator (\ref{JACID}) in the last line.
Since the exterior derivative of the full curvature can be written as
 \be
  \mc{D}_{[\mu} \mc{F}_{\nu\rho]} \ = \ \mc{D}_{[\mu} F_{\nu\rho]}+\widehat\partial(\mc{D}_{[\mu}B_{\nu\rho]})\;,
 \ee
we have shown that it is $\widehat\partial$ exact. Therefore,
if we define the field strength of the 2-form as
\be\label{FirstcalH}
 {\cal H}_{\mu\nu\rho} \ = \ 3\Big(
 \mc{D}_{[\mu} B_{\nu\rho]}
 -A_{[\mu}\bullet \partial_{\nu}A_{\rho]}+\fracs{1}{3}A_{[\mu}\bullet\big[A_\nu,A_{\rho]}\big]_{\rm E}+\cdots \Big)\;,
\ee
we obtain the modified Bianchi identity
\be
3\,\mc{D}_{[\mu}\mc{F}_{\nu\rho]} \ = \ \widehat\partial\, {\cal H}_{\mu\nu\rho}\;. \label{Bianchi3}
\ee
Since the left-hand side is manifestly gauge covariant, this relation shows that ${\cal H}_{\mu\nu\rho}$
is gauge covariant up to possibly $\widehat\partial$ closed terms, which are indicated
by dots in (\ref{FirstcalH}).
A fully gauge covariant 3-form curvature can be constructed by adding a 3-form potential
$C_{\mu\nu\rho}\in \mathfrak{C}(\fracs{1}{2})$ as follows
\be\label{CSLF}
 \mc{H}_{\mu\nu\rho} \ = \ 3\Big(
 \mc{D}_{[\mu} B_{\nu\rho]}
 -A_{[\mu}\bullet \partial_{\nu}A_{\rho]}+\fracs{1}{3}A_{[\mu}\bullet\big[A_\nu,A_{\rho]}\big]_{\rm E}\Big)
 +\widehat\partial C_{\mu\nu\rho}\;,
\ee
or, restoring explicit SL(3)$\times$SL(2) index notation,
\be\label{explcitH}
\bsp
 \mc{H}_{\mu\nu\rho\, m} \ = \ 3\Big(
 \mc{D}_{[\mu} B_{\nu\rho]m} -\epsilon_{ijm}\epsilon_{\alpha\beta}A_{[\mu}{}^{i\alpha}\ptl_\nu A_{\rho]}{}^{j\beta}+\fracs{1}{3}\epsilon_{ijm}\epsilon_{\alpha\beta}A_{[\mu}{}^{i\alpha}\big[A_\nu,A_{\rho]}\big]_{\rm E}^{j\beta}\Big)
 +\ptl_{m\alpha}C_{\mu\nu\rho}{}^\alpha\; .
 \end{split}
 \ee
As before, we will also write
 \be
  {\cal H}_{\mu\nu\rho} \ = \ H_{\mu\nu\rho} +\widehat\partial C_{\mu\nu\rho}\;,
 \ee
denoting by $H$ the naive but not gauge covariant field strength.

Let us now determine the gauge variation of $C_{\mu\nu\rho}$ that makes this curvature gauge covariant.
To this end it is again convenient to first compute the transformation of ${\cal H}_{\mu\nu\rho}$ under
arbitrary variations $\delta A_{\mu}$, $\delta B_{\mu\nu}$ and $\delta C_{\mu\nu\rho}$ and write it covariantly.
The direct variation yields
\be\label{DirectVAR}
\bsp
\delta {\cal H}_{\mu\nu\rho} \ = \ &\,3\Big(\mc{D}_{[\mu}\delta B_{\nu\rho]}-\mathbb{L}_{\delta A_{[\mu}}B_{\nu\rho]}-
\delta A_{[\mu}\bullet \ptl_\nu A_{\rho]}-A_{[\mu}\bullet \ptl_\nu\delta A_{\rho]}\\[0.5ex]
&\qquad
+\fracs{1}{3}\delta A_{[\mu}\bullet\big[A_\nu,A_{\rho]}\big]_{\rm E}
+\fracs{2}{3}A_{[\mu}\bullet\big[\delta A_\nu,A_{\rho]}\big]_{\rm E}\Big)+\widehat\partial\,\delta C_{\mu\nu\rho}\;.
\end{split}
\ee
Our task is now to rewrite this in terms of covariant objects.
In order to organize this computation in a transparent form let us first note that the variation of ${\cal H}_{\mu\nu\rho}$
is already determined by the Bianchi identity (\ref{Bianchi3}) up to $\widehat\partial$ exact terms.
Indeed, writing the variation of the right-hand side of this equation in terms of the variation
of the left-hand side, using (\ref{deltacalF}),  we compute
 \be
 \bsp
 \widehat\partial (\delta {\cal H}_{\mu\nu\rho}) \ =\ &3\,\mc{D}_{[\mu}\,\delta \mc{F}_{\nu\rho]}-3\, \mathbb{L}_{\delta A_{[\mu}}\mc{F}_{\nu\rho]}
 \ = \  3\,\mc{D}_{[\mu}\big(\,2\,\mc{D}_\nu\delta A_{\rho]}+\widehat\partial(\Delta B_{\nu\rho]})\big)-3\, \mathbb{L}_{\delta A_{[\mu}}\mc{F}_{\nu\rho]}\\[1ex]
 \ = \ &-3\,\mathbb{L}_{\mc{F}_{[\mu\nu}}\delta A_{\rho]} -3\, \mathbb{L}_{\delta A_{[\mu}}\mc{F}_{\nu\rho]}
 +3\,\widehat\partial(\mc{D}_{[\mu}\,\Delta B_{\nu\rho]}) \\[1ex]
 \ = \ &\, \widehat\partial\left(3\,\mc{D}_{[\mu}\Delta B_{\nu\rho]}-3\,\delta A_{[\mu}\bullet\mc{F}_{\nu\rho]}\right)\;, \\
 \end{split}
 \ee
 where we used the commutator of covariant derivatives (\ref{commDA}) and the
 relation (\ref{roundbracketexact}) for the symmetrized generalized Lie derivative.
 Thus, we infer
  \be\label{partoffullvar}
   \delta {\cal H}_{\mu\nu\rho} \ = \ 3\,\mc{D}_{[\mu}\,\Delta B_{\nu\rho]}-3\,\delta A_{[\mu}\bullet\mc{F}_{\nu\rho]}
   +\cdots\;,
  \ee
up to $\widehat\partial$ closed terms.  Next, we determine these terms, which are $\widehat\partial$ exact, explicitly
by comparing with (\ref{DirectVAR}). To this end we insert $\Delta B$ defined in (\ref{DELTAB}) into (\ref{partoffullvar}),
which yields after a quick computation
 \be
 \begin{split}
  3\,\mc{D}_{[\mu}\,\Delta B_{\nu\rho]}-3\,\delta A_{[\mu}\bullet\mc{F}_{\nu\rho]} \ = \ &\,
  3\,{\cal D}_{[\mu}(\delta B_{\nu\rho]})-3\,\delta A_{[\mu}\bullet \partial_{\nu}A_{\rho]}
  -3 A_{[\mu}\bullet \partial_{\nu}\,\delta A_{\rho]}\\[1ex]
  &+3 A_{[\mu}\bullet \mathbb{L}_{A_{\nu}}\delta A_{\rho]}
  -3\,\delta A_{[\mu}\bullet \widehat\partial B_{\nu\rho]}\;.
 \end{split}
 \ee
Comparing now to (\ref{DirectVAR}), with the E-bracket in the second line written out according to
(\ref{ASYmmStuff}), one finds
\be
 \bsp
\delta &{\cal H}_{\mu\nu\rho} \ = \
3\,\mc{D}_{[\mu}\,\Delta B_{\nu\rho]}-3\,\delta A_{[\mu}\bullet\mc{F}_{\nu\rho]} \\[1ex]
&-3\, \mathbb{L}_{\delta A_{[\mu}}B_{\nu\rho]}+3\,\delta A_{[\mu}\bullet\widehat\partial B_{\nu\rho]}+\delta A_{[\mu}\bullet[A_\nu,A_{\rho]}]_{\rm E}
+A_{[\mu}\bullet\mathbb{L}_{\delta A_\nu}A_{\rho]}-2A_{[\mu}\bullet\mathbb{L}_{A_\nu}\delta A_{\rho]}\;.
\end{split}
 \ee
The first term in the second line can be rewritten by means of the
magic identity (\ref{magic}),
 \be
  \mathbb{L}_{\delta A_{\mu}} B_{\nu\rho} \ = \ \delta A_{\mu}\bullet \widehat\partial B_{\nu\rho}
  +\widehat\partial\big(\delta A_{\mu}\bullet B_{\nu\rho}\big)\;,
 \ee
which yields
 \be\label{STEP142435}
 \bsp
\delta &H_{\mu\nu\rho} \ = \
3\,\mc{D}_{[\mu}\,\Delta B_{\nu\rho]}-3\,\delta A_{[\mu}\bullet\mc{F}_{\nu\rho]} \\[1ex]
&-3\, \widehat\partial\big(\delta A_{\mu}\bullet B_{\nu\rho}\big) +\delta A_{[\mu}\bullet[A_\nu,A_{\rho]}]_{\rm E}+A_{[\mu}\bullet\mathbb{L}_{\delta A_\nu}A_{\rho]}-2A_{[\mu}\bullet\mathbb{L}_{A_\nu}\delta A_{\rho]}\;.
\end{split}
 \ee
Finally, we can write all terms in the second line in a $\widehat\partial$ exact form,  using the following
lemma for any $A_\mu,A_\nu,C\in\mathfrak{A}(\fracs{1}{6})$:
\be
\bsp
[A_\mu,A_\nu]\bullet C-A_{[\mu}\bullet\mathbb{L}_C A_{\nu]}-2A_{[\mu}\bullet\mathbb{L}_{A_{\nu]}}C \ &= \ \mathbb{L}_{A_{[\mu}}(A_{\nu]}\bullet C)-A_{[\mu}\bullet\mathbb{L}_C A_{\nu]}-A_{[\mu}\bullet\mathbb{L}_{A_{\nu]}}C\\[1ex]
\ &= \ \mathbb{L}_{A_{[\mu}}(A_{\nu]}\bullet C)-A_{[\mu}\bullet\widehat\partial(A_{\nu]}\bullet C)\\[1ex]
\ &= \ \widehat\partial(A_{[\mu}\bullet(A_{\nu]}\bullet C))\; .
\label{3identity}
\end{split}
\ee
Here we used the distributivity (\ref{distribution}), the relation (\ref{roundbracketexact})
for the symmetrized generalized Lie derivative and, in the last step, the magic identity (\ref{magic}).
Specializing now to $C=\delta A_{\rho}$ we infer that the last
line in (\ref{STEP142435}) takes the form of a total $\widehat\partial$ derivative.
We have shown
\be
\delta \mc{H}_{\mu\nu\rho} \  = \ 3\, \mc{D}_{[\mu}\Delta B_{\nu\rho] }-3\,\delta A_{[\mu}\bullet\mc{F}_{\nu\rho]}
+\widehat\partial (\Delta C_{\mu\nu\rho})\, ,
\label{varyfieldH}
\ee
with the covariant variation of the 3-form
\be
\bsp
\Delta C_{\mu\nu\rho}\ \equiv \
\delta C_{\mu\nu\rho}-3\,\delta A_{[\mu}\bullet B_{\nu\rho]}+ A_{[\mu}\bullet (A_\nu\bullet \delta A_{\rho]})\;.
\end{split}
\ee

We are now ready to determine the explicit gauge transformations of the 3-form.
Specializing (\ref{varyfieldH}) to the gauge variation under
the $\Lambda$ transformations given in (\ref{DeltaBLam}) and (\ref{deltaAgauge})
we compute
 \be
 \begin{split}
  \delta_{\Lambda}{\cal H}_{\mu\nu\rho} \ &= \
  3\,{\cal D}_{[\mu}\big(\Lambda\bullet {\cal F}_{\nu\rho]}\big)-3\,{\cal D}_{[\mu}\Lambda\bullet {\cal F}_{\nu\rho]}
  +\widehat\partial(\Delta_{\Lambda} C_{\mu\nu\rho})\\[1ex]
  \ &= \ 3\,\Lambda\bullet {\cal D}_{[\mu}{\cal F}_{\nu\rho]}+\widehat\partial(\Delta_{\Lambda} C_{\mu\nu\rho})\\[1ex]
  \ &= \ \Lambda\bullet \widehat\partial{\cal H}_{\mu\nu\rho}+\widehat\partial(\Delta_{\Lambda} C_{\mu\nu\rho})\;,
 \end{split}
 \ee
where we used the Bianchi identity (\ref{Bianchi3}). Defining the covariant $\Lambda$ variation
of $C$ to be
 \be
  \Delta_{\Lambda}C_{\mu\nu\rho} \ \equiv \ \Lambda\bullet {\cal H}_{\mu\nu\rho}\;,
 \ee
it follows with the magic identity (\ref{magic}) for the generalized Lie derivative that the gauge variation
takes the covariant form
 \be\label{DELTALamC}
   \delta_{\Lambda}{\cal H}_{\mu\nu\rho} \ = \  \mathbb{L}_{\Lambda}{\cal H}_{\mu\nu\rho}\;,
 \ee
as required. Next, we turn to the gauge symmetry (\ref{DELXIB}), (\ref{DELXIA}) parametrized by $\Xi_{\mu}$,
which leaves ${\cal H}$ invariant provided the 3-form transforms as
 \be
  \Delta_{\Xi}C_{\mu\nu\rho}  \ = \ 3\,{\cal F}_{[\mu\nu}\bullet \Xi_{\rho]}\;.
 \ee
Indeed, we then find with (\ref{varyfieldH})
 \be
 \begin{split}
  \delta_{\Xi} {\cal H}_{\mu\nu\rho} \ &= \ 3\big[{\cal D}_{[\mu},{\cal D}_{\nu}\big]\Xi_{\rho]}
  +3\,\widehat\partial\Xi_{[\mu}\bullet {\cal F}_{\nu\rho]}+\widehat\partial\big(3\,{\cal F}_{[\mu\nu}\bullet \Xi_{\rho]}\big)\\[1ex]
  \ &= \ -3\, \mathbb{L}_{{\cal F}_{[\mu\nu}}\Xi_{\rho]}
  +3\,\widehat\partial\Xi_{[\mu}\bullet {\cal F}_{\nu\rho]}+\widehat\partial\big(3\,{\cal F}_{[\mu\nu}\bullet \Xi_{\rho]}\big)
   \ = \ 0\;,
 \end{split}
 \ee
using again the magic identity (\ref{magic}) in the last step.
Finally, the 3-form potential $C_{\mu\nu\rho}$ has its own associated gauge symmetry
with 2-form parameter $\Theta_{\mu\nu}$, which acts on the fields as
 \be
   \Delta_{\Theta} C_{\mu\nu\rho} \ = \ 3\,{\cal D}_{[\mu}\Theta_{\nu\rho]}\;,
   \qquad
   \Delta_{\Theta}B_{\mu\nu} \ = \ -\widehat\partial\Theta_{\mu\nu}\;, \qquad
   \delta_{\Theta} A_{\mu} \ = \ 0\;.
 \ee
Gauge invariance of the 3-form curvature then follows immediately with (\ref{varyfieldH}) and
the commutativity of ${\cal D}_{\mu}$ and $\widehat\partial$.

Up to now we have presented all technical details of
the proofs, which make repeatedly use of the identities of the Cartan-like calculus
developed in sec.~2. In the next and the following subsections
we will not give all proofs in similar technical detail
as they largely follow the same scheme.

\subsection{4-form potentials}
We now define a covariant field strength for the 3-form introduced above,
which in turn forces us to introduce 4-form potentials. In complete parallel to the above discussion we
do so by requiring a Bianchi identity for the 3-form field strength of the 2-form.
An additional subtlety is that, as one can quickly see, ${\cal D}_{[\mu}{\cal H}_{\nu\rho\sigma]}$
is not even zero up to $\widehat\partial$ exact terms.
This is due to the Chern-Simons terms in ${\cal H}_{\mu\nu\rho}$.
Rather, we have a Bianchi identity of the form\footnote{This is analogous to the
Chern-Simons modification familiar in string theory, leading to the
modified Bianchi identity ${\rm d}{H} =-{\rm tr}(F\wedge F)$ in presence of
Yang-Mills gauge fields.}
\be
4\,\mc{D}_{[\mu}\mc{H}_{\nu\rho\sigma]}+3\,\mc{F}_{[\mu\nu}\bullet\mc{F}_{\rho\sigma]}
\ = \ \widehat\partial \mc{J}_{\mu\nu\rho\sigma}\;, \label{Bianchi4}
\ee
for some 4-form field strength $\mc{J}_{\mu\nu\rho\sigma}\in\mathfrak{C}(\fracs{1}{2})$ to be determined,
for which we also write
\be\label{CALLJ}
\mc{J}_{\mu\nu\rho\sigma}  \ \equiv \  J_{\mu\nu\rho\sigma}+\widehat\partial D_{\mu\nu\rho\sigma}\;,
\ee
with the newly introduced 4-form potential $D_{\mu\nu\rho\sigma}\in\mathfrak{C}(\fracs{1}{2})$.
Inserting the definition of ${\cal H}$ and ${\cal F}$ we obtain, after a somewhat tedious computation
using in particular (\ref{3identity}) specialized to $C=\partial A$,
\be
\bsp
J_{\mu\nu\rho\sigma} \ = \ &\,4\,\mc{D}_{[\mu}C_{\nu\rho\sigma]}+3\,\widehat\partial B_{[\mu\nu}\bullet B_{\rho\sigma]}-6\, \mc{F}_{[\mu\nu}\bullet B_{\rho\sigma]}\\
&+4\,A_{[\mu}\bullet(A_\nu\bullet\ptl_\rho A_{\sigma]})-A_{[\mu}\bullet(A_\nu\bullet[A_\rho,A_{\sigma]}]_{\rm E}) \; .
\end{split}
\ee
Again, this form is only determined by (\ref{Bianchi4}) up to $\widehat\partial$ closed terms,
but we will see that any such ambiguity can be absorbed into $D_{\mu\nu\rho\sigma}$.
Since the left-hand side of the Bianchi identity is manifestly gauge covariant, it follows
that $J$ is gauge covariant up to $\widehat\partial$ exact terms and hence that $\mc{J}$
is fully gauge covariant upon assigning a suitable gauge transformation to the 4-form $D_{\mu\nu\rho\sigma}$.

In order to determine the gauge transformations that make ${\cal J}$ fully gauge covariant,
again we first give its general variation under arbitrary $\delta A_{\mu}$, $\delta B_{\mu\nu}$,
$\delta C_{\mu\nu\rho}$ and $\delta D_{\mu\nu\rho\sigma}$, which can be written as
\be\label{deltacalJ}
\bsp
\delta {\cal J}_{\mu\nu\rho\sigma} \ = \ 4\,\mc{D}_{[\mu}\,\Delta C_{\nu\rho\sigma]}-4\,\delta A_{[\mu}\bullet\mc{H}_{\nu\rho\sigma]}-6\, \mc{F}_{[\mu\nu}\bullet \Delta B_{\rho\sigma]}+\widehat\partial\Delta D_{\mu\nu\rho\sigma}\;, 
\end{split}
\ee
upon defining the covariant variation of $D_{\mu\nu\rho\sigma}$ as follows 
 \be
  \Delta D_{\mu\nu\rho\sigma} \ \equiv \ \delta D_{\mu\nu\rho\sigma}
  -4\,\delta A_{[\mu}\bullet C_{\nu\rho\sigma]}+3\,B_{[\mu\nu}\bullet(\delta B_{\rho\sigma]}+2A_\rho\bullet\delta A_{\sigma]})+A_{[\mu}\bullet(A_\nu\bullet(A_\rho\bullet\delta A_{\sigma]}))\,.
 \ee
We can now use this relation in order to show that ${\cal J}_{\mu\nu\rho\sigma}$
is gauge covariant under $\Lambda$ transformations provided we set
 \be
  \Delta_\Lambda D_{\mu\nu\rho\sigma} \ = \ \Lambda \bullet \mc{J}_{\mu\nu\rho\sigma}\;.
 \ee
Indeed, inserting this, (\ref{deltaAgauge}), (\ref{DeltaBLam}) and (\ref{DELTALamC}) into (\ref{deltacalJ})
we obtain
 \be
  \begin{split}
    \delta_{\Lambda}{\cal J}_{\mu\nu\rho\sigma} \ &= \ 4\,{\cal D}_{[\mu}(\Lambda\bullet
    {\cal H}_{\nu\rho\sigma]})-4\,{\cal D}_{[\mu}\Lambda \bullet {\cal H}_{\nu\rho\sigma]}
    -6\,{\cal F}_{[\mu\nu}\bullet (\Lambda\bullet {\cal F}_{\rho\sigma]})
    +\widehat\partial(\Lambda\bullet {\cal J}_{\mu\nu\rho\sigma}) \\[1ex]
    \ &= \ \Lambda\bullet (4\,{\cal D}_{[\mu}{\cal H}_{\nu\rho\sigma]})
     -6\,{\cal F}_{[\mu\nu}\bullet (\Lambda\bullet {\cal F}_{\rho\sigma]})
    +\widehat\partial(\Lambda\bullet {\cal J}_{\mu\nu\rho\sigma})\\[1ex]
    \ &= \ \Lambda\bullet \widehat\partial{\cal J}_{\mu\nu\rho\sigma}+\widehat\partial(\Lambda\bullet
    {\cal J}_{\mu\nu\rho\sigma})
    -3\,\Lambda\bullet ({\cal F}_{[\mu\nu}\bullet {\cal F}_{\rho\sigma]})
    -6\,{\cal F}_{[\mu\nu}\bullet (\Lambda\bullet {\cal F}_{\rho\sigma]})\;,
  \end{split}
 \ee
where we used the Bianchi identity (\ref{Bianchi4}) in the last line.
With the associativity-type relation (\ref{AAAprop}) we infer that the last two terms
in here are zero. The first two terms in the last line combine into the generalized Lie derivative
by the magic identity (\ref{magic}), hence we have shown
that ${\cal J}$ transforms covariantly,
 \be
  \delta_{\Lambda}{\cal J}_{\mu\nu\rho\sigma} \ = \ \mathbb{L}_{\Lambda}{\cal J}_{\mu\nu\rho\sigma}\;.
 \ee
Similarly, it is straightforward to verify that ${\cal J}_{\mu\nu\rho\sigma}$ is also gauge invariant
under the gauge transformations parametrized by $\Xi$ and $\Theta$, which act on
the 4-form as
 \be
  \begin{split}
   \Delta_\Xi D_{\mu\nu\rho\sigma} \ &= \ -4\,\Xi_{[\mu}\bullet\mc{H}_{\nu\rho\sigma]}\;, \\[1ex]
   \Delta_\Theta D_{\mu\nu\rho\sigma} \ &= \ 6\,\mc{F}_{[\mu\nu}\bullet\Theta_{\rho\sigma]}\;.
  \end{split}
\ee
To show the invariance one has to use in particular the property (\ref{BBprop}).

Finally, the 4-form $D_{\mu\nu\rho\sigma}$ has an associated gauge symmetry parametrized by a 3-form parameter
$\Omega_{\mu\nu\rho}\in\mathfrak{D}(\fracs{2}{3})$,
 \be
  \Delta_{\Omega}D_{\mu\nu\rho\sigma} \ = \ 4\,{\cal D}_{[\mu}\Omega_{\nu\rho\sigma]}\;.
 \ee
This leaves the field strength ${\cal J}_{\mu\nu\rho\sigma}$ invariant
provided this symmetry acts on the lower-form potentials as
\be
\delta_\Omega A_\mu \ = \ 0\;, \qquad
 \Delta_\Omega B_{\mu\nu} \ = \ 0\;, \qquad
 \Delta_\Omega C_{\mu\nu\rho} \ = \ -\widehat\partial\Omega_{\mu\nu\rho}   \;,
\ee
which follows immediately with (\ref{deltacalJ}).

\subsection{5-form potentials}
We complete the tensor hierarchy (needed for the SL$(3)\times {\rm SL}(2)$ EFT)
by introducing the 5-form potentials, starting again from the non-trivial Bianchi identity, which
here reads
\be
5\,\mc{D}_{[\mu}\mc{J}_{\nu\rho\sigma\tau]}+10\,\mc{F}_{[\mu\nu}\bullet\mc{H}_{\rho\sigma\tau]}
\ = \ \widehat\partial {\cal K}_{\mu\nu\rho\sigma\tau}\; ,\label{Bianchi5}
\ee
with the field strengths ${\cal K}_{\mu\nu\rho\sigma\tau}$ for the 4-form to be determined.
As before we also write
\be\label{CALLK}
\mc{K}_{\mu\nu\rho\sigma\tau} \ = \ K_{\mu\nu\rho\sigma\tau}+\widehat\partial E_{\mu\nu\rho\sigma\tau}\; .
\ee
with a 5-form potential $E_{\mu\nu\rho\sigma\tau,m\alpha}\in\mathfrak{E}(\fracs{5}{6})$
that drops out of the Bianchi identity but is needed for the 5-form curvature to be fully
gauge covariant.
Inserting the above definitions of the field strengths on the left-hand side of (\ref{Bianchi5}) one computes
for $K$ (up to $\widehat\partial$ exact terms)
\be
\bsp
K_{\mu\nu\rho\sigma\tau} \ = \ \, &5\,\mc{D}_{[\mu}D_{\nu\rho\sigma\tau]}+15\, B_{[\mu\nu}\bullet\mc{D}_\rho B_{\sigma\tau]}-10\, \mc{F}_{[\mu\nu}\bullet C_{\rho\sigma\tau]}\\[1ex]
&+30\,B_{[\mu\nu}\bullet(-A_\rho\bullet\ptl_\sigma A_{\tau]}+\fracs{1}{3}A_\rho\bullet[A_\sigma,A_{\tau]}]_E)\\[1ex]
&-5\,A_{[\mu}\bullet (A_\nu\bullet (A_\rho\bullet\ptl_\sigma A_{\tau]}))+A_{[\mu}\bullet (A_\nu\bullet (A_\rho\bullet[A_\sigma,A_{\tau]}]_E))\;.
\end{split}
\ee
The general variation takes the covariant form
 \be
 \begin{split}
  \delta {\cal K}_{\mu\nu\rho\sigma\tau} \ = \ \, &
  5\, {\cal D}_{[\mu}\Delta D_{\nu\rho\sigma\tau]}-5\,\delta A_{[\mu}\bullet {\cal J}_{\nu\rho\sigma\tau]}
  -10\,{\cal F}_{[\mu\nu}\bullet \Delta C_{\rho\sigma\tau]}\\[1ex]
  &-10\,{\cal H}_{[\mu\nu\rho}\bullet \Delta B_{\sigma\tau]}
  +\widehat\partial(\Delta E_{\mu\nu\rho\sigma\tau})\;,
 \end{split}
 \ee
where
 \be
 \begin{split}
  \Delta E_{\mu\nu\rho\sigma\tau} \ = \ \, &\delta E_{\mu\nu\rho\sigma\tau}-5\,\delta A_{[\mu}\bullet 
  D_{\nu\rho\sigma\tau]}-10\,\delta B_{[\mu\nu}\bullet C_{\rho\sigma\tau]}\\[1ex] 
  & -15\, B_{[\mu\nu}\bullet(\delta A_\rho\bullet B_{\sigma\tau]})
  -10\, (A_{[\mu}\bullet \delta A_\nu)\bullet C_{\rho\sigma\tau]}\\[1ex]
 & 
 +10\, B_{[\mu\nu}\bullet (A_\rho\bullet(A_\sigma\bullet\delta A_{\tau]}))+A_{[\mu}\bullet (A_\nu\bullet (A_\rho\bullet (A_\sigma\bullet \delta A_{\tau]})))\:,
  \end{split}
 \ee
The 5-form field strength then transforms covariantly under $\Lambda$ by setting
 \be
  \Delta _{\Lambda}E_{\mu\nu\rho\sigma\tau} \ = \ \Lambda\bullet {\cal K}_{\mu\nu\rho\sigma\tau} \;.
 \ee
Similarly, it is invariant under the previously discussed gauge symmetries parametrized by
$\Xi$, $\Theta$, $\Omega$, acting on the 5-form as
 \be
  \Delta E_{\mu\nu\rho\sigma\tau} \ = \ -5\,{\cal J}_{[\mu\nu\rho\sigma}\bullet \Xi_{\tau]}
-10\,{\cal H}_{[\mu\nu\rho}\bullet \Theta_{\sigma\tau]}+10\,{\cal F}_{[\mu\nu}\bullet \Omega_{\rho\sigma\tau]} \;.
 \ee
Finally, the 5-form is associated to a new gauge symmetry, with 4-form parameter
$\Upsilon_{\mu\nu\rho\sigma\, m\alpha}\in\mathfrak{E}(\fracs{5}{6})$,
\be
 \Delta_\Upsilon E_{\mu\nu\rho\sigma\tau} \  = \ 5\,\mc{D}_{[\mu}\Upsilon_{\nu\rho\sigma\tau]}\; .
\ee
It leaves all field strengths invariant provided it acts on the lower-form potentials as
\be
\delta_\Upsilon A_\mu \ = \ \delta_\Upsilon B_{\mu\nu} \ = \
\delta_\Upsilon C_{\mu\nu\rho} \ = \ 0\,,\ \quad
\delta_\Upsilon D_{\mu\nu\rho\sigma} \ = \ - \widehat\partial\Upsilon_{\mu\nu\rho\sigma}\;.
\ee

\medskip

For the convenience of the reader we summarize this section on the tensor hierarchy by
giving the action of all gauge
symmetries and the form of the Bianchi identities. The form potentials $A$, $B$, $C$, $D$
and $E$ transform as
 \be
  \begin{split}
    \delta A_\mu \ &= \ \mathcal{D}_\mu \Lambda - \widehat\partial\, \Xi_{\mu}\;, \\[1ex]
    \Delta B_{\mu\nu} \ &= \  2\,\mc{D}_{[\mu}\Xi_{\nu]}
    +\Lambda\bullet {\cal F}_{\mu\nu}-\widehat\partial\Theta_{\mu\nu}\;, \\[1ex]
    \Delta C_{\mu\nu\rho} \ &= \ 3\,{\cal D}_{[\mu}\Theta_{\nu\rho]}
    +\Lambda\bullet {\cal H}_{\mu\nu\rho}
    + 3\,{\cal F}_{[\mu\nu}\bullet \Xi_{\rho]}
    -\widehat\partial\Omega_{\mu\nu\rho} \\[1ex]
    \Delta D_{\mu\nu\rho\sigma} \ &= \ 4\,{\cal D}_{[\mu}\Omega_{\nu\rho\sigma]}
    +\Lambda \bullet \mc{J}_{\mu\nu\rho\sigma}
    - 4\, \mc{H}_{[\mu\nu\rho}\bullet\,\Xi_{\sigma]}  + 6\,\mc{F}_{[\mu\nu}\bullet\,\Theta_{\rho\sigma]}
    - \widehat\partial\Upsilon_{\mu\nu\rho\sigma}\;, \\[1ex]
    \Delta E_{\mu\nu\rho\sigma\tau} \ &= \ 5\,\mc{D}_{[\mu}\Upsilon_{\nu\rho\sigma\tau]}
    +\Lambda\bullet {\cal K}_{\mu\nu\rho\sigma\tau}-5\,{\cal J}_{[\mu\nu\rho\sigma}\bullet \Xi_{\tau]} \\[0.5ex]
  &\qquad -10\,{\cal H}_{[\mu\nu\rho}\bullet \Theta_{\sigma\tau]}+10\,{\cal F}_{[\mu\nu}\bullet \Omega_{\rho\sigma\tau]}
  +\cdots \;.
   \end{split}
  \ee
Here, in the last equation, we indicated by dots a term that is immaterial in all relations
discussed so far, but would appear as the gauge parameter of the 6-form if we continued the
construction of the hierarchy. For our present purposes it is, however, sufficient to stop here.
The field strengths of these potentials, defined in (\ref{improvedF}), (\ref{CSLF}),
(\ref{CALLJ}) and (\ref{CALLK}) are fully covariant under these symmetries and satisfy the following Bianchi
identities
 \be\label{collectBianchi}
  \begin{split}
   3\,\mc{D}_{[\mu}\mc{F}_{\nu\rho]} \ &= \ \widehat\partial\, {\cal H}_{\mu\nu\rho}\;, \\[1ex]
   4\,\mc{D}_{[\mu}\mc{H}_{\nu\rho\sigma]}+3\,\mc{F}_{[\mu\nu}\bullet\mc{F}_{\rho\sigma]}
   \ &= \ \widehat\partial \mc{J}_{\mu\nu\rho\sigma}\;,  \\[1ex]
   5\,\mc{D}_{[\mu}\mc{J}_{\nu\rho\sigma\tau]}+10\,\mc{F}_{[\mu\nu}\bullet\mc{H}_{\rho\sigma\tau]}
    \ &= \ \widehat\partial {\cal K}_{\mu\nu\rho\sigma\tau}\;.
   \end{split}
  \ee

\section{The exceptional field theory action}
In this section we define the complete dynamics of the SL$(3)\times {\rm SL}(2)$ 
exceptional field theory. We first define the various terms of the (pseudo-)action: 
the kinetic terms, the potential terms (i.e.~carrying only internal derivatives)
and finally a topological Chern-Simons like action that is needed 
for compatibility with the first-order duality relations to be imposed at the level 
of the field equations.

\subsection{Kinetic terms}
We start by giving the total bosonic  field content, which consists of 
 \be\label{propFIELD}
  \{\, g_{\mu\nu},\, {\cal M}_{MN},\, A_{\mu}{}^{i\alpha},\, B_{\mu\nu\, m},\, 
  C_{\mu\nu\rho}{}^{\alpha}\,,\;D_{\mu\nu\rho\sigma}{}^{m}\,\}\;. 
 \ee 
Formally, we may also keep the 5-forms  $E_{\mu\nu\rho\sigma\tau\,m\alpha}$
in order to make gauge covariance of all curvatures manifest, although 
we will see that the 5-forms and their variations drop out of the action. 
In here, all first five fields enter with a kinetic term, while the 4-form $D$ is topological 
in that it 
only enters via topological terms and as modifications of curvatures. The action reads 
\be
 S \ = \ \int {\rm d}^8 x\, {\rm d}^6 Y e\big(\widehat{R}+\mc{L}_{\rm{kin}}+e^{-1}\mc{L}_{\rm{top}}- V(\mc{M},g)\big)\;, 
\ee
whose various terms we will define in the following.

We begin with the Einstein-Hilbert term, which can be defined in terms of the `achtbein' $e_{\mu}{}^{a}$
that carries density weight $\lambda(e_{\mu}{}^{a}) = \tfrac{1}{6}$, 
 \be
  S_{\rm EH} \ \equiv  \ \int {\rm d}^8 x\, {\rm d}^6 Y e\,e_{a}{}^{\mu} e_{b}{}^{\nu}\,
  \widehat{R}_{\mu\nu}{}^{ab}\;. 
 \ee
Here the Riemann tensor is computed in the standard fashion, except that all partial derivatives are 
replaced by $A_{\mu}$-covariant derivatives and its definition contains an 
improvement term, 
 \be
  \widehat{R}_{\mu\nu}{}^{ab} \ \equiv  \ R_{\mu\nu}{}^{ab}+\mc{F}_{\mu\nu}{}^M e^{\rho[a}\ptl_M e_\rho{}^{b]}\;, 
\ee
which is necessary for local Lorentz invariance.  With $e_{\mu}{}^{a}$ carrying weight 
$\tfrac{1}{6}$ its determinant $e$ carries weight $\frac{4}{3}$, while $\widehat{R}$ has weight zero, 
so that the total Einstein-Hilbert Lagrangian has weight one. 
This is the right weight needed for gauge invariance, as in this case the Lagrangian 
varies into a total derivative under $\Lambda^M$ transformations, c.f.~the discussion 
around (\ref{STEPInt}). 

Next we turn to the kinetic term of the scalar matrix (or `generalized metric') ${\cal M}$, 
which lives in the coset space 
 \be
  \frac{{\rm SL}(3)}{{\rm SO}(3)} \times \frac{{\rm SL}(2)} {{\rm SO}(2)}\;, 
 \ee 
encoding $7$ physical degrees of freedom. Because of this product structure of the 
duality group we have two generalized metrics, the SL$(3)$ and SL$(2)$ 
valued matrices ${\cal M}_{ij}$ and ${\cal M}_{\alpha\beta}$, respectively. 
Often it is convenient to represent them as a matrix in the $(3,2)$ representation, 
\be\label{SL2SL3Deco}
  \mc{M}_{MN} \ \equiv \ \mc{M}_{i\alpha,j\beta} \ = \ \mc{M}_{ij}\mc{M}_{\alpha\beta}\;. 
\ee  
The matrices here all satisfy ${\rm det}\,{\cal M}=1$, which is compatible 
with the gauge symmetries for density weight $\lambda=0$. 
The manifestly gauge invariant kinetic term is then given by 
\be
 {\cal L}_{{\rm kin},{\cal M}} \ = \ 
 \frac{1}{4}\big( \mc{D}^\mu\mc{M}^{ij}\,\mc{D}_\mu\mc{M}_{ij}
 +\mc{D}^\mu\mc{M}^{\alpha\beta}\,\mc{D}_\mu\mc{M}_{\alpha\beta}\big)\;, 
\ee 
where the coefficients will be determined below. It is again straightforward to 
see that the Lagrangian has the correct total weight:
the inverse metric $g^{\mu\nu}$ implicit in the contraction of indices has weight $-\frac{1}{3}$
which combines with the weight $\frac{4}{3}$ of $e$ to a total weight of one 
needed for gauge invariance. 

The kinetic terms for the remaining three (tensor-)fields in (\ref{propFIELD}) can similarly be 
written in a manifestly gauge invariant fashion, 
 \be\label{KINTERMS}
\mc{L}_{{\rm kin,tensor}} \  = \  
-\frac{1}{4} \mc{M}_{MN}\mc{F}^{\mu\nu M}\mc{F}_{\mu\nu}{}^N
-\frac{1}{12} \mc{M}^{mn}\mc{H}^{\mu\nu\rho}{}_m\mc{H}_{\mu\nu\rho\, n}-\frac{1}{96} 
\mc{M}_{\alpha\beta}\mc{J}^{\mu\nu\rho\sigma,\alpha}\mc{J}_{\mu\nu\rho\sigma}{}^\beta\;, 
\ee
 in terms of the covariant curvatures defined in (\ref{explicitF}), (\ref{explcitH}) and (\ref{CALLJ}). 
 It is again straightforward to verify that the density weights determined in the previous section 
 from the consistency of the tensor hierarchy are precisely the correct ones that make
 the action corresponding to this Lagrangian gauge invariant.

\subsection{Potential terms}
We now turn to the potential terms that are characterized by using only `internal' derivatives 
$\partial_M$. Its form is determined by $\Lambda^M$ gauge invariance (up to one free coefficient 
that, however, is universal in all EFTs) and reads in the present case 
\be\label{FInalV}
\bsp
V\ =\ &-\fracs{1}{4}\mc{M}^{MN}\ptl_M\mc{M}^{kl}\ptl_N\mc{M}_{kl}-\fracs{1}{4}\mc{M}^{MN}\ptl_M\mc{M}^{\alpha\beta}\ptl_N\mc{M}_{\alpha\beta}+\fracs{1}{2}\mc{M}^{MN}\ptl_M\mc{M}^{KL}\ptl_K\mc{M}_{LN}\\
&-\fracs{1}{2}\ptl_M\mc{M}^{MN}g^{-1}\ptl_N g-\fracs{1}{4}\mc{M}^{MN}g^{-1}\ptl_M g g^{-1} \ptl_N g-\fracs{1}{4}\mc{M}^{MN}\ptl_M  g^{\mu\nu} \ptl_N g_{\mu\nu}\; , 
\end{split}
\ee
where we used the decomposition (\ref{SL2SL3Deco}) of ${\cal M}_{MN}$ into SL$(3)$ and SL$(2)$ 
matrices, with the standard notation ${\cal M}^{-1\,ij}\equiv {\cal M}^{ij}$, etc. 
Note that, in contrast to the EFT of simple duality groups,  
the first two terms cannot be written in the form $\mc{M}^{MN}\ptl_M\mc{M}^{KL}\ptl_N\mc{M}_{KL}$, 
but this is consistent since the form given is SL$(3)\times {\rm SL}(2)$ invariant. 
In order to bring the potential into a more geometric form we may introduce internal curvatures and 
covariant derivatives and define 
 \be
  \nabla_M g_{\mu\nu} \ = \ \partial_M g_{\mu\nu}-\frac{1}{4}(e^{-1}\partial_Me) g_{\mu\nu}\;, 
 \ee
which transforms covariantly. Up to total derivatives, the potential terms may then be written 
in the form given in the first line of (\ref{actionIntro}), where the generalized Ricci scalar ${\cal R}$
can be computed by taking the variational derivative w.r.t.~the vielbein determinant, 
 \be
  {\cal R} \ = \ \frac{\delta}{\delta e}\Big(-eV-\frac{1}{4}e{\cal M}^{MN}\nabla_Mg^{\mu\nu}\nabla_N g_{\mu\nu}\Big)\;, 
 \ee  
where we note that, despite appearance, the expression in parenthesis depends only on $e$, 
not the full metric, and so the variation is well-defined.  One may also construct ${\cal R}$ 
geometrically, defining connections and curvatures, in analogy 
to DFT \cite{Siegel:1993th,Hohm:2010xe}, but we will not do so here.

Let us now return to the expression (\ref{FInalV}) and confirm 
the $\Lambda^M$ gauge invariance directly by computing  
the `non-covariant' variation of each term. More precisely, this variation is defined as  
$\Delta_{\rm nc}=\delta_\Lambda-\mathbb{L}_\Lambda$, and we have to verify that 
the total variation of the potential combines into a total $\partial_M$ derivative. 
The density weight of the action of $\mathbb{L}_\Lambda$ is determined by the requirement 
that $\Delta_{\rm nc}$ contains only second derivatives of the 
gauge parameter (i.e.~$\partial\partial\Lambda$ terms). 
Let us illustrate this for ${\cal M}^{kl}$, whose gauge variation can be read off from (\ref{GenLie31}), 
 \be
  \delta_{\Lambda}{\cal M}^{kl} \ = \ \Lambda^N\partial_N{\cal M}^{kl}
  -2\,{\cal M}^{p(k}\,\partial_{p\gamma}\Lambda^{l)\gamma}
  +\tfrac{2}{3}\,\partial_N\Lambda^N {\cal M}^{kl}\;, 
 \ee
where we recalled that the density weight is $\lambda=0$ for ${\cal M}^{ij}\in {\rm SL}(3)$.  
This determines the gauge variation of $\partial_M{\cal M}^{kl}$, which has to be 
compared with its Lie derivative, 
 \be\label{LiepartialDER}
 \begin{split}
  \mathbb{L}_{\Lambda}\big(\partial_M{\cal M}^{kl}) \ \equiv \ &\,
  \Lambda^N\partial_N\big(\partial_M{\cal M}^{kl}) +\partial_M\Lambda^N \partial_N{\cal M}^{kl}
  -2\,\partial_M{\cal M}^{p(k}\,\partial_{p\gamma}\Lambda^{l)\gamma} \\
  &+\big(\lambda(\partial {\cal M})+\tfrac{1}{6}+\tfrac{2}{3}\big)\partial_N\Lambda^N\partial_M{\cal M}^{kl}\;. 
 \end{split}
 \ee 
Here we used (\ref{COVVarDown}), (\ref{GenLie31}) for the definition of the 
generalized Lie derivative and the section constraint (\ref{Zsection}).  
One then finds for the non-covariant variation 
 \be
\bsp
\Delta_{\rm nc}(\ptl_M\mc{M}^{kl}) \ \equiv \ &\,\ptl_M(\delta_\Lambda\mc{M}^{kl})
-\mathbb{L}_\Lambda(\ptl_M\mc{M}^{kl})\\
\ =\ &-2\,\mc{M}^{p(k}\,\ptl_{p\gamma}\ptl_M\Lambda^{l)\gamma}
+\fracs{2}{3}\mc{M}^{kl}\ptl_M\ptl_{j\beta}\Lambda^{j\beta}\;,
\end{split}
\ee
where the weight is determined to be $\lambda=-\tfrac{1}{6}-\tfrac{2}{3}$, so that 
the density term in the second line of (\ref{LiepartialDER}) vanishes. 
A similar formula holds for $\Delta_{\rm nc}(\ptl_M\mc{M}_{kl})$, where we note that now the density 
weight is $\lambda=-\tfrac{1}{6}+\tfrac{2}{3}$. 
By a completely analogous computation one finds for ${\cal M}^{\alpha\beta}$
\be
\Delta_{\rm nc}(\ptl_M\mc{M}^{\alpha\beta})\ =\ -2\,\mc{M}^{\gamma(\alpha}\,\ptl_M\ptl_{k\gamma}\Lambda^{|k|\beta)}+\mc{M}^{\alpha\beta}\ptl_M\ptl_{k\gamma}\Lambda^{k\gamma}\;.
\ee

It is now easy to see that all covariant terms in the variation of the potential combine 
into total derivatives. For instance, in the first term in (\ref{FInalV}) the weights of 
$\ptl_M\mc{M}^{kl}$ and $\ptl_M\mc{M}_{kl}$ add up to $-\tfrac{1}{3}$, which combines 
with the weight $\tfrac{4}{3}$ of the vielbein determinant $e$ to a total weight of $1$, 
exactly as needed for gauge invariance, see (\ref{STEPInt}). 
Thus, it remains to verify the cancellation of all non-covariant variations $\Delta_{\rm nc}$. 
To this end one has to use that the current 
$(J_K)_L{}^M\equiv\mc{M}^{MN}\ptl_K\mc{M}_{LN}$, which decomposes as 
 \be
   \mc{M}^{k\alpha,l\delta}\ptl\mc{M}_{j\beta,l\delta} \ = \ 
   \mc{M}^{kl}\ptl\mc{M}_{jl}\, \delta^\alpha_\beta+\mc{M}^{\alpha\delta}\ptl  
    \mc{M}_{\beta\delta}\, \delta^k_j\;, 
\ee
takes values in the Lie algebra $\frak{sl}(3)\oplus \frak{sl}(2)$. 
Consequently, the invariance of the $Z$ tensor (\ref{Ztensor}) 
implies identities like
\be
Z^{L(M}{}_{PQ}J_L{}^{R)}-Z^{MR}{}_{(P|L}J_{|Q)}{}^{L} \ = \ 0\;.
\ee
The invariance of the potential now follows by direct computation. 
(For more details see, for instance, the E$_{6(6)}$ case 
discussed in \cite{E6}.)

\subsection{Topological terms}
Finally, the action requires terms that are topological (or of Chern-Simons type) 
in the sense that they
can be defined using only the form fields, not the external metric $g_{\mu\nu}$ 
nor the internal generalized metric ${\cal M}_{MN}$.  
Most conveniently, this term is defined by viewing the 8-dimensional 
`external' space as the boundary of a 9-dimensional space, on which the 
topological term takes the form of a manifestly gauge invariant total derivative 
term. As such, it effectively reduces to an 8-dimensional action that is 
gauge invariant (albeit not manifestly) up to boundary terms. 
We find for the 9-dimensional form of the action, written in terms of 
the gauge covariant curvatures ${\cal F}$, ${\cal H}$ and ${\cal J}$,  
\be
\begin{split}
S_{\rm top} \ = \ 
\kappa\int {\rm d}^9 x \,{\rm d}^6 Y\, \epsilon^{\mu_1\cdots \mu_9}
\Big[&\mc{J}_{\mu_1\cdots\mu_4}\bullet\mc{D}_{\mu_5}\mc{J}_{\mu_6\ldots \mu_9}
+4\, \mc{J}_{\mu_1\cdots\mu_4}\bullet\big(\mc{F}_{\mu_5\mu_6}\bullet\mc{H}_{\mu_7\cdots\mu_9}\big)\\[0.5ex]
&-\fracs{8}{9}\,\mc{H}_{\mu_1\cdots\mu_3}\bullet\big(\mc{H}_{\mu_4\cdots\mu_6}
\bullet\mc{H}_{\mu_7\cdots \mu_9}\big)\Big]\;, \label{topological}
\end{split}
\ee
where by slight abuse of notation we momentarily denote by $\mu,\nu,\ldots$ 
9-dimensional indices, and the overall normalization $\kappa$ will be determined below. 
Restoring explicit index notation and writing out the tensor operations $\bullet$ 
the action reads
\be
\begin{split}
S_{\rm top} \ = \ 
\kappa\int {\rm d}^9 x \,{\rm d}^6 Y\, \epsilon^{\mu_1\cdots \mu_9}
\Big[&\,  \epsilon_{\alpha\beta}\big(\mc{J}_{\mu_1\cdots\mu_4}{}^{\alpha}\,
\mc{D}_{\mu_5}\mc{J}_{\mu_6\ldots \mu_9}{}^{\beta}
+ 4 \mc{J}_{\mu_1\cdots\mu_4}{}^{\alpha}\,\mc{F}_{\mu_5\mu_6}{}^{m\beta}\, \mc{H}_{\mu_7\cdots\mu_9\, m}\big)\\[0.5ex]
&-\fracs{8}{9}\,\epsilon^{mnk}\, \mc{H}_{\mu_1\cdots\mu_3\, m} \, \mc{H}_{\mu_4\cdots\mu_6\, n}\,
\mc{H}_{\mu_7\cdots \mu_9\, k} \Big]\;. \label{topological22}
\end{split}
\ee
Let us note that the action is indeed manifestly $\Lambda^M$ gauge invariant. 
Since the curvatures employed here are gauge covariant by construction 
it only remains to verify that the $\bullet$ operations above lead to scalar
densities of weight 1, as needed for gauge invariance of the action. 
This is indeed the case, as can be inferred from table 2. 
For instance, in the leading term we have ${\cal J}\in \mathfrak{C}(\frac{1}{2})$ and so $\bullet$ maps 
$\mathfrak{C}(\frac{1}{2})\times \mathfrak{C}(\frac{1}{2})$ into $\mathfrak{S}(1)$. 

Our task is now to verify that this topological action is a total derivative. 
We prove this by showing that it varies into a total derivative under arbitrary variations 
of the tensor fields. This proof requires a subtle interplay of the covariant variations of 
field strengths and the Bianchi identities of the tensor hierarchy. 
We illustrate this by first considering the variation only under 
$\Delta D_{\mu\nu\rho\sigma}=\delta D_{\mu\nu\rho\sigma}$, setting $\delta A=\delta B=\delta C=0$, 
under which $\delta {\cal J}_{\mu\nu\rho\sigma}=\widehat\partial(\Delta D_{\mu\nu\rho\sigma})$, see (\ref{deltacalJ}), 
while all other curvatures are inert. 
We then compute for the variation of the Lagrangian corresponding to (\ref{topological}) 
 \be
 \begin{split}
  \delta{\cal L}_{\rm top} \ &= \  \epsilon^{\mu_1\cdots \mu_9}
  \Big[\,\delta {\cal J}_{\mu_1\ldots\mu_4}\bullet {\cal D}_{\mu_5} {\cal J}_{\mu_6\cdots \mu_9}
  +{\cal J}_{\mu_1\ldots \mu_4}\bullet \mc{D}_{\mu_5}(\delta {\cal J}_{\mu_6\ldots \mu_9}) \\
  &\qquad \qquad\qquad  
  +4\, \delta \mc{J}_{\mu_1\cdots\mu_4}\bullet (\mc{F}_{\mu_5\mu_6}\bullet\mc{H}_{\mu_7\cdots\mu_9})\,\Big] \\[0.5ex]
  \ &= \ \epsilon^{\mu_1\cdots \mu_9} \Big[\, {\cal D}_{\mu_5}({\cal J}_{\mu_1\ldots\mu_4}
  \bullet \delta {\cal J}_{\mu_6\ldots\mu_9})+2\,\delta \mc{J}_{\mu_1\ldots\mu_4}\bullet ({\cal D}_{\mu_5}
  \mc{J}_{\mu_6\ldots\mu_9}+2\,\mc{F}_{\mu_5\mu_6}\bullet\mc{H}_{\mu_7\cdots\mu_9})\Big]\\[0.5ex]
  \ &= \ \epsilon^{\mu_1\cdots \mu_9} \Big[\, {\cal D}_{\mu_5}({\cal J}_{\mu_1\ldots\mu_4}
  \bullet \widehat\partial(\Delta D_{\mu_6\ldots\mu_9}))+\tfrac{2}{5}\,\widehat\partial(\Delta D_{\mu_1\ldots\mu_4})
  \bullet \widehat\partial{\cal K}_{\mu_5\ldots \mu_9}\Big] \;, 
 \end{split}
 \ee 
where we collected a total derivative term (recalling that the $\bullet$ operation is 
antisymmetric in the first term, as is manifest in (\ref{topological22}) due to the contraction 
with $\epsilon_{\alpha\beta}$),  
and we used the Bianchi identity (\ref{Bianchi5}) in the last step. 
The last term in here is a total $\partial_M$ derivative, see the discussion around (\ref{totalbarpart}), 
 and can hence be ignored 
since we still assume that the $Y$-space has no boundary. 
On the contrary, the total $x$-derivative given by the first term reduces the 
variation to that of an 8-dimensional action, i.e., 
 \be
 \begin{split}
  \delta S_{\rm top} \ &= \ \kappa \int {\rm d}^8 x \,{\rm d}^6 Y\, \epsilon^{\mu_1\cdots \mu_8}\,
  {\cal J}_{\mu_1\ldots\mu_4}  \bullet \widehat\partial(\Delta D_{\mu_5\ldots\mu_8}) \\[0.5ex]
   \ &= \ -\kappa \int {\rm d}^8 x \,{\rm d}^6 Y\, \epsilon^{\mu_1\cdots \mu_8}\,
  \widehat\partial(\Delta D_{\mu_1\ldots\mu_4}) \bullet  {\cal J}_{\mu_5\ldots\mu_8} \;, 
 \end{split}
 \ee 
using the antisymmetry of $\bullet$ in the last step.   
Similarly, one can work out the 8-dimensional form of the total variation 
using the covariant variations (\ref{deltacalF}), (\ref{varyfieldH}) and 
(\ref{deltacalJ}) of $\mc{F}$, $\mc{H}$ and $\mc{J}$, respectively, 
and employing the Bianchi identities (\ref{collectBianchi}). 
One finally finds for the total variation 
\be
\bsp
 \delta S_{\rm top} 
 \ = \ \, \kappa\int {\rm d}^8 x\, {\rm d}^6 Y\, \epsilon^{\mu_1\ldots\mu_8} 
 \Big[ \,& 4\,\mc{J}_{\mu_1\ldots \mu_4}\bullet(\delta A_{\mu_5}\bullet \mc{H}_{\mu_6\ldots\mu_8})\\
 & +6\,\Delta B_{\mu_1\mu_2}\bullet(\mc{F}_{\mu_3\mu_4}\bullet\mc{J}_{\mu_5\ldots\mu_8}
 -\fracs{4}{9}\mc{H}_{\mu_3\ldots\mu_5}\bullet\mc{H}_{\mu_6\ldots\mu_8})\\[0.5ex]
& +4\,\Delta C_{\mu_1\ldots\mu_3}\bullet(\mc{D}_{\mu_4}\mc{J}_{\mu_5\ldots\mu_8}
+4\,\mc{F}_{\mu_4\mu_5}\bullet\mc{H}_{\mu_6\ldots\mu_8}) \\
&-\widehat\partial \Delta D_{\mu_1\ldots\mu_4}\bullet\mc{J}_{\mu_5\ldots\mu_8}\Big]\;. \label{varitop}
\end{split}
\ee
Note that the variation of the 5-form potential is absent, showing that it drops out of the theory. 

We close this section by explaining how, thanks to the topological terms, 
the field equations are consistent with the self-duality relation present in type IIB. 
Specifically, the 4-form potentials $D_{\mu\nu\rho\sigma}{}^{m}$ do not carry kinetic terms,  
but due to their presence inside covariant field strengths and topological terms 
their variation yields (a projection of) the self-duality constraints of the 
3-forms $C_{\mu\nu\rho}{}^{\alpha}$. These in turn encode the degrees of freedom 
of the self-dual 4-form of type IIB, as we shall discuss below. 
Consider the variation of the Lagrangian, whose relevant parts 
consist of the kinetic terms in (\ref{KINTERMS}) and the topological terms in 
(\ref{topological22}), w.r.t.~$D_{\mu\nu\rho\sigma}{}^{m}$, 
 \be
 \begin{split}
  \delta_D {\cal L} 
  \ &= \ 
  -\tfrac{1}{48}\,e{\cal M}_{\alpha\beta} \mc{J}^{\mu\nu\rho\sigma,\alpha}\,\delta
  \mc{J}_{\mu\nu\rho\sigma}{}^{\beta}
  -\kappa \,\epsilon^{\mu\nu\rho\sigma\lambda_1\ldots\lambda_4}\,
  \epsilon_{\alpha\beta}\,(\widehat\partial\Delta D_{\mu\nu\rho\sigma})^{\alpha}
  {\cal J}_{\lambda_1\ldots\lambda_4}{}^{\beta}\\[0.5ex] 
  \ &= \ \delta D_{\mu\nu\rho\sigma}{}^{m}\Big[\,
  \tfrac{1}{48}\,\epsilon^{\beta\gamma}\partial_{m\gamma}
  \big(e{\cal M}_{\alpha\beta}\mc{J}^{\mu\nu\rho\sigma,\alpha}\big)
  -\epsilon^{\beta\gamma}\partial_{m\gamma}\big(\kappa\epsilon_{\alpha\beta}
  \epsilon^{\mu\nu\rho\sigma\lambda_1\ldots\lambda_4}\mc{J}_{\lambda_1\ldots\lambda_4}{}^{\alpha}\big)\,\Big]\;.  
 \end{split}
 \ee 
Here we used (\ref{varitop}) for the variation of the topological term, and we integrated by parts
in the second line. 
Thus, the field equations for $D_{\mu\nu\rho\sigma}{}^{m}$ read 
 \be
  \epsilon^{\beta\gamma}\partial_{m\gamma}\Big[\, 
  \tfrac{1}{48}\,e\,{\cal M}_{\alpha\beta}\,\mc{J}^{\mu\nu\rho\sigma,\alpha}
 - \kappa\,\epsilon_{\alpha\beta}\,
  \epsilon^{\mu\nu\rho\sigma\lambda_1\ldots\lambda_4}\,\mc{J}_{\lambda_1\ldots\lambda_4}{}^{\alpha}\,\Big] \ = \ 0\;. 
 \ee 
This is a projected self-duality relation. It is projected, because it appears only under the 
differential operator $\epsilon^{\beta\gamma}\partial_{m\gamma}$. 
The action does not imply the full set of self-duality relations, and therefore 
we have to impose the complete self-duality relations by hand, 
 \be\label{EFTSELFDUAL}
  \tfrac{1}{48}\,{\cal M}_{\alpha\beta}\,\mc{J}^{\mu\nu\rho\sigma,\beta}
 \ = \ - \kappa\,\epsilon_{\alpha\beta}\,
  e^{-1}\epsilon^{\mu\nu\rho\sigma\lambda_1\ldots\lambda_4}\,\mc{J}_{\lambda_1\ldots\lambda_4}{}^{\beta}\;, 
 \ee
to be 
imposed at the level of the field equations after varying the (pseudo-)action.  
Let us emphasize again that it is only consistent to impose the self-duality relations 
due to the topological terms in the action. 
Note that consistency of the self-duality relations determines $\kappa$ to be 
 \be
  \kappa \ = \ \frac{1}{2(24)^2}\;. 
 \ee

\section{External diffeomorphisms}
So far we dealt exclusively with the `internal' generalized diffeomorphisms 
generated by $\Lambda^M(x,Y)$ and their higher-form descendants 
emerging in the tensor hierarchy. These gauge symmetries are made completely 
manifest thanks to the novel calculus introduced above. 
Here we turn to the equally important symmetry of `external' generalized 
diffeomorphisms generated by the 8 parameters $\xi^{\mu}(x,Y)$, 
which is a non-manifest symmetry (that, accordingly, fixes all relative coefficients in the action). 
We first discuss the gauge algebra and then the invariance of the action. 

\subsection{Gauge algebra of external diffeomorphisms} 

We start by defining the external diffeomorphisms and confirming their consistency 
by proving closure of the gauge algebra. 
The external diffeomorphisms act on the external and internal metrics as   
\be\label{deltaMdeltag}
 \bsp
 \delta_\xi\mc{M}_{MN} \ = \ &\, \xi^\mu\mc{D}_\mu\mc{M}_{MN}\;, \\
  \delta_\xi g_{\mu\nu} \ = \ 
  &\, \xi^\rho\mc{D}_\rho g_{\mu\nu}+\mc{D}_\mu\xi^\rho g_{\rho\nu}+\mc{D}_\nu\xi^\rho g_{\rho\mu}\;. 
 \end{split}
 \ee
This takes the same form as conventional infinitesimal diffeomorphisms, except that all 
derivatives are covariant w.r.t.~the connection $A_{\mu}$ of the separate (internal) diffeomorphism
symmetry. Here we treat the parameter $\xi^{\mu}$ as a scalar of weight zero, 
hence ${\cal D}_{\mu}\xi^{\rho}=\partial_{\mu}\xi^{\rho}-A_{\mu}{}^{M}\partial_{M}\xi^{\rho}$. 
For the gauge vectors the minimal covariant choice for the gauge transformations is 
 \be\label{delta0A}
  \delta_{\xi}^0 A_{\mu}{}^{M} \ = \ \xi^{\nu}{\cal F}_{\nu\mu}{}^{M}\;, 
 \ee
with the covariant field strength (\ref{improvedF}). It turns out that in the full EFT an extra 
term is required, but in order to streamline the proof of closure let us consider this 
minimal form first. Recalling the definition (\ref{COvDER}) of the covariant derivative 
we compute for the closure on ${\cal M}$, 
 \be\label{CLOSUREstep1}
  \begin{split}
   \big[\,\delta^0_{\xi_1}, \delta^0_{\xi_2}\,\big]{\cal M}_{MN} \ &= \ 
   \xi_2^{\mu}{\cal D}_{\mu}\big(\xi_1^{\nu}{\cal D}_{\nu}{\cal M}_{MN}\big)
   -\xi_2^{\mu}\,\mathbb{L}_{\delta^0_{\xi_1}A_{\mu}}{\cal M}_{MN}-(1\leftrightarrow 2) \\
   \ &= \ 2\,\xi_{[2}^{\mu}{\cal D}_{\mu}\xi_{1]}^{\nu}\,{\cal D}_{\nu}{\cal M}_{MN}
   +\xi_2^{\mu}\xi_1^{\nu}\big[{\cal D}_{\mu},{\cal D}_{\nu}\big]{\cal M}_{MN}
   +2\,\xi_{[2}^{\mu}\,\mathbb{L}_{\xi_{1]}^{\nu}{\cal F}_{\mu\nu}}{\cal M}_{MN}\;. 
  \end{split}
 \ee   
We now rewrite the last term in the second line, using the form (\ref{COVVarDown}) 
of the generalized Lie derivative. Specifically, we pull out the $\xi^{\nu}_1$ from 
the argument of $\mathbb{L}$ and collect the extra terms to find (leaving the 
symmetrization in $M,N$ implicit), 
 \be\label{quickcheck}
 \begin{split}
  2\,\xi_{[2}^{\mu}\,\mathbb{L}_{\xi_{1]}^{\nu}{\cal F}_{\mu\nu}}{\cal M}_{MN} \ = \ 
  &\,2\,\xi_2^{\mu}\xi_1^{\nu}\,\mathbb{L}_{{\cal F}_{\mu\nu}}{\cal M}_{MN}
  +2\,\partial_{M}\big(\xi_2^{\mu}\xi_1^{\nu}\big){\cal F}_{\mu\nu}{}^{K}{\cal M}_{KN}\\
  &-2\,Z^{PQ}{}_{MK}\partial_P\big(\xi_2^{\mu}\xi_1^{\nu}\big) {\cal F}_{\mu\nu}{}^{K} {\cal M}_{QN}
  +\tfrac{1}{6}\partial_K \big(\xi_2^{\mu}\xi_1^{\nu}\big){\cal F}_{\mu\nu}{}^{K}{\cal M}_{MN}\;, 
 \end{split}
 \ee 
where we employed the antisymmetry of ${\cal F}_{\mu\nu}$ in order to make 
the antisymmetrization $(1\leftrightarrow 2)$ manifest. 
Using next the commutator (\ref{commDA}) of covariant derivatives in the second term
of the second line of (\ref{CLOSUREstep1}), one finds that this changes the coefficient of the 
term with $\mathbb{L}_{{\cal F}_{\mu\nu}}$ in (\ref{quickcheck}) so that in total 
 \be
  \begin{split}
   \big[\,\delta^0_{\xi_1}, \delta^0_{\xi_2}\,\big]{\cal M}_{MN} \ &= \ 
   \ 2\,\xi_{[2}^{\mu}{\cal D}_{\mu}\xi_{1]}^{\nu}\,{\cal D}_{\nu}{\cal M}_{MN}
   +\xi_2^{\mu}\xi_1^{\nu}\,\mathbb{L}_{{\cal F}_{\mu\nu}}{\cal M}_{MN}
  +2\,\partial_{M}\big(\xi_2^{\mu}\xi_1^{\nu}\big){\cal F}_{\mu\nu}{}^{K}{\cal M}_{KN}\\
  &\qquad -2\,Z^{PQ}{}_{MK}\partial_P\big(\xi_2^{\mu}\xi_1^{\nu}\big) {\cal F}_{\mu\nu}{}^{K} {\cal M}_{QN}
  +\tfrac{1}{6}\partial_K \big(\xi_2^{\mu}\xi_1^{\nu}\big){\cal F}_{\mu\nu}{}^{K}{\cal M}_{MN}\;. 
 \end{split}
\ee 
It is now easy to see upon inspection of the definition (\ref{COVVarDown}) of the 
generalized Lie derivative that this combines into 
 \be
    \big[\,\delta^0_{\xi_1}, \delta^0_{\xi_2}\,\big]{\cal M}_{MN} \ = \ 
   \ 2\,\xi_{[2}^{\mu}{\cal D}_{\mu}\xi_{1]}^{\nu}\,{\cal D}_{\nu}{\cal M}_{MN}
   +\mathbb{L}_{\xi_2^{\mu}\xi_1^{\nu}{\cal F}_{\mu\nu}}{\cal M}_{MN}\;. 
 \ee
The first term on the right-hand side takes the form of a local $\xi^{\mu}$ transformation 
(\ref{deltaMdeltag}), while the second term is a field-dependent $\Lambda^M$ 
diffeomorphism. Thus, we proved closure, 
 \be
   \big[\,\delta^0_{\xi_1}, \delta^0_{\xi_2}\,\big]{\cal M}_{MN} \ = \ \delta_{\xi_{12}}{\cal M}_{MN}
   +\delta_{\Lambda_{12}^{(0)}}{\cal M}_{MN}\;, 
 \ee
where 
 \be\label{Effpara}
 \xi_{12}^\mu\ \equiv \ \xi_2^\nu\mc{D}_\nu\xi_1^\mu-\xi_1^\nu\mc{D}_\nu\xi_2^\mu\;, \qquad 
  \Lambda_{12}^{(0)M}\ \equiv \ \xi_2^\mu\xi_1^\nu\mc{F}_{\mu\nu}{}^{M}\;. 
 \ee 
 
Next we verify closure of the vector transformations  (\ref{delta0A}), which illustrates 
once more the subtle interplay of the various identities of the tensor hierarchy. 
In fact, the 2-form potential and its associated gauge parameter play a crucial role 
in establishing closure. We compute, in index-free notation and using the covariant 
variation (\ref{deltacalF}) of the 2-form curvature, 
 \be\label{Aalgebracomp}
 \begin{split}
  \big[\delta^0_{\xi_1},\delta^0_{\xi_2}\big]A_{\mu} \ &= \ \xi_2^{\nu}\Big(
  2\,{\cal D}_{[\nu}\,\big(\xi_1^{\rho}{\cal F}_{|\rho|\mu]}\big)
  +\widehat\partial(\Delta_{\xi_1}B_{\nu\mu})\Big)-(1\leftrightarrow 2)\\[0.5ex]
   \ &= \ 
    2\,\xi_{[2}^{\nu}{\cal D}_{\nu}\xi_{1]}^{\rho}\,{\cal F}_{\rho\mu}
   -2\,\xi_{[2}^{\nu}{\cal D}_{\mu}\xi_{1]}^{\rho}\, {\cal F}_{\rho\nu}
   +3\,\xi_2^{\nu}\xi_1^{\rho}{\cal D}_{[\mu}{\cal F}_{\nu\rho]}+\xi_2^{\nu}\xi_1^{\rho}{\cal D}_{\mu}{\cal F}_{\nu\rho}
   +2\,\xi_{[2}^{\nu}\widehat\partial(\Delta_{\xi_{1]}}B_{\nu\mu})  \\[0.5ex]
     \ &= \ 
    \xi_{12}^{\rho}\,{\cal F}_{\rho\mu}
   -{\cal D}_{\mu}\big(\xi_2^{\nu}\xi_1^{\rho}\big){\cal F}_{\rho\nu}
   +\xi_2^{\nu}\xi_1^{\rho}\,\widehat\partial{\cal H}_{\mu\nu\rho}+\xi_2^{\nu}\xi_1^{\rho}{\cal D}_{\mu}{\cal F}_{\nu\rho}
   +2\,\xi_{[2}^{\nu}\widehat\partial(\Delta_{\xi_{1]}}B_{\nu\mu}) \\[0.5ex]
   \ &= \ \xi_{12}^{\rho}\,{\cal F}_{\rho\mu}+ {\cal D}_{\mu}\big(\xi_2^{\nu}\xi_1^{\rho} {\cal F}_{\nu\rho}\big)
   +\xi_2^{\nu}\xi_1^{\rho}\,\widehat\partial{\cal H}_{\mu\nu\rho}
    +2\,\xi_{[2}^{\nu}\widehat\partial(\Delta_{\xi_{1]}}B_{\nu\mu})
   \;, 
 \end{split} 
 \ee 
where we implemented the antisymmetrization in $(1\leftrightarrow 2)$, inserted $\xi_{12}$ and 
used the Bianchi identity (\ref{Bianchi3}) in the third line. 
The first two terms on the right-hand side are precisely the $\xi^{\mu}$ and $\Lambda^M$ 
gauge transformations of $A_{\mu}$, defined in (\ref{delta0A}) and (\ref{deltaAgauge}), respectively, 
w.r.t.~the parameters in (\ref{Effpara}). It remains to manipulate the final two terms 
on the right-hand side. The third term in the last line can be written as\footnote{Note that writing  
$\widehat\partial\xi_1^{\rho}$ is, strictly speaking, an abuse of notation. We simply mean 
by this the partial derivative $\partial_M$ acting on $\xi_1^{\rho}$, with its SL$(3)\times {\rm SL}(2)$ indices contracted 
as if it was acting on ${\cal H}$, i.e., the index structure is 
\be
 \big( \xi_2^{\nu}\widehat\partial\xi_1^{\rho}{\cal H}_{\mu\nu\rho}\big)^{i\alpha}
 \ \equiv \ \epsilon^{ijk}\epsilon^{\alpha\beta}\xi_2^{\nu}\partial_{j\beta}\xi_1^{\rho}{\cal H}_{\mu\nu\rho k}\;.  
\ee} 
 \be\label{INTMani}
 \begin{split}
   \xi_2^{\nu}\xi_1^{\rho}\,\widehat\partial{\cal H}_{\mu\nu\rho} \ &= \ 
   -\widehat\partial \big(\xi_2^{\nu}\xi_1^{\rho}\,{\cal H}_{\mu\nu\rho}\big)
   +2\, \xi_2^{\nu}\xi_1^{\rho}\,\widehat\partial{\cal H}_{\mu\nu\rho}
   +2\,\xi_{[2}^{\nu}\,\widehat\partial\xi_{1]}^{\rho}\,{\cal H}_{\mu\nu\rho} \\[0.5ex]
   \ &= \ -\widehat\partial \big(\xi_2^{\nu}\xi_1^{\rho}\,{\cal H}_{\mu\nu\rho}\big)
   +2\,\xi_{[2}^{\nu}\,\widehat\partial\big(\xi_{1]}^{\rho}\,{\cal H}_{\mu\nu\rho}\big)\;. 
 \end{split}  
 \ee  
The first term in here can be interpreted as a field-dependent $\Xi_{\mu}$ transformation, c.f.~(\ref{DELXIA}), 
with parameter 
 \be\label{EffXI}
  \Xi_{12\mu} \ = \ \xi_2^{\nu}\xi_1^{\rho}\,{\cal H}_{\mu\nu\rho}\;. 
 \ee
The second term in (\ref{INTMani}) cancels against the last term in (\ref{Aalgebracomp}) provided we set 
 \be
  \Delta_{\xi}B_{\mu\nu} \ = \ \xi^{\rho}{\cal H}_{\mu\nu\rho}\;. 
 \ee
Thus we have shown 
 \be
     \big[\delta^0_{\xi_1},\delta^0_{\xi_2}\big]A_{\mu} \ = \ \big(\delta_{\xi_{12}}+\delta_{\Lambda_{12}^{(0)}}
     +\delta_{\Xi_{12}}\big) A_{\mu}\;, 
  \ee   
with effective parameters (\ref{Effpara}), (\ref{EffXI}). 

Before discussing the exterior diffeomorphisms for the remaining form fields 
of the tensor hierarchy, let us complete the vector gauge transformations 
(\ref{delta0A}). In fact, as mentioned above, although the gauge algebra closes for this 
minimal covariant choice of the gauge transformations, gauge invariance of the full EFT requires a 
further covariant term: 
 \be\label{fulldelA}
  \delta_\xi A_\mu{}^M \ = \ \xi^\nu\mc{F}_{\nu\mu}{}^M+\mc{M}^{MN}g_{\mu\nu}\ptl_N\xi^\nu\;.
 \ee
The extra term is universal for all EFTs.  
Let us verify that with this modification the gauge algebra still closes. 
For the closure on the generalized metric ${\cal M}_{MN}$  
one finds with (\ref{CLOSUREstep1}) the following additional contribution 
 \be\label{EXtraCLOsure}
  -\xi_2^{\mu}\, \mathbb{L}_{{\cal M}^{\bullet K}g_{\mu\nu}\partial_K\xi_{1}^{\nu}}{\cal M}_{MN}-(1\leftrightarrow 2)
  \ = \ \mathbb{L}_{-2\,{\cal M}^{\bullet K}g_{\mu\nu}\xi_{[2}^{\mu}\partial_K\xi_{1]}^{\nu}}{\cal M}_{MN}\;. 
 \ee 
Here we moved $\xi_2^{\mu}$ inside the argument of the generalized Lie derivative. 
In order to verify that this does not lead to extra correction terms one may use 
the explicit form (\ref{COVVarDown}) of the generalized Lie derivative to show 
that all such terms with derivatives of $\xi_2^{\mu}$ vanish as a consequence of the 
$(1\leftrightarrow 2)$ antisymmetrization.\footnote{We note that this requires 
using that the tensor $Z$ is invariant under the group action by ${\cal M}$, which leads to identities 
such as 
\be
 Z^{PQ}{}_{MK} {\cal M}^{KL}{\cal M}_{QN} \ = \ Z^{QL}{}_{KN}{\cal M}^{KP}{\cal M}_{QM}\;. 
\ee} The term on the right-hand side of (\ref{EXtraCLOsure})
can be interpreted as a field-dependent $\Lambda^M$ gauge transformation. 
Thus we still have closure, with the complete effective gauge parameters of internal and 
external diffeomorphisms given by 
 \be\label{finalparam}
  \begin{split}
    \Lambda_{12}^M\ &\equiv \ \xi_2^\mu\xi_1^\nu\mc{F}_{\mu\nu}{}^{M}
    -2\,{\cal M}^{MN}g_{\mu\nu}\,\xi_{[2}^{\mu}\partial^{}_N\xi_{1]}^{\nu}\;, \\
    \xi_{12}^\mu\ &\equiv \ \xi_2^\nu\mc{D}_\nu\xi_1^\mu-\xi_1^\nu\mc{D}_\nu\xi_2^\mu\;. 
   \end{split}
  \ee  
Similarly, one may verify closure on the gauge vector $A_{\mu}$ according to the 
same parameters. 

We close this subsection by giving the form of the generalized diffeomorphisms 
on the higher forms, whose closure can be verified in analogy to the above 
discussions, making repeated use of the Bianchi and variational identities of the tensor hierarchy. 
One finds in terms of the covariant variations,  
 \be
 \begin{split}
   \Delta_\xi B_{\mu\nu} \ & = \ \xi^\rho\mc{H}_{\rho\mu\nu}\;, \\
   \Delta_\xi C_{\mu\nu\rho}  \ &= \ \xi^\sigma\mc{J}_{\sigma\mu\nu\rho}\;, \\
    \Delta_\xi D_{\mu\nu\rho\sigma} \ &= \ \xi^\tau\mc{K}_{\tau\mu\nu\rho\sigma}\;. 
 \end{split}
 \ee
These transformations close w.r.t.~the parameters (\ref{finalparam}) and
 \be
  \begin{split}
    \Xi_{12,\mu} \ & \equiv \ \xi_2^\nu\xi_1^\rho\mc{H}_{\nu\rho\mu}\;, \\
    \Theta_{12,\mu\nu} \ & \equiv \ \xi_2^\rho\xi_1^\sigma\mc{J}_{\rho\sigma\mu\nu}\;, \\
    \Omega_{12,\mu\nu\rho} \ &\equiv \ \xi_2^\sigma\xi_1^\tau\mc{K}_{\sigma\tau\mu\nu\rho}\;. 
   \end{split}
  \ee

\subsection{Gauge invariance} 
We now compute the gauge variation of the (pseudo-)action under 
external diffeomorphisms. To this end we have to compute the 
gauge variation of the field strengths, where for the moment 
we will only consider the minimal gauge variation (\ref{delta0A}) of $A_{\mu}$. 
Starting with the 2-form curvature we compute:
 \be
  \begin{split}
   \delta^0_{\xi}{\cal F}_{\mu\nu} \ &= \ 2\,{\cal D}_{[\mu}\,\delta^0_{\xi}A_{\nu]}+\widehat\partial(\Delta_{\xi} B_{\mu\nu}) \\[0.5ex]
   \ &= \ 2\, {\cal D}_{[\mu}\big(\xi^{\rho}{\cal F}_{|\rho|\nu]}\big)+\widehat\partial\big(\xi^{\rho}{\cal H}_{\mu\nu\rho}\big) \\[0.5ex]
   \ &= \ 2\, {\cal D}_{[\mu}\xi^{\rho}\,{\cal F}_{|\rho|\nu]}+\xi^{\rho}{\cal D}_{\rho}{\cal F}_{\mu\nu}
   -3\,\xi^{\rho}{\cal D}_{[\mu}{\cal F}_{\nu\rho]}+\widehat\partial\big(\xi^{\rho}{\cal H}_{\mu\nu\rho}\big)\;. 
  \end{split}
 \ee   
The first two terms here take the form of a 
conventional Lie derivative w.r.t.~$\xi^{\mu}$, except that all partial derivatives are 
replaced by gauge covariant derivatives. Such (generalized) Lie derivatives can 
be defined for any tensor and will henceforth be denoted by ${\cal L}_{\xi}$.  
Using the Bianchi identity in the third term we get a cancellation against one contribution 
of the last term, so that in total, restoring index notation, 
 \be\label{deltaxiF}
  \delta^0_{\xi}{\cal F}_{\mu\nu}{}^{i\alpha} \ = \ {\cal L}_{\xi}{\cal F}_{\mu\nu}{}^{i\alpha}
  +\epsilon^{ijk}\epsilon^{\alpha\beta}\partial_{j\beta}\xi^{\rho}\,{\cal H}_{\mu\nu\rho,k}\;.
 \ee 
Similarly, it is straightforward, using the covariant variations and the Bianchi identities, 
to prove the analogous relations for the higher field strengths, 
 \be\label{delta0VAR}
  \begin{split}
   \delta^0_{\xi}{\cal H}_{\mu\nu\rho,m} \ &= \ {\cal L}_{\xi}{\cal H}_{\mu\nu\rho,m}
   +\partial_{m\alpha}\xi^{\lambda}\,{\cal J}_{\lambda\mu\nu\rho}{}^{\alpha}\;, \\
   \delta^0_{\xi}{\cal J}_{\mu\nu\rho\sigma}{}^{\alpha} \ &= \ {\cal L}_{\xi}{\cal J}_{\mu\nu\rho\sigma}{}^{\alpha}
   +\epsilon^{\alpha\beta}\partial_{m\beta}\xi^{\lambda}\,{\cal K}_{\lambda\mu\nu\rho\sigma}{}^{m}\;. 
  \end{split}
 \ee  
 
It is easy to see that a Lagrangian built with determinant $e$ times a scalar (w.r.t.~the
Lie derivatives ${\cal L}_{\xi}$) is gauge invariant. Thus it remains to collect the 
`non-invariant' terms. Let us illustrate how the cancellation works. 
Consider the variation of the Yang-Mills term under $\delta_{\xi}^0$, for which by the preceding discussion 
up to total derivatives only 
the second term in (\ref{deltaxiF}) gives a non-vanishing contribution,  
 \be\label{anomalyTerm}
  \delta^0_{\xi}\big(-\tfrac{1}{4}e{\cal M}_{i\alpha,j\beta}{\cal F}^{\mu\nu\,i\alpha}{\cal F}_{\mu\nu}{}^{j\beta}\big)
  \ = \ -\tfrac{1}{2}e{\cal M}_{ij}{\cal M}_{\alpha\beta}\epsilon^{jpq}\epsilon^{\beta\gamma}
  \partial_{p\gamma}\xi^{\rho}\,{\cal F}^{\mu\nu\,i\alpha}{\cal H}_{\mu\nu\rho,q}\;. 
 \ee 
(In order not to clutter the equations, here and in the following we suppress the integration symbol.) 
In order to cancel these terms we have to complete the 
vector gauge transformation to the full (\ref{fulldelA}), which leads to additional variations, 
denoted by $\delta'$ in the following,  
that precisely cancel the above terms. 
For instance, using (\ref{varyfieldH}) we infer that the variation of ${\cal H}_{\mu\nu\rho}$ 
receives the following additional contribution, 
 \be\label{COntri} 
 \begin{split}
  \delta_{\xi}{\cal H}_{\mu\nu\rho,m} \ &= \ \delta_{\xi}^0{\cal H}_{\mu\nu\rho,m}
  -3\,\epsilon_{mnk}\epsilon_{\alpha\beta}\,\delta'_{\xi} A_{[\mu}{}^{n\alpha}{\cal F}_{\nu\rho]}{}^{k\beta}\\
  \ &= \  \delta_{\xi}^0{\cal H}_{\mu\nu\rho,m}-3\,\epsilon_{mnk}\epsilon_{\alpha\beta}
  {\cal M}^{n\alpha,l\gamma}\,\partial_{l\gamma}\xi^{\lambda} \,g_{\lambda[\mu} {\cal F}_{\nu\rho]}{}^{k\beta}\;. 
 \end{split} 
 \ee 
The extra variation of the ${\cal H}^2$ term in (\ref{KINTERMS}) then precisely cancels (\ref{anomalyTerm}), 
which in turn fixes the relative coefficient between these two terms.\footnote{In order to verify this cancellation 
one has to use standard identities like 
${\cal M}^{il}{\cal M}^{jp}{\cal M}^{kq}\epsilon_{lpq}=\epsilon^{ijk}$, which is equivalent to $\det {\cal M}=1$.}
Moreover, the Yang-Mills term receives an additional variation from (\ref{fulldelA})
in the form $\delta{\cal F}_{\mu\nu}=2{\cal D}_{[\mu}\delta A_{\nu]}$, 
one contribution of which is 
cancelled by the variation from the Einstein-Hilbert term, as explained in detail in \cite{UDuality}, 
while  the remaining term is cancelled against terms in the variation of the potential. 

Apart from the variation of the ${\cal H}^2$ kinetic term in (\ref{COntri}), which cancelled 
the extra term (\ref{anomalyTerm}) in the variation of the Yang-Mills term, 
due to (\ref{delta0VAR}) 
its variation also yields an anomalous term in complete parallel to that in (\ref{anomalyTerm}),  
 \be\label{result1}
\bsp
 &\delta^0_\xi\big(-\fracs{1}{12} e\mc{M}^{mn}\mc{H}^{\mu\nu\rho}{}_{m}\mc{H}_{\mu\nu\rho,n}\big) \ = \ 
  -\fracs{1}{6}e\mc{M}^{mn}\ptl_{m\alpha}\xi^\sigma 
  \mc{H}^{\mu\nu\rho}{}_n\mc{J}_{\sigma\mu\nu\rho}{}^\alpha\;. 
 \end{split}
 \ee
Next, we compute the variation of the ${\cal J}^2$ term, using the complete gauge variations of ${\cal J}$
obtained with (\ref{deltacalJ}), 
 \be\label{FULLdeltaJ}
  \delta_{\xi}{\cal J}_{\mu\nu\rho\sigma}{}^{\alpha} \ = \  {\cal L}_{\xi}{\cal J}_{\mu\nu\rho\sigma}{}^{\alpha}
   +\epsilon^{\alpha\beta}\partial_{m\beta}\xi^{\lambda}\,{\cal K}_{\lambda\mu\nu\rho\sigma}{}^{m}
  -4\,{\cal M}^{m\alpha,n\beta}\partial_{n\beta}\xi^{\lambda} g_{\lambda[\mu}
  {\cal H}_{\nu\rho\sigma]m}\;. 
 \ee
This leads to the following variation of the kinetic term 
  \be
 \begin{split}
 \delta_\xi\big(-\tfrac{1}{96} e\mc{M}_{\alpha\beta}\mc{J}_{\mu\nu\rho\sigma}{}^\alpha\mc{J}^{\mu\nu\rho\sigma,\beta}\big)
 \ = \ \,
 &-\tfrac{1}{48} e\mc{M}_{\alpha\beta}\epsilon^{\beta\gamma}\ptl_{m\gamma}\xi^\lambda 
 \mc{K}_{\tau\mu\nu\rho\sigma}{}^m\mc{J}^{\mu\nu\rho\sigma,\alpha}\\
 &+\tfrac{1}{12} e\mc{M}^{mn}\ptl_{m\alpha}\xi^\sigma 
 \mc{H}^{\mu\nu\rho}{}_n\mc{J}_{\sigma\mu\nu\rho}{}^\alpha  \;.   \label{variJJ}
 \end{split}
 \ee
Finally, we have to consider the variation of the topological term. 
Using its general variation (\ref{varitop}) and inserting $\delta_\xi A_\mu,\Delta_\xi B_{\mu\nu},\Delta_\xi C_{\mu\nu\rho},\Delta_\xi D_{\mu\nu\rho\sigma}$ we obtain 
\be\label{result3}
\bsp
\delta_\xi {\cal L}_{\rm top} \ = \ \,&\kappa\, \epsilon^{\mu_1\ldots\mu_8}\Big[\,4\,
\epsilon_{\alpha\beta}\mc{J}_{\mu_1\ldots\mu_4}{}^\alpha\,\xi^{\nu}\mc{F}_{\nu\mu_5}{}^{m\beta}\,\mc{H}_{\mu_6\mu_7\mu_8,m}+6\,\epsilon_{\alpha\beta}\,\xi^{\nu}\mc{H}_{\nu \mu_1\mu_2,m}\,\mc{F}_{\mu_3\mu_4}{}^{m\beta}\,
\mc{J}_{\mu_5\ldots\mu_8}{}^\alpha\\[0.5ex]
&\qquad\qquad  +4\, \epsilon_{\alpha\beta}\,\mc{J}_{\mu_1\ldots\mu_4}{}^\alpha\mc{M}^{m\beta,k\gamma}g_{\nu\mu_5}
\ptl_{k\gamma}\xi^{\nu}\mc{H}_{\mu_6\mu_7\mu_8,m}
+\xi^{\nu}\mc{K}_{\nu \mu_1\ldots\mu_4}\ptl_{m\alpha}\mc{J}_{\mu_5\ldots\mu_8}{}^\alpha
\\[0.5ex]
&\qquad\qquad  +4\,\epsilon_{\alpha\beta}\,\xi^\nu\mc{J}_{\nu\mu_1\mu_2\mu_3}{}^\alpha\big(\mc{D}_{\mu_4}
\mc{J}_{\mu_5\ldots\mu_8}{}^\beta+4\,\mc{F}_{\mu_4\mu_5}{}^{m\beta}\,\mc{H}_{\mu_6\mu_7\mu_8,m}\big)
\Big]\\[0.5ex]
\ = \ &\kappa\int\epsilon^{\mu_1\ldots\mu_8}
\Big[\,4\,\epsilon_{\alpha\beta}\,\mc{J}_{\mu_1\ldots\mu_4}{}^\alpha\mc{M}^{m\beta,k\gamma}
g_{\nu \mu_5}\ptl_{k\gamma}\xi^\nu\mc{H}_{\mu_6\mu_7\mu_8,m}
-\ptl_{m\alpha}\xi^{\nu} \mc{K}_{\nu \mu_1\ldots\mu_4}{}^m\mc{J}_{\mu_5\ldots\mu_8}{}^\alpha\Big]\;. 
\end{split}
\ee
Here we combined the ${\cal J}{\cal F}{\cal H}$ terms and used the last of the Bianchi identities (\ref{collectBianchi})
to rewrite them as a $\widehat\partial{\cal K}$ term, after which we integrate by parts with $\widehat\partial$.
In all of these manipulations we make use of Schouten identities according to which antisymmetrization in 
nine external indices gives zero. 
 
We are now ready to collect the left-over terms from the gauge variation of the fields of the tensor hierarchy.  
With (\ref{result1}), (\ref{variJJ}) and (\ref{result3}) the total variation is given by 
 \be
\begin{split}
&\delta_\xi {\cal L} \ = \ -\tfrac{1}{48} e\mc{M}_{\alpha\beta}\epsilon^{\beta\gamma} \ptl_{m\gamma}\xi^\tau \mc{K}_{\tau\mu\nu\rho\sigma}{}^m
\mc{J}^{\mu\nu\rho\sigma,\alpha}
-\tfrac{1}{12} e\mc{M}^{mn}\ptl_{m\alpha}\xi^\sigma \mc{H}^{\mu\nu\rho}{}_n\mc{J}_{\sigma\mu\nu\rho}{}^\alpha\\[0.5ex]
&\;\;\;\;\;
+\kappa\epsilon^{\mu_1\ldots\mu_8}\big(\,4\,\epsilon_{\alpha\beta}\,\mc{J}_{\mu_1\dots\mu_4}{}^\alpha\,
\mc{M}^{m\beta,k\gamma}g_{\nu\mu_5}\ptl_{k\gamma}\xi^\nu\mc{H}_{\mu_6\mu_7\mu_8,m}
-\ptl_{m\alpha}\xi^\nu \mc{K}_{\nu\mu_1\ldots\mu_4}{}^m\mc{J}_{\mu_5\ldots\mu_8}{}^\alpha\big)\;. \label{variexttot}
\end{split}
\ee
This is non-zero, but the terms cancel if we impose the self-duality relation (\ref{EFTSELFDUAL}). 
This is sufficient in order to prove the gauge invariance of the second-order equations supplemented 
by the self-duality constraint. 

In order to complete the proof of gauge invariance 
it thus remains to verify the gauge invariance of the duality constraint (\ref{EFTSELFDUAL}), which reads 
 \be\label{calOO}
  {\cal O}^{\mu_1\ldots\mu_4}{}_{\alpha} \ \equiv  \
  {\cal M}_{\alpha\beta}{\cal J}^{\mu_1\ldots\mu_4,\beta}
  +\tfrac{1}{4!}\epsilon_{\alpha\beta}e^{-1}\epsilon^{\mu_1\ldots\mu_4\nu_1\ldots\nu_4}
  {\cal J}_{\nu_1\ldots\nu_4}{}^{\beta}\ =\ 0\;. 
 \ee  
We now compute its gauge variation under external diffeomorphisms, using (\ref{FULLdeltaJ}), 
to find 
 \be
 \begin{split}
  \delta_{\xi}& {\cal O}^{\mu_1\ldots\mu_4}{}_{\alpha} \ = \ {\cal L}_{\xi} {\cal O}^{\mu_1\ldots\mu_4}{}_{\alpha}\\[0.5ex]
  &+{\cal M}_{\alpha\beta}\epsilon^{\beta\gamma}\partial_{m\gamma}\xi^{\lambda} g_{\lambda\tau}
  {\cal K}^{\tau\mu_1\ldots\mu_4,m}
  -\tfrac{1}{3!}\epsilon_{\alpha\beta}{\cal M}^{mn}{\cal M}^{\beta\gamma}
  e^{-1}\epsilon^{\mu_1\ldots\mu_4\nu_1\ldots\nu_4}\partial_{n\gamma}\xi^{\lambda} 
  g_{\lambda\nu_1}{\cal H}_{\nu_2\nu_3\nu_4 m}
   \\[0.5ex]
  &-\tfrac{1}{4!}e^{-1}\epsilon^{\mu_1\ldots\mu_4\nu_1\ldots\nu_4}\partial_{m\alpha}\xi^{\lambda}
  {\cal K}_{\lambda\nu_1\ldots\nu_4}{}^{m}
  -4\,{\cal M}^{mn}\partial_{n\alpha}\xi^{[\mu_1}{\cal H}^{\mu_2\mu_3\mu_4]}{}_m\;. 
 \end{split} 
 \ee 
The covariant Lie derivative term in the first line is zero for ${\cal O}^{\mu_1\ldots\mu_4}{}_{\alpha}=0$, 
i.e., it is zero on-shell.   
The remaining terms in each line cancel against each other upon using the 
duality relation between 2-forms and 4-forms, 
 \be\label{35duality}
  {\cal H}^{\mu_1\mu_2\mu_3}{}_{m} \ \equiv \ \tfrac{1}{5!}\,{\cal M}_{mn}
  e^{-1}\epsilon^{\mu_1\mu_2\mu_3\nu_1\ldots\nu_5}{\cal K}_{\nu_1\ldots\nu_5}{}^{n} 
  \;. 
 \ee
Thus, the self-duality constraint is gauge invariant on-shell provided this duality 
relation is part of the field equations. One may indeed argue that this duality relation  
follows by combining the integrability condition of the self-duality constraint and 
the second-order equations of the pseudo-action as follows. 
Acting with ${\cal D}_{\mu}$ on (\ref{calOO}) we obtain the integrability condition 
 \be\label{inTEGRA} 
 \begin{split}
  0 \ &= \ {\cal D}_{\mu_1}(e\, {\cal O}^{\mu_1\ldots\mu_4}{}_{\alpha} )\\[0.5ex]
   \ &= \ 
  {\cal D}_{\mu_1}\big(e\,{\cal M}_{\alpha\beta}{\cal J}^{\mu_1\ldots\mu_4,\beta} \big)
  +\tfrac{1}{5!}\epsilon_{\alpha\beta}\epsilon^{\mu_1\ldots \mu_4\nu_1\ldots\nu_4}
  \big(\widehat\partial{\cal K}_{\mu_1\nu_1\ldots\nu_4}
  -10\,{\cal F}_{\mu_1\nu}\bullet {\cal H}_{\nu_2\nu_3\nu_4}\big)^{\beta}\;, 
 \end{split}
 \ee
where we used the last Bianchi identity in (\ref{collectBianchi}) in order to rewrite 
the covariant exterior derivative of ${\cal J}$. On the other hand, 
varying the pseudo-action w.r.t.~the 3-forms also yields a second-order equation:
 \be
 \begin{split}
\mc{D}_\lambda&\big(e\,\mc{M}_{\alpha\beta}\mc{J}^{\lambda\mu\nu\rho,\beta})+2\,\ptl_{m\alpha}
\big(e\,\mc{M}^{mn}\mc{H}^{\mu\nu\rho}{}_n\big)\\[0.5ex]
&+12\,\epsilon^{\mu\nu\rho\lambda_1\dots\lambda_5}(-\fracs{4\kappa}{5}\ptl_{m\alpha}\mc{K}_{\lambda_1\dots\lambda_5}{}^m-8\kappa\epsilon_{\alpha\beta}\mc{F}_{\lambda_1\lambda_2}{}^{m\beta}\mc{H}_{\lambda_3\lambda_4\lambda_5,m}
\big) \ = \ 0 \;,  \label{eomC2}
\end{split}
\ee
where the second term in the first line originates from the variation of $C_{\mu\nu\rho}$ inside the 3-form 
curvature ${\cal H}$, c.f.~(\ref{varyfieldH}). 
Comparing (\ref{inTEGRA}) with (\ref{eomC2}) we observe a mismatch in terms with ${\cal K}$ 
and ${\cal H}$, both under a derivative.  
The combined  field equations thus imply 
 \be
  \partial_{m\alpha}\Big(e{\cal M}^{mn}{\cal H}^{\mu\nu\rho}{}_{n}
  -\tfrac{1}{5!} \epsilon^{\mu\nu\rho\sigma_1\ldots\sigma_5}{\cal K}_{\sigma_1\dots\sigma_5}{}^{m}\Big) \ = \ 0\;, 
 \ee 
or, bringing the constant $\epsilon$ tensor to the other side and employing  
an index-free notation, 
 \be
  \widehat\partial\big(\star {\cal M}{\cal H}^{(3)}-{\cal K}^{(5)}\big) \ = \ 0\;. 
 \ee 
The tensor in parenthesis is thus $\widehat\partial$ closed. 
Assuming the validity of a Poincar\'e lemma
we conclude that this tensor is $\widehat\partial$ exact, 
so that 
 \be\label{DualitySTEP}
     \star {\cal M}{\cal H}^{(3)}-{\cal K}^{(5)} \ = \ \widehat\partial\Omega^{(5)}\;, 
 \ee
for some 5-form $\Omega^{(5)}$. Recalling the definition of the 5-form curvature, 
${\cal K}^{(5)}=K^{(5)}+\widehat\partial E^{(5)}$, we observe that upon redefining 
$E^{(5)}\rightarrow E^{(5)}+\Omega^{(5)}$ the right-hand side of (\ref{DualitySTEP}) 
can be set to zero. In fact, as we saw above, the 5-form potential and its variations drop out of all equations 
and play only a formal role in making gauge invariance manifest.
Thus, it can be redefined arbitrarily and so we obtain 
  \be\label{35Duality}
     \star {\cal M}{\cal H}^{(3)}-{\cal K}^{(5)} \ = \ 0\;, 
 \ee
or, equivalently, the full unprojected duality relations (\ref{35duality}). 
However, we should recall the subtleties involved in establishing the  
Poincar\'e lemma just assumed. 
In fact, the Poincar\'e lemma can only be derived 
before picking a particular solution of the section constraint; more precisely, 
in the case at hand the Poincar\'e lemma is only valid for the M-theory solution of the section constraint. 
In the case that the duality relation (\ref{35Duality}) does not follow from the other equations 
it has to be imposed by hand as part of the definition of the theory, for which the 
self-duality constraint (\ref{calOO}) is then gauge invariant. 
This completes our discussion of gauge invariance under external diffeomorphisms.

\section{Embedding of conventional supergravity}
In this section we discuss the embedding of $D=11$ and type IIB supergravity 
into the SL$(3)\times {\rm SL}(2)$ exceptional field theory, upon picking the  
appropriate solution of the section constraint. This requires a Kaluza-Klein type 
decomposition of coordinates and tensor indices in a $8+3$ or $8+2$ split, respectively, 
but \textit{without} truncation of the coordinate dependence. 
As such, the theories resulting from EFT by reducing the coordinate dependence 
upon solving the section constraint 
are on-shell fully equivalent to either $D=11$ or type IIB supergravity.

\subsection{Embedding of $D=11$ supergravity}
We start by recalling the bosonic field content of $D=11$ supergravity, which consists 
of the 11-dimensional metric $G$ and a 3-form gauge potential $A^{(3)}$:
 \be
  D=11 \text{\; field content\;}:\qquad G_{MN}\;, \;\;A_{MNK}\;, 
 \ee 
where (in this subsection only) we denote the $D=11$ spacetime indices by $M,N,\ldots$.\footnote{There is no 
danger of confusing these indices with fundamental $(3,2)$ indices of SL$(3)\times {\rm SL}(2)$ as we will always 
write out the SL$(3)$ and ${\rm SL}(2)$ indices individually.} 
For the comparison with EFT it is convenient to also introduce a dual 6-form potential $A^{(6)}$
that is related to the 3-form via the duality relation to the field strength $F^{(4)}={\rm d}A^{(3)}$, 
 \be\label{dualityREL}
  F^{(7)} \ = \ \star F^{(4)}\;, \qquad F^{(7)} \ = \ {\rm d}A^{(6)}+A^{(3)}\wedge {\rm d}A^{(3)}\;. 
 \ee 
Here we defined the 7-form field strength, which requires a Chern-Simons modification 
by $A^{(3)}$ in order for the duality relation to be compatible with the $D=11$ supergravity 
equations. Indeed the integrability condition of the duality relation yields, by ${\rm d}^2=0$, 
precisely the second order equation of motion for $A^{(3)}$. 

Let us now discuss the fields originating from these upon a $8+3$ decomposition of the tensor indices, 
writing 
 \be 
  M \ = \ (\mu,m)\;, \qquad {\rm etc.}\;, 
 \ee  
as would be appropriate for Kaluza-Klein compactification to $D=8$. Let us stress again, however, 
that the coordinate dependence will be untouched and so we merely reformulate $D=11$ 
supergravity in a manner appropriate for the comparison with EFT. Note that 
this decomposition leads to a manifestly SL$(3)$ covariant formulation, with SL$(3)$ 
indices $m,n,\ldots$, the group being a subgroup of the internal diffeomorphism group. 
The $D=11$ metric gives rise to 
 \be\label{METRICD=8}
  G_{MN}\,:\qquad g_{\mu\nu}\;, \quad A_{\mu}{}^{m}\;, \quad  G_{mn}\;, 
 \ee 
where $g_{\mu\nu}$ is the (external) 8-dimensional spacetime metric, $G_{mn}$
the internal metric (encoding part of the scalars in $D=8$) and $A_{\mu}{}^{m}$
are the Kaluza-Klein vectors. Next, the 3-form gives rise to 
 \be\label{3FORM}
  A_{MNK}\,: \qquad A_{\mu\nu\rho}\;, \quad A_{\mu\nu m}\;,\quad 
  A_{\mu mn} \ \equiv \ \widetilde{A}_{\mu}{}^{k}\epsilon_{mnk}\;, \quad A_{mnk} \ \equiv \ \widetilde{A}\, \epsilon_{mnk}\;.
 \ee 
Here we used the three-dimensional epsilon symbol of the internal space in 
order to lower the number of SL$(3)$ indices. 
Finally, the 6-form potential decomposes as 
 \be\label{SIXFORM}
  A_{MNKLPQ}\;: \qquad A_{\mu\nu\rho mnk} \ \equiv \ \widetilde{A}_{\mu\nu\rho}\, \epsilon_{mnk}\;, \quad
  A_{\mu\nu\rho\sigma mn} \ \equiv \ \widetilde{A}_{\mu\nu\rho\sigma}{}^{k}\,\epsilon_{mnk}\;. 
 \ee 
Note that, in $D=8$ language, 
there are no lower forms than 3-forms since for such fields the total antisymmetry in the internal 
SL$(3)$ indices implies that they vanish identically. 
Moreover, in principle there are also 5-forms $A^{(5)}_m$ and a singlet 6-form $A^{(6)}$.  
However, the former are on-shell dual, via (\ref{dualityREL}), to the vectors $\widetilde{A}_{\mu}{}^{m}$ 
already encoded in the fields (\ref{3FORM}) originating from the 3-form, which will enter 
with a kinetic term. Hence these fields can be eliminated. 
Similarly, the singlet 6-form is on-shell dual to the scalar $\widetilde{A}$ and can 
also be eliminated. Note that also the 4-form $\widetilde{A}_{\mu\nu\rho\sigma}{}^{m}$ 
is on-shell dual to a field that enters with a kinetic term, namely the 2-forms $A_{\mu\nu m}$. 
It turns out to be necessary, however, to keep the 4-forms as separate but non-propagating 
fields that enter without a kinetic term. Rather, its presence in the Chern-Simons like topological 
couplings plays an important role in guaranteeing the consistency with the first-order duality 
relations. This mechanism is a general feature of the tensor hierarchy in gauged supergravity. 

The above decomposition shows that the reformation of $D=11$ supergravity based on a 
$8+3$ split of fields and coordinates exhibits a manifest SL$(3)$ symmetry, reflecting 
the internal diffeomorphism invariance. The SL$(2)$, on the other hand, is hidden. 
More precisely, this symmetry is not actually present in $D=11$ supergravity, but emerges only
upon genuine torus reduction to $D=8$. Indeed, in order to embed $D=11$ supergravity 
into EFT we have to embed the three-dimensional  derivatives $\partial_m$ according to 
 \be
  \partial_m\;\rightarrow \; \partial_{m\alpha} \ = \ (\partial_{m1},\partial_{m2}) \ \equiv \ 
  (\partial_m,0)\;, 
 \ee
solving the section constraint by singling out one SL(2) direction and hence breaking this symmetry, 
see the discussion in the introduction.  The only way to solve the section constraint so that 
it preserves the full duality group is to set $\partial_{m\alpha}=0$, which of course is equivalent 
to dimensional reduction. 

Next, we match the field content of $D=11$ supergravity in the $8+3$ split 
with that of EFT summarized in (\ref{propFIELD}). 
Although the SL$(2)$ symmetry is broken we can still reorganize 
the above fields into SL$(2)$ multiplets.  
First, the SL$(2)$ singlet external metric $g_{\mu\nu}$ matches that in (\ref{METRICD=8}). 
The scalars from (\ref{METRICD=8}) and (\ref{3FORM}) encoded in EFT correspond to 
 \be
  {\cal M}_{mn}\,,\; {\cal M}_{\alpha\beta}\,: \qquad  (\,G_{mn}\,,\;\widetilde{A}\,)\;. 
 \ee 
The EFT scalar matrices encode $5$ degrees of freedom in ${\cal M}_{mn}$ 
and 2 degrees of freedom in ${\cal M}_{\alpha\beta}$,
both satisfying $\det {\cal M}=1$, 
giving a total of 7, which precisely matches the $6+1$ scalar degrees of freedom 
in supergravity. The vector components from 
(\ref{METRICD=8}) and (\ref{3FORM}) are 
 \be
  A_{\mu}{}^{m\alpha} \,: \qquad (\,A_{\mu}{}^{m}\,,\; \widetilde{A}_{\mu}{}^{m}\,)\;, 
 \ee
which perfectly matches the vector field content of EFT.   
The 2-forms are directly identified with those in (\ref{3FORM}), 
 \be
  B_{\mu\nu m} \,: \qquad A_{\mu\nu m}\;. 
 \ee 
The 3-forms are collected from (\ref{3FORM}) and (\ref{SIXFORM}) to combine as 
 \be\label{SL2threeforms}
  C_{\mu\nu\rho}{}^{\alpha} \,: \qquad (\, A_{\mu\nu\rho}\,,\; \widetilde{A}_{\mu\nu\rho}\,)\;.
 \ee 
Finally, the 4-forms are directly identified with those in (\ref{SIXFORM}), 
 \be
  D_{\mu\nu\rho\sigma}{}^{m} \,: \qquad \widetilde{A}_{\mu\nu\rho\sigma}{}^{m}\;. 
 \ee 
Summarizing, we see that all bosonic physical fields of $D=11$ supergravity 
are encoded in the EFT fields. Moreover, these also include topological fields 
that do not enter with a kinetic term, here the 4-forms.  
It is important to verify that we propagate the right number of degrees of freedom and 
thus do not over-count. In fact,  the 3-forms (\ref{SL2threeforms}) are subject 
to a self-duality constraint which originates from (\ref{dualityREL}) in $D=11$
upon performing the $8+3$ split. This is, however, perfectly consistent 
with the self-duality relation (\ref{EFTSELFDUAL}) in EFT, and so we indeed 
describe the correct number of degrees of freedom. Note that the presence of 
the topological $D_{\mu\nu\rho\sigma}{}^{m}$ was necessary in order 
to obtain field equations that are compatible with the self-duality constraint and 
hence with $D=11$ supergravity. 

In the above discussion we have shown that the fields of $D=11$ supergravity 
match those of EFT (subjected to the appropriate solution of 
the section constraint). The discussion was schematic as we did not display the 
precise field redefinitions needed in order to relate both sets of fields, 
and we did not verify the detailed match of the field equations. 
In fact, there are laborious Kaluza-Klein-like field redefinitions needed 
that mix the various tensor fields in order to bring the gauge symmetries into 
a canonical form. In the E$_{6(6)}$ EFT the match with $D=11$
supergravity has been verified in all detail for the bosonic sector and the match for type IIB to a large extent \cite{E6}. 
In the E$_{7(7)}$ EFT the match with $D=11$ supergravity is largely contained 
in the original work of de Wit--Nicolai \cite{deWit:1986mz} and the more recent  work \cite{SUSYE7}, 
including fermions in the supersymmetric form. Thus there is little doubt that 
here it works out similarly, but 
we leave a more detailed verification for the SL$(3)\times {\rm SL}(2)$ EFT for future work.

\subsection{Embedding of type IIB}
Let us now turn to the embedding of type IIB supergravity, whose bosonic field content is given by 
the 10-dimensional metric $G$, two scalar fields (the dilaton $\phi$ and the RR zero-form $C_0$
that may be combined into the axion-dilaton $\tau=C_0+ie^{-\phi}$ or, equivalently,  into 
an SL$(2)/{\rm SO}(2)$ coset matrix ${\cal M}_{i'j'}$), an SL(2) doublet $A^{(2)i'}$ of two forms and 
a self-dual 4-form $A^{(4)}$,  
 \be
  \text{Type IIB field content:}  \qquad
  G_{MN}\;, \quad {\cal M}_{i'j'} \in {\rm SL}(2,\mathbb{R})\;, 
  \quad A_{MN}{}^{i'}\;, \quad A_{MNKL}\;, 
 \ee
where now (and in this subsection only) $M,N,\ldots$ 
denote $D=10$ spacetime indices and $i',j'=1,2$ denote SL$(2)$ indices. 
The self-duality constraint of the 4-form is given by 
 \be\label{IIBselfdual}
  \star F^{(5)} \ = \ F^{(5)}\;, \qquad 
  F^{(5)} \ \equiv \ {\rm d}A^{(4)}-\tfrac{1}{2}\epsilon_{i'j'}A^{(2)i'}\wedge {\rm d}A^{(2)j'}\;.
 \ee  

Next we perform the $8+2$ splitting of tensor indices, writing 
 \be 
  M \ = \ (\mu\,,\;\alpha)\;, \quad \alpha=1,2\;. 
 \ee 
For the metric this yields
 \be\label{IIBMetric}
 G_{MN}\,:\qquad g_{\mu\nu}\;, \quad A_{\mu}{}^{\alpha}\;, \quad  G_{\alpha\beta}\;, 
 \ee 
introducing the Kaluza-Klein vector and the internal scalars. 
The SL(2) valued scalar matrix ${\cal M}_{i'j'}$ decomposes trivially. The 2-forms decompose as 
 \be\label{IIB2form}
  A_{MN}{}^{i'} \,:\qquad A_{\mu\nu}{}^{i'} \ \equiv \ \epsilon^{i'j'}\widetilde{A}_{\mu\nu j'}\;, 
  \quad A_{\mu\alpha}{}^{i'} \ \equiv \ \epsilon_{\alpha\beta}
  \widetilde{A}_{\mu}{}^{\beta i'}\;, \quad 
  A_{\alpha\beta}{}^{i'} \ \equiv \  \epsilon_{\alpha\beta}\widetilde{A}^{i'}\;, 
 \ee
where we used again the Levi-Civita symbols $\epsilon_{\alpha\beta}$ and  $\epsilon_{i'j'}$ to   
reduce the number of indices.    
Finally, the 4-form decomposes as 
 \be\label{IIB4form}
  A_{MNKL} \,:\qquad A_{\mu\nu\rho\sigma}\;, \quad A_{\mu\nu\rho\alpha} \ \equiv \ 
  \epsilon_{\alpha\beta}\widetilde{A}_{\mu\nu\rho}{}^{\beta}\;, \quad
  A_{\mu\nu\alpha\beta} \ \equiv \ \epsilon_{\alpha\beta}\,\widetilde{A}_{\mu\nu}\;.
 \ee
Note that this does not yield forms of degree lower than two as such fields are 
identically zero by having more than two antisymmetrized SL(2) indices.   

In order to embed type IIB into EFT we have to pick the second, inequivalent solution 
of the section constraint. To this end we have to break the manifest SL(3) symmetry of EFT 
to the SL(2) S-duality symmetry of type IIB
by splitting the SL$(3)$ index as $i=(i',3)$. 
The 2-dimensional (internal) derivatives of type IIB can then be embedded into the 
derivatives of EFT as
 \be\label{solEMbedd}
  \partial_{\alpha}\;\rightarrow \; \partial_{m\alpha} \ = \  
  (\partial_{m'\alpha}\,,\; \partial_{3\alpha}) \ \equiv \ (0\,,\; \partial_\alpha )\;, 
 \ee 
which then solves the section constraint as discussed in the introduction. 

We now verify that the EFT field content, upon taking this solution of the section 
constraint and hence breaking SL$(3)$ to SL(2), precisely reproduces the 
field content of type IIB. First, for the scalar components we count 
 \be
  2+5 \quad (\,{\cal M}_{\alpha\beta}\,,\;{\cal M}_{ij}\,)\qquad \Leftrightarrow \qquad 
  3+2+2 \quad (G_{\alpha\beta}\,,\;{\cal M}_{i'j'}\,,\; \widetilde{A}^{i'}\,)\;, 
 \ee 
finding the same number of components. Indeed, in precise analogy to 
dimensional reduction to $D=8$, the scalars reorganize into 
an SL$(3)\times {\rm SL}(2)/{\rm SO}(3)\times {\rm SO}(2)$ coset space
(although here the SL(3) symmetry is actually broken to SL(2)). 
Next, the EFT vector fields are identified as 
 \be
  A_{\mu}{}^{i\alpha} \ \equiv \ (\,A_{\mu}{}^{i'\alpha}\,,\; A_{\mu}{}^{3\alpha}\,)
  \ \cong \ (\, \widetilde{A}_{\mu}{}^{i'\alpha}\,,\; A_{\mu}{}^{\alpha}\,)\;, 
 \ee
combining the vector components from (\ref{IIBMetric}) and (\ref{IIB2form}). 
The 2-forms of EFT are identified as 
 \be
  B_{\mu\nu i} \ \equiv \ (\, B_{\mu\nu i'}\,,\; B_{\mu\nu 3}\, ) \ \cong \ 
  (\, \widetilde{A}_{\mu\nu i'}\,,\; \widetilde{A}_{\mu\nu}\,)\;, 
 \ee 
combining  the 2-forms from (\ref{IIB2form}) and (\ref{IIB4form}). 
The EFT 3-forms can be directly identified with the 3-forms in (\ref{IIB4form}):
 \be
  C_{\mu\nu\rho}{}^{\alpha} \ \cong \ \widetilde{A}_{\mu\nu\rho}{}^{\alpha}\;. 
 \ee 
Finally, we need to identify the 4-forms. Here there seems to be a mismatch, 
because EFT features the three 4-forms $D_{\mu\nu\rho\sigma}{}^{m}$, 
while type IIB has only the single 4-form given in (\ref{IIB4form}). 
It turns out, however, that upon putting the type IIB solution (\ref{solEMbedd}) 
of the section constraint only one of the three 4-forms in EFT survives. 
To see this note that the 4-form $D$ enters in EFT only under the 
differential $\widehat\partial$, as in the field strength ${\cal J}_{\mu\nu\rho\sigma}$ 
in (\ref{CALLJ}) or in the topological terms (as can be seen 
in the variation (\ref{varitop})).  Using the solution (\ref{solEMbedd}) of the 
section constraint we then compute with (\ref{barpartialD}) for $\widehat\partial D$ 
 \be\label{IIBdisappearance}
  (\widehat\partial D_{\mu\nu\rho\sigma})^{\alpha} \ = \ \epsilon^{\alpha\beta}\partial_{m\beta}
  D_{\mu\nu\rho\sigma}{}^{m} \ = \ 
  \epsilon^{\alpha\beta}\left( \partial_{m'\beta}D_{\mu\nu\rho\sigma}{}^{m'}
  +\partial_{\beta}D_{\mu\nu\rho\sigma}{}^3\right)
  \ = \  \epsilon^{\alpha\beta}\partial_{\beta}D_{\mu\nu\rho\sigma}{}^{3}\;. 
 \ee 
We thus see that only a single 4-form survives in the theory, in 
precise agreement with the field content of type IIB. 
We finally note that in type IIB the 3-forms from (\ref{IIB4form})
are subject to the self-duality constraint originating from the 
self-duality (\ref{IIBselfdual}) of the original 4-form. 
This is again precisely consistent with EFT which postulates 
the self-duality relation (\ref{EFTSELFDUAL}).

Above we have matched the fields of type IIB with those of EFT subjected 
to the second solution of the section constraint. As for $D=11$ supergravity 
this match is somewhat schematic as 
we have not given the precise field redefinitions relating both sets of fields, 
nor have we verified the match of the equations of motion on both sides. 
Again, there is little doubt that this works out in complete parallel to the 
larger duality groups already investigated in the literature, and we 
leave the detailed verification for future work.

\subsection{Remarks on F-theory interpretation}
Let us briefly comment on a possible relation to F-theory, which  
geometrizes the SL(2) of type IIB so that one may ask whether EFT can 
be viewed as an implementation of F-theory. In fact, F-theory has originally been argued for in order 
to explain the duality symmetries of type II strings in a unified geometric way \cite{Vafa:1996xn}. 
For instance, compactifying type IIB and type IIA 
on a 2-torus to $D=8$, the resulting duality group SL$(3,\mathbb{Z})\times {\rm SL}(2,\mathbb{Z})$
has seemingly different origins from the point of view of type IIB or type IIA/M-theory. 
In type IIB, the SL$(3,\mathbb{Z})$ is an enhancement of the SL$(2,\mathbb{Z})$ S-duality 
present in $D=10$, while the SL$(2,\mathbb{Z})$ originates from the diffeomorphisms 
on the 2-torus. In M-theory it is the other way around: the SL$(3,\mathbb{Z})$ originates 
from the diffeomorphisms on a 3-torus, which is the original 2-torus times the M-theory circle, 
while the second SL$(2,\mathbb{Z})$ is a `hidden' symmetry that cannot be understood 
from the symmetries of $D=11$ supergravity before compactification. 
It would clearly be desirable to have a framework in which all these symmetries have a common geometrical 
origin. 

This suggests to think of type IIB as originating, for instance, from a 12-dimensional theory compactified on a two-torus, 
where the S-duality group is the diffeomorphism group of the torus and the axion-dilaton 
$\tau$ is its complex structure.\footnote{In order to geometrize the U-duality symmetries present 
below $D=10$ or $9$ even higher-dimensional spacetimes are needed, as for instance the 
14 dimensions discussed here.}
There are many reasons why this picture cannot be correct in 
any naive sense --- the obvious one being that there simply are no Lorentz invariant 
supersymmetric theories beyond 11 dimensions. Another obstacle is to explain what 
happens to the third degree of freedom of the internal two-dimensional metric, the overall volume, 
that should accompany the complex structure $\tau$. In fact, truncating this degree of freedom 
by hand, setting the volume to a constant, breaks diffeomorphism invariance. 
In other respects the field content of type IIB also does not fit a 12-dimensional interpretation 
in that, for instance, a 4-form in $D=12$ would lead to more fields in $D=10$ than 
just a 4-form. 

For the SL(3)$\times {\rm SL}(2)$ covariant EFT constructed in this paper these obstacles 
are circumvented. The SL(3)$\times {\rm SL}(2)$ symmetries are all on the same footing, 
represented by \textit{generalized diffeomorphisms} on an extended 
6-dimensional space. Because of this, the submatrix of the generalized metric in 
PSL(2)$\subset {\rm SL}(3)$\footnote{Here PSL(2)$\equiv {\rm SL}(2)/\{\pm 1\}$, where one mods out the overall sign 
of ${\cal M}_{i'j'}$ since Im$\,\tau>0$.} 
can be parametrized by $\tau\in\mathbb{H}$ as 
 \be
  {\cal M}_{i'j'} \ = \ \frac{1}{{\rm Im}\,\tau}
    \begin{pmatrix}    |\tau|^2 & -{\rm Re}\,\tau \\[0.5ex]
    -{\rm Re}\,\tau & 1  \end{pmatrix}\;. 
 \ee   
This metric has determinant 1 and so carries only two degrees of freedom, but now this \textit{is}
consistent with  the \textit{generalized} notion of diffeomorphisms, as discussed in this paper. 
Moreover, as we saw in the previous subsection, the field content matches type IIB in 
general. This is possible, because the theory is not a diffeomorphism invariant theory 
in 14 dimensions. It does have a 14-dimensional group of generalized diffeomorphisms  
but these are split as $8+6$ in such a way that they do not reorganize into 
14-dimensional conventional diffeomorphisms (although they do combine either into
10- or 11-dimensional conventional diffeomorphisms plus tensor gauge transformations 
for the appropriate solutions of the section constraint). 
Finally, although here we discussed only the bosonic theory, there is no doubt that it
can also be made supersymmetric, as has been done for the E$_{7(7)}$ and E$_{6(6)}$ 
cases \cite{SUSYE7,Musaev:2014lna}. 

It should be emphasized that in the modern view of F-theory the extra two dimensions play 
an auxiliary role in that no fields depend 
on the coordinates corresponding to this torus. Rather, one considers compactifications on 
a space that is a  2-torus which is fibered over a base manifold in the sense that 
$\tau$ depends on the coordinates of the base. 
(This dependence is usually such that 
$\tau$ is only defined up to SL$(2,\mathbb{Z})$ transformations. For instance, at locations corresponding 
to D7 branes $\tau \rightarrow \tau+1$.) This auxiliary nature of the extra dimensions is also in line with that in EFT:
although the section constraint implies that fields never depend on more coordinates than 
present in supergravity it does allow for non-standard compactification ansaetze, 
with a non-trivial dependence of the generalized metric on the internal coordinates. 
 
The interesting question therefore is whether the formalism of EFT could be useful 
in analyzing certain F-theory compactifications. For instance, one often uses the 
M-theory/F-theory duality, performing an M-theory compactification followed by a T-duality 
transformation mapping it to type IIB \cite{Denef:2008wq,Grimm:2010ks}. 
As in EFT these dualities as well as the mapping from 
M-theory to type IIB are manifest one may wonder whether EFT provides a technical 
simplification. Moreover, one may speculate that the necessary SL$(2,\mathbb{Z})$ transformations 
at the locations of 7-branes can be captured in `non-geometric' spaces of the type 
appearing in DFT, see \cite{Hohm:2012gk,Hohm:2013bwa,Berman:2014jba},\footnote{See also 
\cite{Berman:2014hna}, where it has been argued that spaces that are singular in conventional 
geometry become non-singular in EFT.} 
possibly permitting transformations $\tau\rightarrow -\frac{1}{\tau}$ characteristic of 
non-perturbative phenomena. 
It should be stressed, however, that F-theory is meant to capture non-perturbative type IIB string theory
more generally,  for instance describing gauge fields corresponding to   
enhanced gauge symmetries such as E$_8$. Most likely, such effects cannot be 
seen directly in the EFTs constructed so far, but it would be interesting to see whether 
EFT can play a technically useful role for F-theory analogously to that of 11-dimensional supergravity 
for M-theory. Clearly, this requires the construction of explicit examples.

\section{Conclusions}
In this paper we constructed the EFT for the duality group SL$(3)\times {\rm SL}(2)$, based on a
$8+6$ dimensional generalized spacetime. Compared to the previous constructions of EFTs for larger 
duality groups, the main technical novelty of our investigation is 
the systematic construction of the tensor hierarchy beyond 1- and 2-forms. 
To this end we developed a novel Cartan-like tensor calculus, based on a covariant 
differential operator $\widehat\partial$ acting on specific  SL$(3)\times {\rm SL}(2)$ representation spaces, 
which is intriguingly analogous to that of standard differential forms. 
To our knowledge such a calculus has not been investigated in the mathematical literature 
and so it would be interesting to further elucidate its properties. In particular, 
it should be beneficial to study the $\widehat\partial$ cohomology, whose subtleties we discussed 
in the main text. There is no general Poincar\'e lemma for the strongly constrained theory
and it would be interesting to understand the significance of this observation,  
perhaps shedding some light on the geometric meaning of the section constraint. 
Moreover, this calculus should have straightforward extensions to the duality groups for which 
the corresponding EFTs so far have been constructed for the internal sector (e.g.~E$_{5(5)}={\rm SO}(5,5)$ and 
E$_{4(4)}={\rm SL}(5)$ \cite{Berman:2010is,Berman:2011pe}). 

There are several potential applications of the  SL$(3)\times {\rm SL}(2)$ EFT.
Most importantly, it is an efficient starting point for non-trivial compactifications to  
$D=8$. In fact, it has recently been shown how compactifications on a large class 
of curved internal manifolds can be described very efficiently in EFT in the form 
of generalized Scherk-Schwarz compactifications \cite{Hohm:2014qga} 
(extending earlier results in DFT \cite{Hohm:2011cp,Aldazabal:2011nj,Geissbuhler:2011mx}). 
For the present theory they  
would be governed by  SL$(3)\times {\rm SL}(2)$ valued 
$6\times 6$ `twist' matrices. They may provide an interesting playground for 
non-trivial (possibly non-geometric or F-theory like) compactifications 
as toy models for more involved reductions to lower dimensions. 
We leave such investigations for future work.

\subsection*{Acknowledgements}
It is a pleasure to thank Barton Zwiebach for collaboration in an early stage of this work 
and Henning Samtleben for helpful discussions. We would also like to thank  
Thomas Grimm, Ashoke Sen, Wati Taylor and Barton Zwiebach 
for explanations and discussions on the notion of `F-theory'.  

This work is supported by the U.S. Department of Energy (DoE) under the cooperative research agreement DE-FG02-05ER41360. The work of O.H. is supported by a DFG Heisenberg fellowship.

\providecommand{\href}[2]{#2}\begingroup\raggedright\endgroup

\end{document}